# An Introduction to Adjoints and Output Error Estimation in Computational Fluid Dynamics

by

Steven M. Kast

Ph.D., Aerospace Engineering
University of Michigan

The original version of these notes can be found in the dissertation

*Methods for Optimal Output Prediction in Computational Fluid Dynamics*

Steven M. Kast
University of Michigan
May 2016

# TABLE OF CONTENTS

# CHAPTER I

# Introduction

In recent years, the use of adjoint vectors in Computational Fluid Dynamics (CFD) has seen a dramatic rise. Their utility in numerous applications, including design optimization, data assimilation, and mesh adaptation has sparked the interest of both researchers and practitioners alike.

In many of these fields, the concept of an adjoint is explained differently, with various notations and motivations employed. Further complicating matters is the existence of two seemingly different types of adjoints – "continuous" and "discrete" – as well as the more formal definition of adjoint operators employed in linear algebra and functional analysis. These issues can make the fundamental concept of an adjoint difficult to pin down.

In these notes, we hope to clarify some of the ideas surrounding adjoint vectors and to provide a useful reference for both continuous and discrete adjoints alike.

In particular, we focus on the use of adjoints within the context of output-based mesh adaptation, where the goal is to achieve accuracy in a particular quantity (or *output*) of interest by performing targeted adaptation of the computational mesh. While this is our application of interest, the ideas discussed here apply directly to design optimization, data assimilation, and many other fields where adjoints are employed.

# CHAPTER II

# Adjoints and Error Estimation

We begin with a discussion of adjoint vectors in the context of steady-state partial differential equations. We first derive adjoints in both a discrete and continuous context, then show how they can be used to compute output error estimates and perform output-based mesh adaptation.

For additional information, see Becker and Rannacher [1], Hartmann and Houston [9], Giles and Pierce [6], and Estep [3], among others.

## 2.1 Discrete Adjoints

Imagine we have a differential equation of the form

$$Lu = f\,, \tag{2.1}$$

where $L$ is some linear differential operator (e.g. $L \equiv \frac{\partial}{\partial x}$ or $L \equiv \nabla^2$), $f$ is a prescribed source term, and $u$ is the unknown solution, defined on some domain $\Omega$. For most operators $L$ (combined with appropriate boundary conditions), this equation would be difficult to solve analytically, and we must instead approximate the solution numerically.

After discretizing the above equation with an appropriate numerical method (such as a finite difference or finite element method), we will arrive at a system of equations of the form

$$\mathbf{A}\mathbf{U} = \mathbf{F}\,, \tag{2.2}$$

where $\mathbf{A} \in \mathbb{R}^{N\times N}$ is a matrix representing the operator $L$, $\mathbf{U} \in \mathbb{R}^N$ is a vector of unknowns representing $u$, and $\mathbf{F} \in \mathbb{R}^N$ is a vector of source terms and boundary data.

Now, if we are interested in computing the *entire* solution $\mathbf{U}$ – i.e. in finding all components of the $\mathbf{U}$ vector – then we will effectively need to know the entire $\mathbf{A}^{-1}$ matrix, since we would then compute $\mathbf{U}$ from

$$\mathbf{U} = \mathbf{A}^{-1}\mathbf{F}\,. \tag{2.3}$$

However, what if, instead of the entire solution, we are interested in just a single *component* of $\mathbf{U}$? In that case, would we still need to know the entire $\mathbf{A}^{-1}$ matrix, or would less information suffice?

For example, imagine that we are interested in computing (say) $U_N$, the last entry in the $\mathbf{U}$ vector. This situation is depicted in the diagram below.

$$\underbrace{\begin{bmatrix} U_1 \\ U_2 \\ \vdots \\ U_N \end{bmatrix}}_{\mathbf{U}} = \underbrace{\begin{bmatrix} \bullet & \bullet & \cdots & \bullet \\ \bullet & \bullet & \cdots & \bullet \\ \vdots & \vdots & & \\ \bullet & \bullet & \cdots & \bullet \end{bmatrix}}_{\mathbf{A}^{-1}} \underbrace{\begin{bmatrix} F_1 \\ F_2 \\ \vdots \\ F_N \end{bmatrix}}_{\mathbf{F}} \tag{2.4}$$

By the properties of matrix multiplication, it is clear that to compute $U_N$ we require knowledge of only a single *row* of $\mathbf{A}^{-1}$. (In this case, the last row – as highlighted above.) This row, when combined with the source term $\mathbf{F}$, contains all the information we need to compute our desired "output" $U_N$.

Now, for our purposes, there is another property of this row that is even more important. Imagine, for example, that the first component of the highlighted row were zero. Then if we were to change the source term $F_1$, the value of $U_N$ would not change at all (since $F_1$ would just be multiplied by 0 during the computation of $U_N$). In other words, we could say that our output $U_N$ is *insensitive* to changes in $F_1$. In a similar manner, this logic could be applied to all entries in the highlighted row: the smaller a given entry, the less sensitive $U_N$ is to changes in the corresponding $F$ component, and likewise, the larger a given entry, the more sensitive it is. In summary then, not only does the highlighted row provide the information required to compute $U_N$, it also provides the ***sensitivity*** of $U_N$ to perturbations in the source terms $F$.

This row then – this sensitivity vector – is what is commonly referred to in the literature as the "**dual vector**," the "**output adjoint vector**," or simply, the "**adjoint**." For a given output of interest (in this case $U_N$), it provides the sensitivity of that output to perturbations in the source terms of the governing equations.

In this work, we will denote the adjoint vector by the symbol $\mathbf{\Psi}$ and the corresponding "**output of interest**" by the symbol $J$, as depicted in the diagram below.

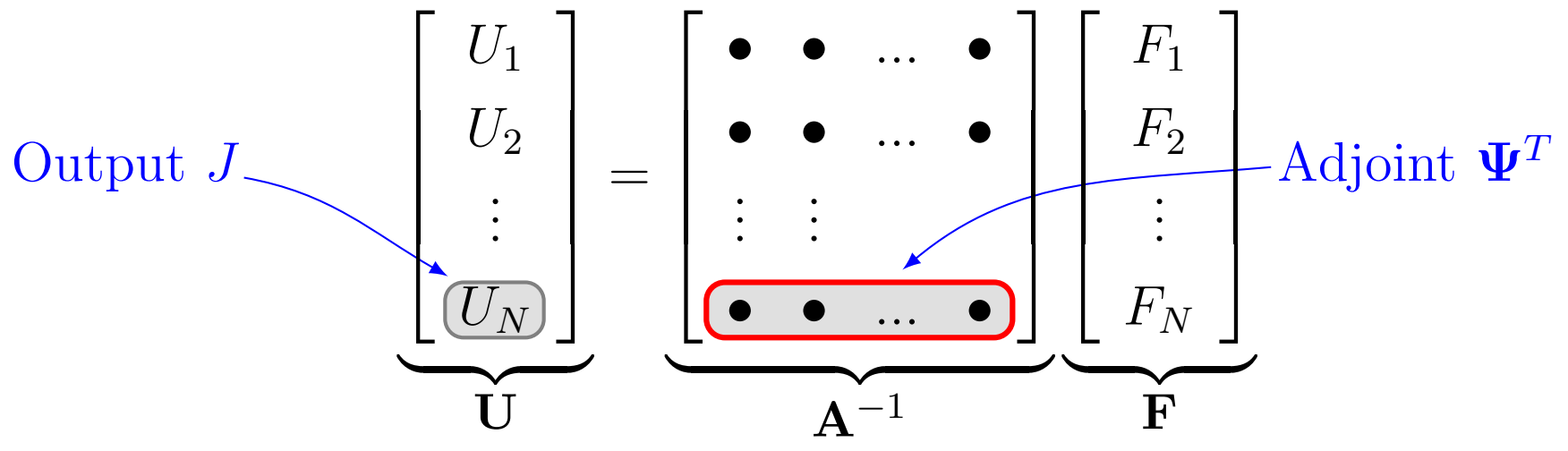

(2.5)

Note that we will use $\mathbf{\Psi}^T$ to refer to the adjoint in row-vector form, whereas $\mathbf{\Psi}$ alone will denote the adjoint in column form.

Using the above notation, an important point is that the output $J$ can be written as the inner product between the adjoint and the source vector, i.e. as

$$J = \mathbf{\Psi}^T \mathbf{F} \ . \tag{2.6}$$

This is known as the "**dual form**" of the output, and is just a mathematical restatement of our earlier claim that the only information required to compute the output is the highlighted row and the source vector $\mathbf{F}$.

The next question is: how should we define the adjoint itself? So far, we have just labeled it as the last row of $\mathbf{A}^{-1}$, but is there a way to define it mathematically?

It turns out that, in order to define the adjoint formally, it will be helpful to first compute the derivative of the output $J$ with respect to $\mathbf{U}$. In this case, since $J = U_N$, we can write its derivative with respect to $\mathbf{U}$ as

$$\frac{\partial J}{\partial \mathbf{U}} \equiv \begin{bmatrix} \frac{\partial J}{\partial U_1} & \frac{\partial J}{\partial U_2} & \ldots & \frac{\partial J}{\partial U_N} \end{bmatrix} = [0 \;\; 0 \; \ldots \; 1] \ . \tag{2.7}$$

This is just the Cartesian row vector with the last entry nonzero. In the present case, we see that the adjoint $\mathbf{\Psi}^T$ can then be defined as

$$\mathbf{\Psi}^T = \frac{\partial J}{\partial \mathbf{U}} \mathbf{A}^{-1} \ . \tag{2.8}$$

Here, multiplying $\mathbf{A}^{-1}$ by the Cartesian row vector simply picks off the last row of $\mathbf{A}^{-1}$ and calls it the adjoint, $\mathbf{\Psi}^T$.

So far, we do not seem to have gained much from the above derivations. However, an important fact is that, while this $J = U_N$ example may seem trivial, both Eqn. 2.8 (the so-called "**adjoint equation**") and Eqn. 2.6 (the "**dual form**") hold *regardless* of what the desired output is.[1] For example, rather than a "single-component" output like $J = U_N$, we might instead be interested in computing an average or sum of certain components of $\mathbf{U}$. One possibility would be the average of all components, i.e.

$$J = \frac{1}{N} \sum_{i=1}^{N} U_i \ . \tag{2.9}$$

In that case, we would have

$$\frac{\partial J}{\partial \mathbf{U}} = \frac{1}{N} [1 \ \ 1 \ ... \ 1] \ , \tag{2.10}$$

and from Eqn. 2.8, applying this vector to $\mathbf{A}^{-1}$ would then tell us that the adjoint of $J$ (i.e. its sensitivity to perturbations in the source terms) is just the corresponding average of all rows of $\mathbf{A}^{-1}$.

In the same way, the sensitivity of *any* scalar output can be represented as a weighted average of the rows of $\mathbf{A}^{-1}$, in accordance with Eqn. 2.8. As we will see, in a CFD simulation where $\mathbf{U}$ may represent the density or momentum of a fluid, these scalar outputs will tend to be quantities of physical importance, such as lift, drag, moment, or heat flux.

### 2.1.1 Alternative (CFD) Notation

Before moving on, let us rewrite the adjoint equation (Eqn. 2.8) in a form typically seen in Computational Fluid Dynamics. In CFD, the governing equations

$$\mathbf{A}\mathbf{U} = \mathbf{F}$$

(i.e. Eqn. 2.2), would typically be written as a set of discrete "residuals," where the residual vector $\mathbf{R} \in \mathbb{R}^N$ is defined as:

$$\mathbf{R}(\mathbf{U}) \equiv \mathbf{A}\mathbf{U} - \mathbf{F} \, . \tag{2.11}$$

The governing equations are therefore satisfied when the residual vector is zero.

[1]Strictly speaking, Eqn. 2.6 (the dual form) holds only if $J$ is a *linear* combination of the components of $\mathbf{U}$, though Eqn. 2.8 (the adjoint equation) holds even in the nonlinear case, since the adjoint is always defined to be a linear sensitivity vector.

Taking the derivative of $\mathbf{R}$ with respect to $\mathbf{U}$, we see that

$$\frac{\partial \mathbf{R}}{\partial \mathbf{U}} = \mathbf{A}\,. \tag{2.12}$$

Thus, we can say that the $\mathbf{A}$ matrix corresponds to the "residual Jacobian" matrix, $\partial\mathbf{R}/\partial\mathbf{U}$. Inserting this Jacobian matrix into the adjoint equation (Eqn. 2.8) then gives

$$\boldsymbol{\Psi}^T = \frac{\partial J}{\partial \mathbf{U}} \frac{\partial \mathbf{R}}{\partial \mathbf{U}}^{-1}\,. \tag{2.13}$$

To obtain a more common form of this equation, we transpose both sides to get

$$\boldsymbol{\Psi} = \frac{\partial \mathbf{R}}{\partial \mathbf{U}}^{-T} \frac{\partial J}{\partial \mathbf{U}}^{T}\,. \tag{2.14}$$

Finally, multiplying both sides by the Jacobian transpose gives

$$\frac{\partial \mathbf{R}}{\partial \mathbf{U}}^{T} \boldsymbol{\Psi} = \frac{\partial J}{\partial \mathbf{U}}^{T}\,. \tag{2.15}$$

This is the adjoint equation often seen in the literature.[2]

By comparing both sides of this equation, we see that in order for the left-hand side to have dimensions of $\partial J/\partial\mathbf{U}$, the adjoint must behave like

$$\boldsymbol{\Psi} \approx \frac{\partial J}{\partial \mathbf{R}}\,. \tag{2.16}$$

Thus, in general, we can say that the adjoint represents the **sensitivity** of an output to perturbations in the **residuals**. (Note that previously we described the adjoint in terms of perturbations to $\mathbf{F}$, but since perturbations in $\mathbf{F}$ lead directly to perturbations in the residuals, these are different ways of saying the same thing.[3])

In the end, Eqn. 2.15 is the same as that in the previous section (Eqn. 2.8), but with one additional benefit. By writing the adjoint equation in terms of a residual Jacobian matrix as opposed to an $\mathbf{A}$ matrix, we have effectively extended its definition to problems with *nonlinear* residuals as well. In that case, although we could not write the residual itself as $\mathbf{R} = \mathbf{A}\mathbf{U} - \mathbf{F}$, we *could* still compute its derivative $\partial\mathbf{R}/\partial\mathbf{U}$ about a given location in state space (just as we could compute the derivative of any nonlinear function – say, a quadratic – at a given point in space). From Eqn. 2.15, this first derivative – this Jacobian matrix – is all that is required to define the adjoint.

[2]Note that the adjoint equation is sometimes defined with a negative sign on $\frac{\partial J}{\partial \mathbf{U}}^T$. This is a convention often used within the field of optimization. In that case, the adjoint would represent the output sensitivity to perturbations on the left-hand rather than right-hand side of the residuals.

[3]Except for a difference in sign, since $\mathbf{F}$ enters the residual with a negative sign.

Thus, for **nonlinear** problems, we would write the adjoint equation as

$$\left.\frac{\partial \mathbf{R}}{\partial \mathbf{U}}^T\right|_{\mathbf{U}} \mathbf{\Psi} = \left.\frac{\partial J}{\partial \mathbf{U}}^T\right|_{\mathbf{U}} . \tag{2.17}$$

where the Jacobian and output linearization are evaluated at a particular state, $\mathbf{U}$.

Finally, we note that, in general, there are several means by which $\frac{\partial \mathbf{R}}{\partial \mathbf{U}}$ and $\frac{\partial J}{\partial \mathbf{U}}$ can be computed. These include finite differencing of the discrete residuals and output, automatic code differentiation, and analytic differentiation. While the former methods are less time-intensive to implement, analytic differentiation is more accurate than finite differencing and typically more efficient than automatic differentiation, making it our preferred method.

### 2.1.2 Adjoint Example

At this point, it is worth giving a practical example. Here, we show results from a simulation of the Navier-Stokes equations around an airfoil. In this case, the flow moves upward and to the right at an angle of attack of 5° and a Reynolds number of 5000, with contours of the $x$-direction momentum shown in Fig. 2.1a. A low-velocity wake forms behind the airfoil, as indicated by the blue region in the figure.

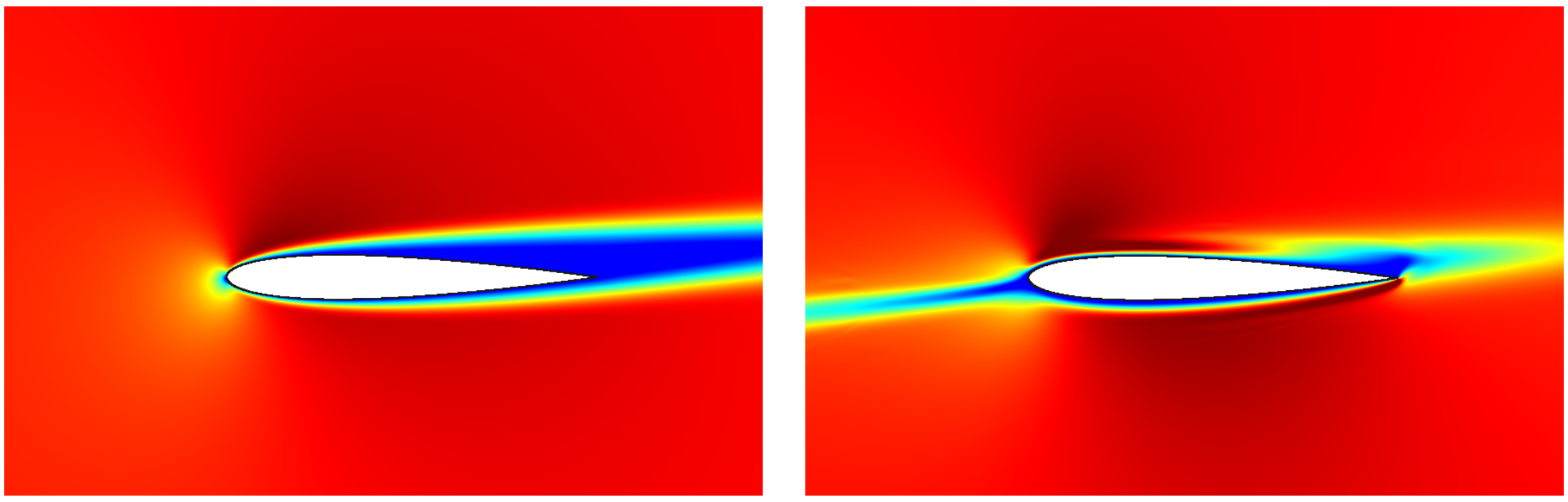

(a) $x$-momentum of fluid (b) cons. of $x$-momentum drag adjoint

Figure 2.1: Contours of (a) $x$-momentum ([blue, dark red] = [-0.035, 0.55]) and (b) the conservation of $x$-momentum component of the drag adjoint ([blue, dark red] = [-1.05, 0.15]) for $Re$ = 5000 flow around an airfoil. The drag is most sensitive to $x$-momentum residual perturbations made in the dark blue/red regions of (b).

If we are interested in the drag on the airfoil, we can define this drag to be our "output" $J$ and can compute a corresponding adjoint from Eqn. 2.17. Looking back at the equations above (e.g. Eqn. 2.5) it is clear that the adjoint is always a vector of the same dimension as the state vector, $\mathbf{U}$. Thus, if we can plot contours of $\mathbf{U}$, we can plot contours of $\boldsymbol{\Psi}$ as well.[4] These contours are shown in Fig. 2.1b. (Specifically, we plot the "conservation of $x$-momentum" component of the adjoint, which represents the sensitivity of the drag to perturbations in the $x$-momentum residuals. The adjoint also has conservation of mass, $y$-momentum, and energy components.)

The dark blue and dark red regions in Fig. 2.1b are those of strong positive and negative output sensitivity, respectively. Any residual perturbations made in these regions would therefore have a large effect on the final drag value, whereas perturbations made in the lighter red regions (where the adjoint is near zero) would have a small influence on the drag.

One interesting feature of the adjoint is the blue region extending leftward from the airfoil leading edge. This adjoint "wake" travels in the opposite direction to the fluid wake, and is representative of a general feature of adjoints: namely, that information in the adjoint problem flows in the ***opposite*** direction as information in the original ("primal") problem. While we will discuss this in more detail later, here, the existence of an adjoint wake on the left makes sense, since any source perturbations made in this region will flow rightward and collide with the airfoil, ultimately having a strong influence on the drag value.

Finally, one potentially suprising fact about the adjoint is that it is ***smooth***. It is not obvious that this should be the case. Upon plotting its contours, we may have expected to see more or less random "spikes" in sensitivity. However, this smoothness can be explained by the fact that, in the end, perturbations in the residuals propagate via physical mechanisms (e.g. acoustic waves or convective motion) on their way to influencing the output. Since most physical modes of propagation are by nature "smooth," this smoothness is reflected in the adjoint. Indeed, as we will discuss later, it turns out that just as the discrete system $\mathbf{AU} = \mathbf{F}$ approximates a continuous (i.e. smooth) primal PDE, the discrete adjoint equations – if properly posed – approximate a **continuous *adjoint* PDE**.

[4]Note that we are using a finite element formulation here, so plotting $\boldsymbol{\Psi}$ requires first multiplying it by the set of finite element basis functions. For non-variational methods the components of $\boldsymbol{\Psi}$ may have to be scaled in other ways before plotting.

### 2.1.3 Applications of Adjoints

While the idea of an adjoint vector is relatively straightforward, this concept lies at the heart of several active fields of research, including design optimization, mesh adaptation, error estimation, data assimilation, and uncertainty quantification. In this section, we discuss briefly the role of adjoints in the context of design optimization and mesh adaptation.

#### 2.1.3.1 Design Optimization

In the field of optimization, the goal is typically to minimize or maximize some output of interest by systematically modifying (or "optimizing") certain parameters of the problem at hand. For example, in aircraft design, the goal may be to minimize the drag of an aircraft by finding the optimal fuselage and/or wing shape at a given angle of attack.

For these problems, where "**gradient-based**" techniques are often employed, the adjoint plays a critical role. This is because any changes in the design parameters (e.g. the aircraft geometry) would result in corresponding changes to the residuals of the governing equations. But since the adjoint provides the sensitivity of an output to perturbations in the residuals, it therefore indicates whether a given design change would cause the output value to increase or decrease. (Most importantly, it provides this sensitivity without any additional solves of the primal problem, which would otherwise be required if a "forward" sensitivity method were used.) Thus, if the goal is to e.g. minimize drag, design changes can be made that – according to the adjoint – lead to a reduction in drag. By iterating this process until the design parameters converge, the "optimal" design that minimizes drag[5] can be found.

#### 2.1.3.2 Error Estimation and Mesh Adaptation

Just as design modifications can be treated as perturbations to the residuals, so can numerical errors. In a CFD simulation, the solution on a given mesh will not satisfy the differential (or "continuous") form of the governing equations exactly. Instead, small **truncation errors** will exist throughout the domain, which can be viewed as source-term perturbations to the local (continuous) **residual**. For a given output of interest, the adjoint then indicates how these residual perturbations translate into errors in the final output value.

[5] At least in a local sense – a global minimum is not guaranteed.

Furthermore, in addition to providing an estimate for the *total* output error, the adjoint also indicates *where* in the domain this error originates from. Roughly speaking, if truncation errors (i.e. residual perturbations) occur in a region where the adjoint is large, then these perturbations are likely contributing a significant amount to the output error. To reduce the output error, we should therefore attempt to reduce these residual perturbations, which can be done via local mesh refinement. Thus, the adjoint – or more accurately, the adjoint multiplied by the local residuals – can serve as an effective **mesh adaptation indicator**. This "**adjoint-weighted residual**" (or "dual-weighted residual") method forms the basis for many output-based adaptation algorithms in the literature.

## 2.2 Continuous Adjoints

In the above section, we have given an overview of the discrete adjoint and its applications within Computational Fluid Dynamics. As mentioned, the discrete adjoint – while useful in its own right – turns out to be an approximation to a more fundamental concept: the continuous adjoint. In this section, we provide an overview of continuous adjoints and their associated operators, which play a prominent role in both theory and practice.

To introduce the relevant ideas, let us return to the differential equation from the previous section:

$$Lu = f\,, \tag{2.18}$$

where $L$ is a linear differential operator, $u$ is the unknown solution, $f$ is a source term, and suitable boundary conditions are imposed.

To make the discussion more concrete, we will initially take $L \equiv a\frac{\partial()}{\partial x}$ and consider the following advection equation,

$$a\frac{\partial u}{\partial x} = f(x) \qquad x \in \Omega \tag{2.19}$$

$$u = 0 \qquad x = x_L\,, \tag{2.20}$$

where $a$ is a positive scalar and a homogeneous Dirichlet condition is imposed on the left side of the domain, defined as $\Omega \equiv [x_L, x_R]$. Here, since the advection speed $a$ is positive, information propagates from left to right, making the specification of a boundary condition on the left well-posed.

Now, imagine that, as in the previous section, we are interested not in solving for the *entire* solution $u$, but instead in computing a certain scalar **output** (or "functional") of the form:

$$J = \int_{\Omega} g(x)\, u(x)\, dx \;\equiv\; (g, u) \;. \tag{2.21}$$

For example, if – as in the discrete adjoint section – we are interested in computing the value of $u$ at a single point (say, $x_p$) then we would choose $g(x) \equiv \delta(x - x_p)$. By the properties of the Dirac delta function, our output of interest would then be

$$J = \int_{\Omega} g(x)\, u(x)\, dx \;=\; \int_{\Omega} \delta(x - x_p)\, u(x)\, dx \;=\; u(x_p)\,. \tag{2.22}$$

Alternatively, if we were interested in computing (e.g.) the average of the solution over the domain, we would simply choose $g(x) \equiv 1$.

For any desired choice of the output $J$ (and hence $g$)[6] the continuous adjoint[7] for that output would satisfy the following differential equation:

$$-a\frac{\partial \psi}{\partial x} = g(x) \qquad\qquad x \in \Omega \tag{2.23}$$

$$\psi = 0 \qquad\qquad x = x_R\,. \tag{2.24}$$

Here, we have labeled the **continuous adjoint** '$\psi$', similar to the notation for the discrete adjoint, $\boldsymbol{\Psi}$.

Looking at the above **adjoint equation** (Eqn. 2.23), we see that it is very similar to the primal equation (Eqn. 2.19), with the only differences being the presence of $g(x)$ rather than $f(x)$, the inclusion of a negative sign in front of $a$, and the imposition of a boundary condition on the *right* rather than the left side of the domain. (Note that these last two differences indicate a **reversal** in the direction of **information flow** compared to the primal problem – as mentioned in Sec. 2.1.2 above.)

Now, what is the significance of this equation, and why should we call $\psi$ an "adjoint"? Consider again the definition of our desired output, from which – in light

[6]Except $g = \delta(x - x_R)$, which lies on the boundary.

[7]Note that the continuous adjoint is sometimes referred to as a ***generalized Green's function*** due to its similarity to standard Green's functions [4].

of Eqn. 2.23 – we can perform the following manipulations:

$$
\begin{aligned}
J &= \int_\Omega g(x)\, u\, dx && \text{(output definition)} \\
&= \int_\Omega -a\frac{\partial \psi}{\partial x}\, u\, dx && \text{(by adjoint equation, Eqn. 2.23)} \\
&= \int_\Omega \psi\, a\frac{\partial u}{\partial x}\, dx - a\psi u\Big|_{x_L}^{x_R} && \text{(integrate by parts)} \\
&= \int_\Omega \psi\, a\frac{\partial u}{\partial x}\, dx && \text{($\psi$ and $u$ BCs = 0)} \\
&= \int_\Omega \psi\, f(x)\, dx\,. && \text{(by primal equation, Eqn. 2.19)} \qquad (2.25)
\end{aligned}
$$

Or, using bracket notation to denote the $L^2$ inner product, we can write simply

$$J = (\psi, f)\;. \qquad (2.26)$$

This is the so-called ***dual form*** of the output, which is analogous to the one derived in the discrete case (i.e. Eqn. 2.6). As in the discrete case, it says that the only information required to compute the desired output is the adjoint $\psi$ and the source term $f$. Furthermore, just as the discrete adjoint $\mathbf{\Psi}$ represents the sensitivity of an output to perturbations in the discrete source term $\mathbf{F}$, the continuous adjoint $\psi$ represents the sensitivity of an output to perturbations in the *continuous* source term $f$.

For the current problem, the continuous adjoint associated with the output $J = u(x_p)$ is plotted in Fig. 2.2. For this output, we have $g = \delta(x - x_p)$, so the solution to the adjoint equation (Eqn. 2.23) is just the step function.[8] From the figure, we see that $\psi$ is nonzero only in the region upstream of $x_p$, since any perturbations made downstream of $x_p$ cannot influence the output value. This confirms our earlier statement that $\psi$ should represent the sensitivity of the output to perturbations made in various regions of the domain.

### 2.2.1 Generalization of the Continuous Adjoint

Now, let us generalize the continuous adjoint to cases beyond advection. Looking back at the dual form derivation (Eqn. 2.25), we see that it hinges on the action performed between steps 2 and 3 – i.e. the integration by parts. The key point is that the operator acting on the adjoint $\psi$ should, after integration by parts, recover the primal differential operator. (E.g, in this case, the $-a\frac{\partial \psi}{\partial x}$ term yields the primal term $a\frac{\partial u}{\partial x}$ after integration by parts.)

[8]Note that the "continuous" adjoint is not necessarily a continuous function itself.

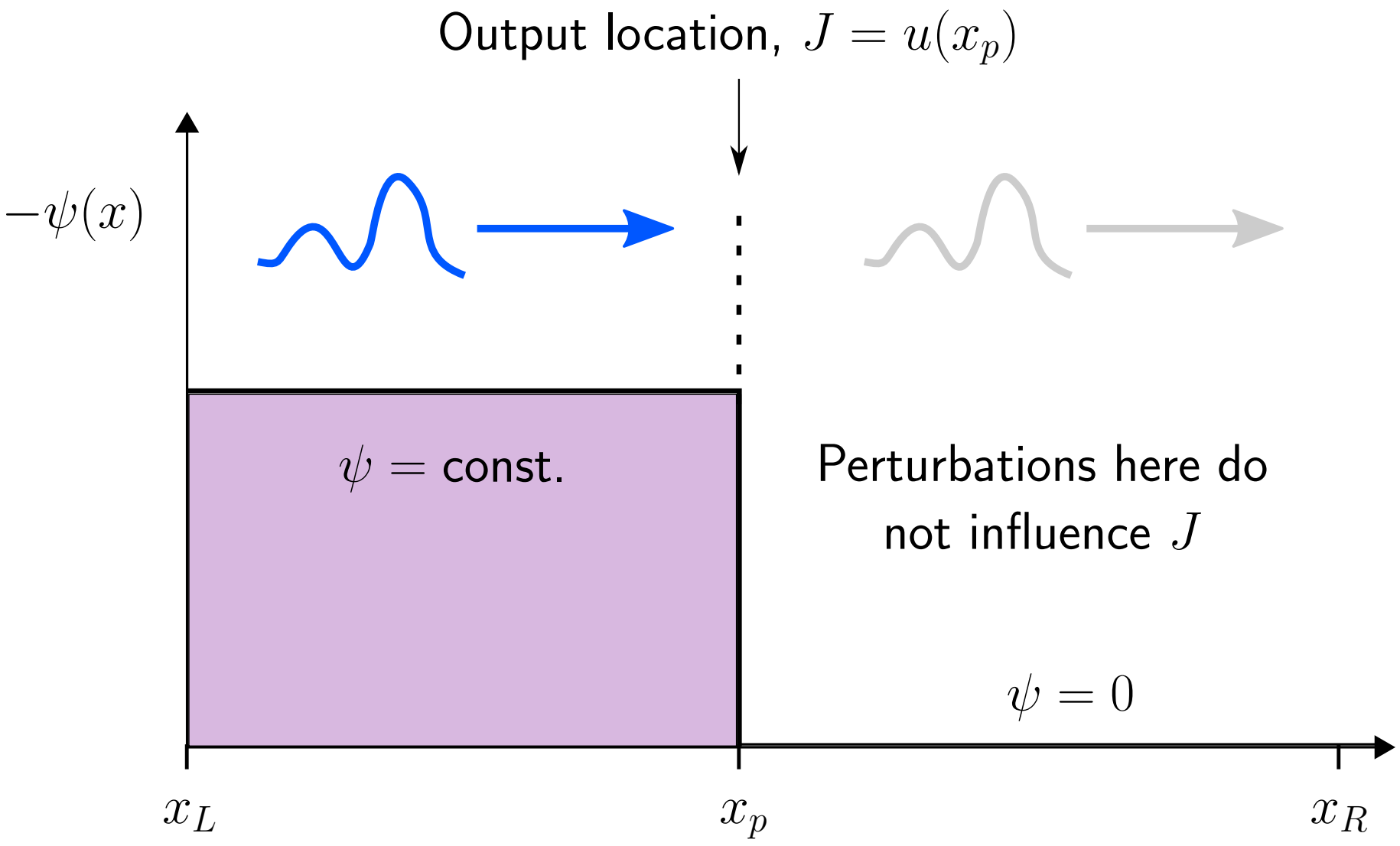

Figure 2.2: 1D advection: The adjoint associated with the output $J = u(x_p)$. Here, primal source perturbations (indicated by the squiggle) travel rightward, so only perturbations made to the left of $x_p$ can influence the output value. Thus, for $x > x_p$, $\psi = 0$. Furthermore, since all perturbations made in $x_L < x < x_p$ will propagate to $x_p$ eventually, the output is equally sensitive to all of them, meaning $\psi =$ constant in this region.

In general, let us denote the operator acting on $\psi$ as $L^*$. Then, for a given primal operator $L$, we need this $L^*$ operator to satisfy the following identity:

$$\int_\Omega (Lu)\,\psi\,dx \;=\; \int_\Omega u\,(L^*\psi)\;dx \qquad\qquad \forall\,\text{suitable } u,\psi \tag{2.27}$$

This says that, if we have a differential term $Lu$ integrated against a function $\psi$, then $L^*$ is the operator that acts on $\psi$ after we integrate by parts enough times to remove all derivatives from $u$.[9] (As suggested above, the integration by parts can also be thought of as occuring in the reverse direction, from $L^*$ to $L$. However, since we typically know $L$ from the primal equation, in practice we integrate the left-hand side

[9]Note that all boundary terms vanish in the integration by parts since we assume homogeneous boundary conditions on $u$ and $\psi$.

of Eqn. 2.27 by parts in order to determine $L^*$.)

For example, if $L \equiv \frac{\partial()}{\partial x}$, only one integration by parts in Eqn. 2.27 is required to remove the derivative on $u$. Since this integration by parts introduces a negative sign, the $L^*$ in Eqn. 2.27 then turns out to be $L^* = -L \equiv -\frac{\partial()}{\partial x}$. However, if we instead had the operator $L \equiv \frac{\partial^2()}{\partial x^2}$, then two integrations by parts would be required, and the two subsequent negative signs would cancel, yielding $L^* = L \equiv \frac{\partial^2()}{\partial x^2}$. This trend holds in general: for any **odd-derivative** terms in $L$, $L^*$ will contain the ***negative*** of these terms, and for any **even-derivative** terms in $L$, $L^*$ will contain the ***same*** (unmodified) terms.[10]

To generalize a bit further, we can write the above relation (Eqn. 2.27) in inner product notation as

$$(Lu, v) \;=\; (u, L^* v) \qquad\qquad \forall u, v \in V \quad , \tag{2.28}$$

where $V$ is a suitable function space over which the above inner product (which could be, e.g., the $L^2$ inner product) is defined. In functional analysis, this relation is known as the **adjoint identity**. For a given operator $L$, it serves as the definition of the so-called ***adjoint operator***, $L^*$. As suggested, this $L^*$ is exactly the operator we need to define the continuous adjoint $\psi$.

We can now summarize the discussion. For a **primal** differential equation

$$Lu = f \qquad\qquad x \in \Omega \tag{2.29}$$

$$\text{primal b.c.} \qquad\qquad x \in \partial\Omega \tag{2.30}$$

and an output of interest $J = (g, u)$, the **continuous adjoint** equation is defined as

$$L^*\psi = g \qquad\qquad x \in \Omega \tag{2.31}$$

$$\text{adjoint b.c.} \qquad\qquad x \in \partial\Omega \tag{2.32}$$

where $L^*$ is the formal adjoint operator defined by Eqn. 2.28, and the adjoint boundary conditions are chosen such that Eqn. 2.28 is satisfied.[11]

With the adjoint problem defined in this way, the following derivation can then

[10]These even-derivative terms are the so-called “self-adjoint” terms.

[11]We will discuss this more later – it is not always *strictly* true when the adjoint boundary conditions are non-homogeneous.

be performed:

$$\begin{aligned} J &= (g, u) && \text{(output definition)} \\ &= (L^*\psi, u) && \text{(by adjoint equation, Eqn. 2.31)} \\ &= (\psi, Lu) && \text{(by adjoint identity, Eqn. 2.28)} \\ &= (\psi, f)\,. && \text{(by primal equation, Eqn. 2.29)} \end{aligned} \tag{2.33}$$

Thus, the above definition of the adjoint problem allows us to write the output in **dual form**, from which it follows that $\psi$ represents the **sensitivity** of the output to perturbations in $f$.

### 2.2.2 Further Generalization of the Continuous Adjoint

So far, we have avoided a detailed discussion of boundary conditions for both the primal and adjoint problems. Our choice of boundary condition for $\psi$ in Eqn. 2.24 happened to work, but we have given little motivation for this choice. In addition, we have assumed homogeneous boundary conditions for the primal problem, which will not always be the case. Finally, we have assumed that the output itself can be represented as an "interior" integral of the form $(g, u)$. However, in practice, the most important outputs – such as lift and drag – are often those defined on the **domain *boundaries***.

In order to address these issues, we need to further generalize our definition of the continuous adjoint. Recall that, in the discrete adjoint section, we initially treated the adjoint as the sensitivity of an output with respect to $\mathbf{F}$, then later generalized it to represent the sensitivity of an output with respect to the residual, $\mathbf{R}$. By following this same path in the definition of the continuous adjoint, we can address the aforementioned issues.

First, as in the discrete case, we can define a **continuous** "**residual**" $r(u)$ to be the function that is zero when the primal differential equation is satisfied, i.e.

$$r(u) \equiv Lu - f\,. \tag{2.34}$$

Now, if the adjoint $\psi$ is to represent the sensitivity of an output $J$ with respect to perturbations in $r(u)$, we require that it satisfy the following equation:

$$J'(\delta u) = \int_{\Omega} \psi\, r'(\delta u)\, d\Omega \qquad \forall\,(\text{permissible})\,\delta u \quad . \tag{2.35}$$

This is the generalized form of the **continuous adjoint equation**, which serves as the definition of $\psi$ regardless of output type, boundary conditions, and even problem nonlinearity. Conceptually, this equation just says that for a given change in the residual, $r'(\delta u)$, the adjoint dictates how this change leads to a change in the output, $J'(\delta u)$. Here, a prime denotes taking the "**first variation**" (or **Fréchet linearization**) of a quantity with respect to $u$. From variational calculus, taking the first variation of e.g. $J$ means considering how some change in the state, $\delta u$ (which is defined over all of $\Omega$ – see Fig. 2.3), would change the corresponding value of $J$.[12]

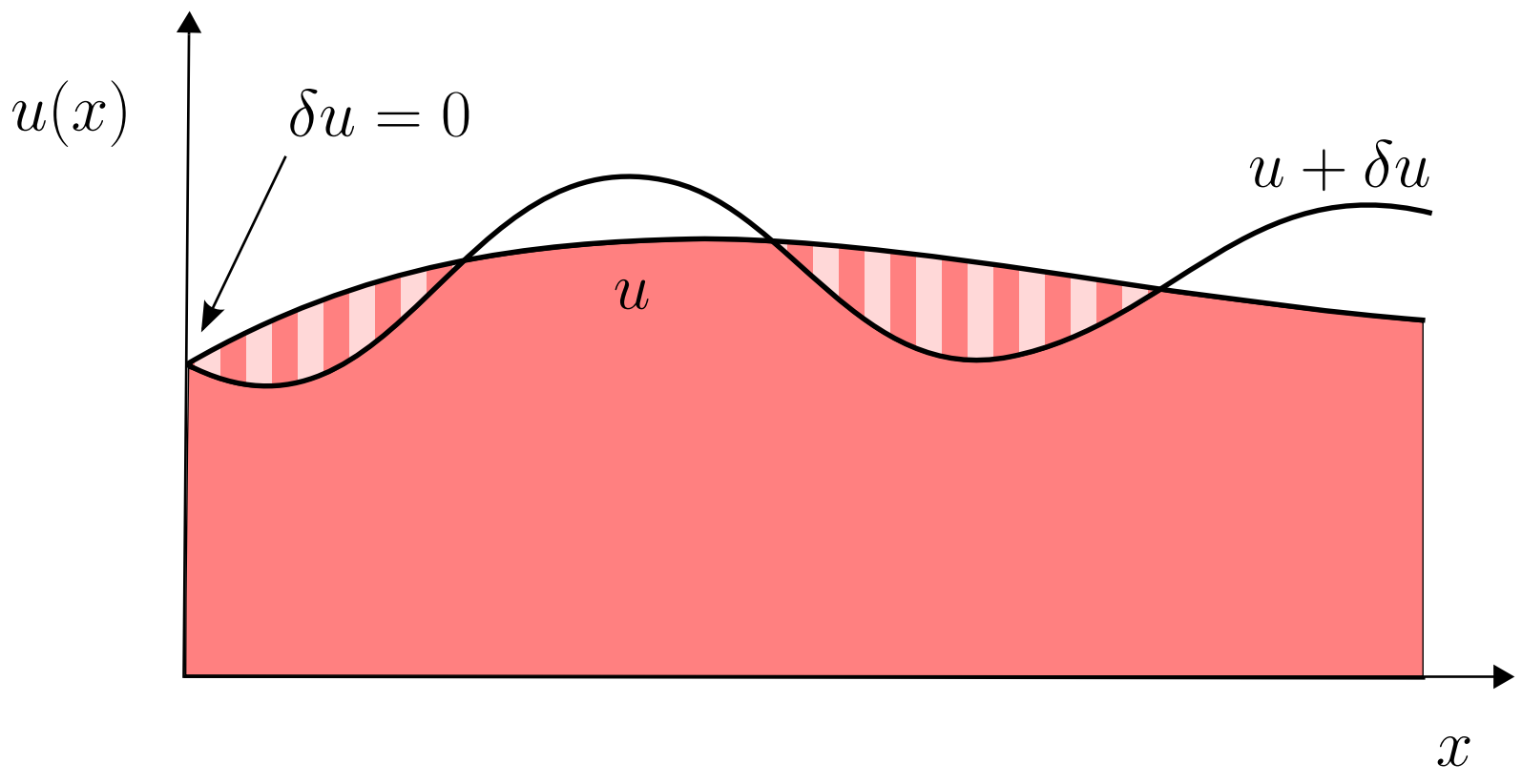

Figure 2.3: Illustration of the variation, $\delta u$. The variation is a function of $x$ and represents a perturbation to the state, $u$. If the value of $u$ is fixed at a boundary by a Dirichlet condition (as it is on the left here) the variation is constrained to be 0 at that boundary.

A similar idea holds for the variation of the residual, $r'(\delta u)$. In practice, for linear problems, computing the variation of the residual simply means inserting a $\delta u$ term in place of the original $u$, and eliminating any terms that have no dependence on $u$. For example, for the advection problem above, we have

$$r(u) \equiv a\frac{\partial u}{\partial x} - f \tag{2.36}$$

[12]Thus, rather than a *rate* of change, $J'(\delta u)$ actually represents a direct change in $J$, which could be written in shorthand as simply $\delta J$.

and would compute its first variation as

$$r'(\delta u) = a\frac{\partial(\delta u)}{\partial x}\,. \tag{2.37}$$

Here, since $f$ does not depend on $u$, it vanishes when the variation is taken.

Finally, an important point is that the variation $\delta u$ must be a "permissible" one, which – in particular – means that it must satisfy any boundary conditions imposed on the primal problem. For example, if a Dirichlet condition is imposed on a given boundary, then the value of $u$ is fixed there, meaning no variation is allowed and hence $\delta u = 0$ there. Figure 2.3 shows an example of what a permissible $\delta u$ might look like in this situation.

**Remark 1.** (**Systems of Equations**) Note that Eqn. 2.35 defines the adjoint in the case of a scalar equaton $r(u)$. If we instead had a system of $N_s$ equations with state components $\mathbf{u} = [\,u_1 \;\; u_2 \;\; \ldots \;\; u_{N_s}\,]^T$ and residual components $\mathbf{r}(\mathbf{u}) = [\,r_1(\mathbf{u}) \;\; r_2(\mathbf{u}) \;\; \ldots \;\; r_{N_s}(\mathbf{u})\,]^T$, then the adjoint would be written as $\boldsymbol{\psi} = [\,\psi_1 \;\; \psi_2 \;\; \ldots \;\; \psi_{N_s}\,]^T$ and the adjoint equation would become:

$$\boxed{J'(\delta\mathbf{u}) = \int_\Omega \boldsymbol{\psi}^T \mathbf{r}'(\delta\mathbf{u})\, d\Omega \qquad \forall\,(\text{permissible})\,\delta\mathbf{u}}\,. \tag{2.38}$$

Here, we use boldface to indicate a vector of $N_s$ state components. Furthermore, the variation is now defined as e.g. $J'(\delta\mathbf{u}) \equiv J'(\delta u_1) + J'(\delta u_2) + \ldots + J'(\delta u_{N_s})$ – in other words, it is the sum of the variations with respect to each component of $\mathbf{u}$. Note that we are being a little loose with notation here: for example, $J'(\delta u_1)$ should be formally written as $J'_{u_1}([\delta u_1\, 0 \ldots 0])$. Here, the $u_1$ subscript means that we are taking the variation of $J$ with respect to $u_1$, which is then a function of $\delta u_1$ only. However, when it is clear from context, we will leave the subscript off and will ignore the zero-valued inputs.

For the sake of simplicity, let us continue with our scalar advection example to show how Eqn. 2.35 is used in practice. We will then provide some further examples.

#### 2.2.2.1 Example: Continuous Adjoint for Steady Advection

Assume a primal advection problem of the form

$$a\frac{\partial u}{\partial x} = f \qquad x \in \Omega \tag{2.39}$$

$$u = u_L \qquad x = x_L \tag{2.40}$$

where $u_L$ is a Dirichlet boundary condition imposed on the left side of the domain. As before, we imagine that we are interested in some output $J(u)$, which will be left unspecified for now.

The variation of the residual, $r'(\delta u)$, is given by Eqn. 2.37. Inserting this into the general adjoint equation (Eqn. 2.35) gives:

$$J'(\delta u) = \int_\Omega \psi\, a \frac{\partial(\delta u)}{\partial x}\, dx \qquad \forall\, \delta u\,. \tag{2.41}$$

To determine the continuous adjoint PDE and boundary conditions, we now integrate by parts, giving

$$J'(\delta u) = \int_\Omega \underbrace{\left(-a\frac{\partial \psi}{\partial x}\right)}_{L^*\psi} \delta u\, dx + \psi\, a\, \delta u\Big|_{x_L}^{x_R} \qquad \forall\, \delta u\,, \tag{2.42}$$

or

$$J'(\delta u) = \int_\Omega (L^*\psi)\, \delta u\, dx + \psi\, a\, \delta u\Big|_{x_L}^{x_R} \qquad \forall\, \delta u\,. \tag{2.43}$$

Next, since the Dirichlet boundary condition is imposed at $x = x_L$, the variation $\delta u$ must be zero there. Eliminating this term in the above equation then leaves

$$J'(\delta u) = \int_\Omega (L^*\psi)\, \delta u\, dx + \psi\, a\, \delta u\Big|_{x_R} \qquad \forall\, \delta u\,. \tag{2.44}$$

Now, if, as before, we assume that our output has the form $J = (g, u)$, then its variaton $J'(\delta u)$ would just be $J'(\delta u) = (g, \delta u)$. Inserting this expression into the left-hand side of the above equation then gives

$$\int_{\Omega} g\,\delta u\,dx = \int_{\Omega} (L^*\psi)\,\delta u\,dx + \psi\,a\,\delta u\Big|_{x_R} \qquad \forall\,\delta u\,. \tag{2.45}$$

We now see that, if the left- and right-hand sides of this equation are to be equal for all $\delta u$, we need $\psi$ to satisfy the following conditions:

$$\boxed{L^*\psi = g \quad \text{and} \quad \psi\big|_{x_R} = 0}\,. \tag{2.46}$$

Inserting these expressions into the right-hand side of Eqn. 2.45 would then leave us with

$$\int_{\Omega} g\,\delta u\,dx = \int_{\Omega} g\,\delta u\,dx \qquad \forall\,\delta u\,, \tag{2.47}$$

which (clearly) holds for all $\delta u$.

Thus, for the output $J = (g, u)$, the constraints in Eqn. 2.46 represent the differential equation and boundary condition that must be satisfied by $\psi$ in order to fulfill the general adjoint equation (Eqn. 2.35). And indeed, looking back at our original example in Eqn. 2.23, this is exactly how we defined the adjoint in that case.

At this point, we have recovered our previous definition of the adjoint for outputs of the form $J = (g, u)$. However, as mentioned, we may also be interested in an output defined along the *boundary* of the domain. It turns out that the specific form of the adjoint equation – in this case Eqn. 2.44 – determines the type of boundary terms that can be included in the output. Whichever output we choose, we need its variation to look similar to the right-hand side of Eqn. 2.44. Otherwise, it would not be possible to choose $\psi$ in such a way that Eqn. 2.44 remains valid for all $\delta u$. In standard terminology, we would say that the output must be ***compatible*** with Eqn. 2.44.

In the present case, since Eqn. 2.44 contains only a right-boundary term (as opposed to a left-boundary term), this means that our output can likewise only involve the state on the right boundary. This makes sense from a logical perspective as well: since we are imposing a Dirichlet condition on the left boundary, we already know the value of $u$ there, so there is no sense in treating it as part of an "unknown" output. This idea holds true in general: ***compatible* outputs** are those whose components have **not already been fixed** by the boundary conditions, and hence, they are the outputs we would naturally be interested in anyway.

For our current problem, the general form of a compatible output is

$$\boxed{J(u) = (g, u) + g_R\, u\big|_{x_R}} \quad , \tag{2.48}$$

where $g_R$ is an arbitrary weight on the right-boundary state (i.e the outgoing flux). The variation of this output, $J'(\delta u)$, is then

$$J'(\delta u) = (g, \delta u) + g_R\, \delta u|_{x_R} . \tag{2.49}$$

To find the corresponding adjoint equation, we insert this $J'(\delta u)$ into Eqn. 2.44, giving

$$\int_\Omega g\, \delta u\, dx + g_R\, \delta u\bigg|_{x_R} = \int_\Omega (L^*\psi)\, \delta u\, dx + \psi\, a\, \delta u\bigg|_{x_R} \qquad \forall\, \delta u . \tag{2.50}$$

We then see that for this equation to hold for all $\delta u$, $\psi$ must satisfy

$$\boxed{L^*\psi = g \quad \text{and} \quad \psi\big|_{x_R} = \frac{g_R}{a}} \ . \tag{2.51}$$

The conclusion is thus that, when the output includes a boundary term, the corresponding boundary condition for the adjoint problem becomes nonhomogeneous.

**Remark 2.** (**Dual Form**) In this section, in addition to the adjoint boundary condition, we have also allowed the ***primal* boundary condition** ($u = u_L$) to be **nonzero**. So far, this has had no influence on the derivation of the adjoint, since this derivation involves only the variation $\delta u$ on the boundary, rather than the actual value there. However, while the primal boundary condition does not influence the adjoint problem itself, it does change how we would express the output in dual form.

To derive the dual form of the output, we start from the output definition in Eqn. 2.48 and perform the following steps:

$$\begin{aligned}
J(u) &= (g, u)\ +\ g_R\, u\big|_{x_R} && \text{(output definition)}\\
&= (L^*\psi, u)\ +\ \psi\, au\big|_{x_R} && \text{(adjoint equation and BC, Eqn. 2.51)}\\
&= (\psi, Lu)\ -\ \psi\, au\big|_{x_L}^{x_R}\ +\ \psi\, au\big|_{x_R} && \text{(integration by parts)}\\
&= (\psi, Lu)\ +\ \psi\, au\big|_{x_L} && \text{(right-boundary term cancellation)}\\
&= (\psi, f)\ +\ \psi\, au_L\big|_{x_L} && \text{(primal equation and BC, Eqn. 2.39)}
\end{aligned} \tag{2.52}$$

Thus, when the primal boundary conditions are nonzero, the dual form of the output

is given by

$$\boxed{J(u) = (\psi, f) + \psi \tilde{f}\big|_{x_L}} \,. \tag{2.53}$$

where $\tilde{f} \equiv a u_L$ is the prescribed boundary flux.

Note that although we now have a boundary term in the dual form, the underlying concept remains the same: the **dual form** involves only the **adjoint** and the ***prescribed data*** associated with the primal problem. In this case, that prescribed data includes both the source term $f$ and the boundary data $\tilde{f} = a u_L$. Note that this is still analogous to the dual form in the discrete case. There, the discrete $\mathbf{F}$ – which we referred to as the "source" vector – in practice contains boundary data as well.

#### 2.2.2.2 Example: Continuous Adjoint for Steady Diffusion

To solidfy the above ideas, we provide another example – a steady diffusion problem. Assume we have the following primal equation

$$-\nu \frac{\partial^2 u}{\partial x^2} = f \qquad x \in \Omega \tag{2.54}$$

$$u = u_L \qquad x = x_L \tag{2.55}$$

$$\frac{\partial u}{\partial x} = u_{x,R} \qquad x = x_R \tag{2.56}$$

where $\nu$ is a positive diffusion coefficient, $u_L$ is a prescribed Dirichlet condition on the left, and $u_{x,R}$ is a prescribed Neumann condition on the right.

The residual is then defined as

$$r(u) \equiv -\nu \frac{\partial^2 u}{\partial x^2} - f, \tag{2.57}$$

and its variation is

$$r'(\delta u) = -\nu \frac{\partial^2 (\delta u)}{\partial x^2} \,. \tag{2.58}$$

We again assume that we are interested in some output $J(u)$, which we will leave unspecified for the moment. Substituting the above residual variation into the general adjoint equation (Eqn. 2.35) gives

$$J'(\delta u) = \int_\Omega \psi \left( -\nu \frac{\partial^2 (\delta u)}{\partial x^2} \right) dx. \tag{2.59}$$

Now integrating by parts (twice) to move all derivatives onto $\psi$ yields:

$$J'(\delta u) \;=\; \int\limits_{\Omega} \underbrace{\left(-\nu\frac{\partial^2\psi}{\partial x^2}\right)}_{L^*\psi} \delta u\, dx \;+\; \nu\frac{\partial\psi}{\partial x}\delta u\Bigg|_{x_L}^{x_R} \;-\; \nu\psi\frac{\partial(\delta u)}{\partial x}\Bigg|_{x_L}^{x_R}. \tag{2.60}$$

Next, since the Dirichlet condition constrains $u$ at the left boundary, the variation $\delta u$ must be zero there. Likewise, since the Neumann condition constrains $\partial u/\partial x$ at the right boundary, the variation of the *derivative* must be zero there. (Note that mathematically, $\delta\left(\partial u/\partial x\right) = \partial\left(\delta u\right)/\partial x$, so we require $\partial\left(\delta u\right)/\partial x = 0$ at $x = x_R$.) Applying these constraints then leaves

$$J'(\delta u) \;=\; \int\limits_{\Omega} L^*\psi\, \delta u\, dx \;+\; \nu\frac{\partial\psi}{\partial x}\delta u\Bigg|_{x_R} \;+\; \nu\psi\frac{\partial(\delta u)}{\partial x}\Bigg|_{x_L}. \tag{2.61}$$

We see now that to be compatible with this equation, our output can include either a "$u$"-type contribution on the right boundary or a "$\partial u/\partial x$"-type contribution on the left boundary. In other words, our output can be of the form

$$\boxed{J(u) = (g, u) \;+\; g_R\, u\Big|_{x_R} \;+\; g_L\,\frac{\partial u}{\partial x}\Bigg|_{x_L}}\,. \tag{2.62}$$

Once again, this makes sense from a logical perspective as well: the output includes only quantities that are *not* already prescribed from the boundary conditions. By taking the variation of this $J(u)$ and comparing it to Eqn. 2.61, we see that the adjoint must satisfy the following conditions:

$$\boxed{L^*\psi = g\,, \qquad \frac{\partial\psi}{\partial x}\Bigg|_{x_R} = \frac{g_R}{\nu}\,, \qquad \text{and} \qquad \psi\Big|_{x_L} = \frac{g_L}{\nu}}\,. \tag{2.63}$$

#### 2.2.2.3 Example: Continuous Adjoint for Nonlinear Burgers Equation

Finally, we note that the adjoint equation (Eqn. 2.35) applies to **nonlinear** problems as well. For these problems, the adjoint represents the sensitivity of an output to residual perturbations *about* some particular state $u$, in the same way that we could compute the slope of a standard nonlinear function (say $f(x) = x^2$) about a particular value of $x$.

In general, both the residual and the desired output may be nonlinear expressions (as would be the case with, e.g. a Navier-Stokes simulation and a lift output), and the variations of these expressions would then represent **linearizations** about some particular state – say, $u_0$. Thus, while for linear problems we just label the variations $J'(\delta u)$ and $r'(\delta u)$, for nonlinear problems we must also indicate the state about which these variations are performed. We can do this by including that state in square brackets. For example, when linearizing about $u_0$, we would write $J'[u_0](\delta u)$ and $r'[u_0](\delta u)$. This notation is common in the literature.

As a simple example, consider the following one-dimensional Burgers equation (with a $u^3$ source term):

$$u\frac{\partial u}{\partial x} + u^3 = f \qquad x \in \Omega \tag{2.64}$$

$$u = u_L \qquad x = x_L \tag{2.65}$$

The residual is defined as

$$r(u) \equiv u\frac{\partial u}{\partial x} + u^3 - f\,, \tag{2.66}$$

and its variation about a particular state $u_0(x)$ is then given by

$$r'[u_0](\delta u) = u_0\frac{\partial(\delta u)}{\partial x} + \delta u\frac{\partial u_0}{\partial x} + 3u_0^2\,\delta u\,. \tag{2.67}$$

Here, the first two terms arise from applying the product rule[13] to the $u\,\partial u/\partial x$ term in Eqn. 2.66, while the last term represents the variation of $u^3$.

Similarly, if our output were some nonlinear quantity, such as the right-boundary flux

$$J(u) = \frac{1}{2}u^2\bigg|_{x_R}\,, \tag{2.68}$$

then its variation could be computed as

$$J'[u_0](\delta u) = u_0\,\delta u\bigg|_{x_R}\,. \tag{2.69}$$

Since $u_0$ is treated as a "frozen" state, both $r'[u_0](\delta u)$ and $J'[u_0](\delta u)$ are now linear in $\delta u$, and the adjoint equations and boundary conditions can be derived as usual from Eqn. 2.35.

[13]The variation obeys the product rule in the same way that a standard derivative does.

## 2.3 Summary: Discrete and Continuous Adjoints

With discrete and continuous adjoints discussed, we conclude with a brief discussion of the connections between these concepts. We have already mentioned some of the similarities between discrete and continuous adjoints, such as how they lead to a "dual form" of an output, and hence represent an output sensitivity to residual perturbations. However, we can draw some further parallels between the discrete and continuous adjoint equations themselves.

We start with a simple side-by-side comparison of these equations. Recall that the discrete adjoint equation is given by Eqn. 2.15 as

$$\frac{\partial \mathbf{R}}{\partial \mathbf{U}}^T \mathbf{\Psi} = \frac{\partial J}{\partial \mathbf{U}}^T .$$

Transposing both sides of this equation and swapping left- and right-hand sides then gives:

$$\frac{\partial J}{\partial \mathbf{U}} = \mathbf{\Psi}^T \frac{\partial \mathbf{R}}{\partial \mathbf{U}} .$$

On the other hand, the continuous adjoint equation (Eqn. 2.38) is defined as

$$J'(\delta \mathbf{u}) = \int_\Omega \boldsymbol{\psi}^T \mathbf{r}'(\delta \mathbf{u}) \, d\Omega .$$

A clear resemblance is seen between these equations – namely, they both involve an output linearization on the left and a residual linearization on the right, related via an adjoint vector.

### 2.3.1 The Discrete Adjoint Operator and Adjoint Consistency

To draw some further connections, let us look more closely at the discrete adjoint equation

$$\frac{\partial \mathbf{R}}{\partial \mathbf{U}}^T \mathbf{\Psi} = \frac{\partial J}{\partial \mathbf{U}}^T .$$

If, for simplicity, we call the residual Jacobian matrix $\mathbf{A}$ and label the output linearization $\mathbf{G}$ (such that, e.g., for a linear output $J \equiv \mathbf{G}^T\mathbf{U}$), then this equation can

be written as

$$\mathbf{A}^T\mathbf{\Psi} = \mathbf{G} \ . \tag{2.70}$$

On the other hand, from Eqn. 2.31, the differential form of the continuous adjoint equation is written as

$$L^*\psi = g.$$

Recall that the adjoint operator in this equation, $L^*$, was defined by the adjoint identity

$$(Lu, v) \ = \ (u, L^*v) \qquad\qquad \forall u, v \in \mathcal{V} \, .$$

A question which then arises is: if the continuous adjoint operator $L^*$ satisfies a so-called "adjoint identity," can the same be said for the operator in the *discrete* adjoint equation – i.e. $\mathbf{A}^T$?

To answer this question, note that we can formulate a discrete version of the above adjoint identity by: (1) replacing $L$ by a discrete operator $\mathbf{A}$, (2) replacing the inner product with a discrete dot product, and (3) replacing $u$ and $v$ by discrete vectors $\mathbf{U}$ and $\mathbf{V}$. We then obtain the following identity

$$\mathbf{AU} \cdot \mathbf{V} = \mathbf{U} \cdot \mathbf{A}^*\mathbf{V} \qquad\qquad \forall\, \mathbf{U}, \mathbf{V} \in \mathbb{R}^N \, , \tag{2.71}$$

which defines a certain matrix $\mathbf{A}^*$ as the **discrete adjoint operator**. To determine $\mathbf{A}^*$, we rewrite the above relation using the formal definition of the dot product (i.e. $\mathbf{u} \cdot \mathbf{v} \equiv \mathbf{u}^T\mathbf{v}$):

$$(\mathbf{AU})^T\mathbf{V} = \mathbf{U}^T(\mathbf{A}^*\mathbf{V}) \qquad\qquad \forall\, \mathbf{U}, \mathbf{V} \in \mathbb{R}^N \tag{2.72}$$

$$\mathbf{U}^T\mathbf{A}^T\mathbf{V} = \mathbf{U}^T\mathbf{A}^*\mathbf{V} \qquad\qquad \forall\, \mathbf{U}, \mathbf{V} \in \mathbb{R}^N \tag{2.73}$$

$$\Longrightarrow \boxed{\mathbf{A}^T = \mathbf{A}^*} \ . \tag{2.74}$$

The formal adjoint operator associated with a matrix $\mathbf{A}$ is therefore just its **transpose**, $\mathbf{A}^T$. Thus, the presence of $\mathbf{A}^T$ in the discrete adjoint equation does in fact "parallel" the continuous adjoint equation, in the sense that both equations involve formal adjoint operators on the left-hand side and output linearizations on the right.

This connection between $\mathbf{A}^T$ and $L^*$ is more than superficial. In fact, just as the discrete primal equation $\mathbf{AU} = \mathbf{F}$ should (ideally) represent a consistent discretization of the continuous primal problem,

$$\mathbf{AU} = \mathbf{F} \quad \Longleftrightarrow \quad Lu = f \quad ,$$

the discrete *adjoint* equation $\mathbf{A}^T\mathbf{\Psi} = \mathbf{G}$ should ideally represent a consistent discretization of the continuous *adjoint* problem,

$$\mathbf{A}^T\mathbf{\Psi} = \mathbf{G} \quad \Longleftrightarrow \quad L^*\psi = g \quad .$$

A discretization which satisfies this property – i.e. whose matrix $\mathbf{A}$ has a transpose that is also a consistent discretization of $L^*$ – is known as ***adjoint consistent***.[14]

While not every primal discretization is adjoint-consistent, a lack of adjoint-consistency can lead to suboptimal convergence rates in outputs of interest. Thus, even if the adjoint equation is never explicitly solved, adjoint-consistent discretizations are often preferable to adjoint-inconsistent ones. Furthermore, if the discrete adjoint equation *is* actually solved using an adjoint-inconsistent discretization, spurious oscillations will appear in the adjoint, limiting its usefulness for applications such as optimization and error estimation. In order to avoid these issues, adjoint-inconsistent discretizations can often be modified to recover adjoint-consistency. A discussion of these issues is provided in e.g. [11, 8, 7].

### 2.3.2 Summary of Adjoint Properties

1. For a **scalar output** of interest, both discrete and continuous adjoints can be used to represent the output in an equivalent "**dual form**" (provided the problem is linear).

2. For both linear and nonlinear problems, the discrete and continuous adjoints represent the **sensitivity** of a given output to perturbations in the governing equations (residuals).

3. The discrete adjoint is just a weighted average of the rows of the inverse Jacobian matrix (i.e. $\mathbf{A}^{-1}$ or $\partial\mathbf{R}/\partial\mathbf{U}^{-1}$).

[14] Technically, both the adjoint operator (along with its corresponding boundary conditions) and the output linearization must be represented in a consistent manner for a method to be considered adjoint-consistent.

4. The continuous adjoint is the solution of a linear "adjoint" PDE involving the adjoint operator $L^*$.

5. Any odd-order derivatives in $L^*$ have the opposite sign as in the primal operator, leading to a **reversal of information flow** in the adjoint problem.

6. The adjoint equations – both discrete and continuous – are **linear** even when the primal problem is nonlinear. For nonlinear problems, the adjoint equations represent a local linearization about a given primal state.

## 2.4 Output-based Error Estimation

With the discussion of adjoints complete, we now turn to one of the primary uses of adjoints: the estimation of numerical errors. The concept of adjoint-based *a posteriori* error estimaton (or simply "**output-based**" **error estimation**) can be treated in either a continuous or discrete context. In this section, we begin with a simplified discussion in a continuous context before moving on to a more general discrete formulation.

### 2.4.1 Continuous Error Estimation

Assume that we are dealing with a linear differential equation of the form

$$Lu = f \qquad x \in \Omega \tag{2.75}$$

$$\text{primal b.c.} \qquad x \in \partial\Omega \tag{2.76}$$

and, for simplicity, that our desired output $J(u)$ can be represented as

$$J(u) = (u, g)\,. \tag{2.77}$$

Now imagine that somehow (e.g. through a numerical method) we have arrived at an approximate solution, $u_H$.[15] If we were to compute our output using this $u_H$, its value would be given by

$$J(u_H) = (u_H, g)\,. \tag{2.78}$$

[15] Assume that $u_H$ is sufficiently smooth to evaluate $r(u_H)$. If $u_H$ were nonsmooth, similar arguments would hold, but the residual would have to be treated in a distributional sense.

A question we might now ask is: how much ***error*** is present in this output? In other words, what is the difference between the exact and approximate outputs, i.e. $\delta J \equiv J(u) - J(u_H)$?

It turns out that the adjoint is helpful in answering this question. Starting from the definition of $\delta J$ and using the relevant adjoint identities, we find:

$$\begin{aligned}
\delta J &= J(u) - J(u_H) \\
&= (u, g) - (u_H, g) && \text{(output definitions)} \\
&= (u - u_H, g) && \text{(linearity of inner product)} \\
&= (u - u_H, L^*\psi) && \text{(adjoint equation, Eqn. 2.31)} \\
&= (L(u - u_H), \psi) && \text{(adjoint identity, Eqn. 2.28)} \\
&= (Lu, \psi) - (Lu_H, \psi) && \text{(linearity of inner product)} \\
&= (f, \psi) - (Lu_H, \psi) && \text{(primal equation, Eqn. 2.75)} \\
&= -(Lu_H - f, \psi) && \text{(linearity of inner product)}
\end{aligned} \tag{2.79}$$

Now, recall that the residual is defined as

$$r(u) \equiv Lu - f\,. \tag{2.80}$$

Thus, the quantity $Lu_H - f$ is just the residual evaluated with the approximate solution, i.e. $r(u_H)$. Since $u_H$ does not satisfy the primal differential equation exactly, this $r(u_H)$ will in general be nonzero, and represents the local truncation error at a given region of the domain.

Replacing $Lu_H - f$ with $r(u_H)$ in Eqn. 2.79 then gives the following expression for the output error:

$$\delta J = -(r(u_H), \psi). \tag{2.81}$$

Or, in integral form, we have

$$\boxed{\delta J = -\int_\Omega \psi\, r(u_H)\, d\Omega} \quad . \tag{2.82}$$

Thus, we see that the amount of error in an output is given by an ***adjoint-weighted*** (or ***"dual-weighted"***) ***residual***. This expression indicates that if nonzero residuals occur in regions where the adjoint is large, then these residuals will con-tribute a relatively large amount to the total output error.[16] Thus, not only does it provide the total output error, it also indicates the regions of the domain that are *responsible* for this output error. For this reason, the above expression will play a critical role in both output error estimation and **adaptive mesh refinement**.

**Remark 3.** (**Dual Form of Error Estimate**) Note that just as we can write the output itself in both primal and dual forms, the output error can be written in dual form as well. For a discussion of this topic, see Appendix A (Sec. A.1).

#### 2.4.1.1 Continuous Error Estimation: Approximate Adjoint

In the current derivation, we have made the assumption of a linear primal problem and a linear interior output. In this context, the above expression for the output error is exact, provided the exact adjoint $\psi$ is used. In practice, we will not have access to this exact adjoint and must instead settle for a numerical approximation, $\psi_h$. The error $\delta J$ above would then become an ***error estimate*** $\delta J_{\text{est}}$, which could be written as

$$\delta J_{\text{est}} = -\int_{\Omega} \psi_h \, r(u_H) \, d\Omega \;\approx\; \delta J \quad . \tag{2.83}$$

The closer $\psi_h$ is to $\psi$, the closer $\delta J_{\text{est}}$ will be to the true output error.

#### 2.4.1.2 Continuous Error Estimation: Finite Element Methods

So far, we have not needed to specify how $\psi_h$ or $u_H$ are obtained. However, since methods of a finite-element type are often employed in the context of error estimation, it is useful to derive a form of the error estimate particular to these schemes.

In general, a finite element method weights the continuous residual $r(u_H)$ by "test functions" $v \in \mathcal{V}_H$, where $\mathcal{V}_H$ may be, e.g., the space of polynomial functions of a certain order $p$. It then enforces orthogonality of the residual with respect to all functions in $\mathcal{V}_H$, so that

$$\int_{\Omega} v \, r(u_H) \, d\Omega \;=\; 0 \qquad\qquad \forall v \in \mathcal{V}_H \, . \tag{2.84}$$

[16] Assuming that minimal error cancellation occurs between different regions within the integral.

Now, if we were to approximate the adjoint *within* $\mathcal{V}_H$ itself, this "$\psi_H$" would necessarily satisfy

$$\int_\Omega \psi_H \, r(u_H) \, d\Omega \; = \; 0 \,, \tag{2.85}$$

since $\psi_H \in \mathcal{V}_H$. We can thus add this term to the error estimate in Eqn. 2.83 with no effect, so that for a finite element method, the following form of the error estimate holds:

$$\delta J_{\text{est}} = - \int_\Omega (\psi_h - \psi_H) \, r(u_H) \, d\Omega \quad . \tag{2.86}$$

From this expression, we see that in order to obtain a useful error estimate for a finite element method, we need to approximate the adjoint $\psi_h$ in a different (typically ***finer***) space than $\mathcal{V}_H$ itself. Otherwise, if we were to take $\psi_h = \psi_H$ in the above formula, we would always obtain an error estimate of zero. Thus, $\psi_h$ is often computed in an **order-enriched** space, though it could also be computed on e.g. a uniformly $h$-refined mesh.

**Remark 4.** (**Output Convergence Rate)** An additional point of interest is that Eqn. 2.86 can be used to predict the output convergence rate associated with a finite element method. First, if we assume that $\psi_h = \psi$ for simplicity, then this expression represents the exact output error. It then says that the output error involves the product of two terms: $(\psi - \psi_H)$ and $r(u_H)$. Therefore, the convergence rate of this output error should (at least) correspond to the sum of the convergence rates of these individual terms. Now, if $\mathcal{V}_H$ is an order-$p$ space, then the quantity $(\psi - \psi_H)$ will typically converge at order $p+1$. Furthermore, for (e.g.) a first-order operator such as advection, the residual $r(u_H)$ will converge at order $p$, since it involves taking one derivative of $u_H$. Summing these respective adjoint and residual rates, we predict that the output error should converge at a rate of

$$\underbrace{(p+1)}_{\text{adjoint}} \; + \underbrace{p}_{\text{residual}} \; = \underbrace{2p+1}_{\text{output}} \quad . \tag{2.87}$$

This is indeed the rate typically observed for methods of (e.g.) a discontinuous Galerkin (DG) type.[17]

[17]Note that although the term $(\psi - \psi_H)$ will only converge at order $p+1$ when $\psi$ is *smooth*, in practice, $2p+1$ output rates are often obtained when $\psi$ is nonsmooth as well. Furthermore, note

#### 2.4.1.3 Continuous Error Estimation: Nonlinear Problems

While we have focused on linear problems so far, the above error estimates carry over to nonlinear problems virtually unmodified, with the caveat that they are no longer exact due to the presence of a linearization error. See Appendix A (Sec. A.3) for a discussion of this topic. For now, we will move on and treat the nonlinear case in the discrete section to follow.

### 2.4.2 Discrete Error Estimation

We now turn to the primary method of error estimation employed in practice – that of discrete adjoint-based error estimation. The ideas described in this section hold for general problems – linear or nonlinear, steady or unsteady – and for arbitrary output types and numerical discretizations.

To start, imagine that we are interested in solving a set of nonlinear equations, which may be written in discrete form as

$$\mathbf{R}_H(\mathbf{U}_H) = \mathbf{0}\,. \tag{2.88}$$

Here, the subscript $H$ denotes a discretization on a given mesh, while $\mathbf{R}_H(\mathbf{U}_H)$ could represent the residuals associated with, for example, the Navier-Stokes equations.

Assume now that we are interested in a particular output, $J_H(\mathbf{U}_H)$, which could represent (say) lift or drag. After solving for $\mathbf{U}_H$ and computing this $J_H(\mathbf{U}_H)$, we may again like to know: how much ***error*** is present in this output?

Ideally, we would like to compute the true output error

$$\delta J = J(\mathbf{U}) - J_H(\mathbf{U}_H)\,. \tag{2.89}$$

However, without knowledge of the exact solution $\mathbf{U}$, this is infeasible. Instead, as a surrogate for $\mathbf{U}$, let us consider a so-called "fine-space" solution $\mathbf{U}_h$, and attempt to compute the **error** ***estimate***

$$\boxed{\delta J_{\mathrm{est}} = J_h(\mathbf{U}_h) - J_H(\mathbf{U}_H)}\,. \tag{2.90}$$

Here, the fine space (which we will denote by $\mathcal{V}_h$) could be a uniformly refined or order-incremented version of the original space, $\mathcal{V}_H$. Thus, the fine space will typically that for diffusion problems the output error is expected to converge at a rate of just $2p$, since $r(u)$ then contains a second-derivative operator and thus converges at order $p-1$.

contain the coarse space, i.e. $\mathcal{V}_H \subset \mathcal{V}_h$.

The fine-space solution $\mathbf{U}_h$ would then satisfy a corresponding set of fine-space equations, written as

$$\mathbf{R}_h(\mathbf{U}_h) = \mathbf{0}\,. \tag{2.91}$$

While we could attempt to solve these equations directly, this would be impractical. Instead, we would like to compute $\delta J_{\text{est}}$ *without* actually solving for $\mathbf{U}_h$.

In order to achieve this, we first define an **injection** of the coarse solution $\mathbf{U}_H$ into the fine space as

$$\mathbf{U}_h^H = \mathbf{I}_h^H \mathbf{U}_H\,. \tag{2.92}$$

Here, $\mathbf{I}_h^H$ is a lossless injection operator which effectively "samples" $\mathbf{U}_H$ on the fine space, such that the resulting $\mathbf{U}_h^H$ has the same dimension as $\mathbf{U}_h$ but contains only coarse-space information.[18] With this $\mathbf{U}_h^H$, we can then define the state perturbation, $\delta\mathbf{U}$, as the difference between the fine and (injected) coarse states:

$$\delta\mathbf{U} = \mathbf{U}_h - \mathbf{U}_h^H\,. \tag{2.93}$$

With these definitions in hand, we now turn to computing $\delta J_{\text{est}}$ (Eqn. 2.90). First, we note that the fine-space output, $J_h(\mathbf{U}_h)$, (which is at this point unknown) can be Taylor-expanded about the coarse-space state $\mathbf{U}_h^H$ as follows:

$$J_h(\mathbf{U}_h) \;=\; J_h(\mathbf{U}_h^H) \;+\; \left.\frac{\partial J_h}{\partial \mathbf{U}_h}\right|_{\mathbf{U}_h^H} \delta\mathbf{U} \;+\; \mathcal{O}(\delta\mathbf{U}^2)\,. \tag{2.94}$$

Dropping the $\mathcal{O}(\delta\mathbf{U}^2)$ remainder then leaves

$$\boxed{J_h(\mathbf{U}_h) \;\approx\; J_h(\mathbf{U}_h^H) \;+\; \left.\frac{\partial J_h}{\partial \mathbf{U}_h}\right|_{\mathbf{U}_h^H} \delta\mathbf{U}}\,. \tag{2.95}$$

Here, the perturbation $\delta\mathbf{U}$ is unknown. In order to determine it, we perform a similar expansion of the fine-space residuals, $\mathbf{R}_h(\mathbf{U}_h) = \mathbf{0}$, about the coarse solution:

$$\mathbf{R}_h(\mathbf{U}_h) \;\approx\; \boxed{\mathbf{R}_h(\mathbf{U}_h^H) \;+\; \left.\frac{\partial \mathbf{R}_h}{\partial \mathbf{U}_h}\right|_{\mathbf{U}_h^H} \delta\mathbf{U} \;=\; \mathbf{0}}\,. \tag{2.96}$$

[18] In a finite element context, this injection would just mean representing $\mathbf{U}_H$ with the fine-space basis functions rather than the coarse-space bases.

From this equation, we can solve for $\delta\mathbf{U}$, giving

$$\delta\mathbf{U} \;=\; -\left.\frac{\partial\mathbf{R}_h}{\partial\mathbf{U}_h}\right|_{\mathbf{U}_h^H}^{-1} \mathbf{R}_h(\mathbf{U}_h^H) \tag{2.97}$$

Inserting this $\delta\mathbf{U}$ back into Eqn. 2.95 then gives

$$J_h(\mathbf{U}_h) \;\approx\; J_h(\mathbf{U}_h^H) \;-\; \underbrace{\left.\frac{\partial J_h}{\partial\mathbf{U}_h}\right|_{\mathbf{U}_h^H} \left.\frac{\partial\mathbf{R}_h}{\partial\mathbf{U}_h}\right|_{\mathbf{U}_h^H}^{-1}}_{\boldsymbol{\Psi}_h^T} \mathbf{R}_h(\mathbf{U}_h^H)\,. \tag{2.98}$$

The terms multiplying the residual in this equation now look familiar. Indeed, looking back at the discrete adjoint equation (Eqn. 2.13), we see that they correspond exactly to the *adjoint* of $J$. Specifically, they correspond to a **fine-space adjoint**, which we can denote by $\boldsymbol{\Psi}_h^T$. By the above definition, this adjoint is computed on the fine-space, $\mathcal{V}_h$, using a linearization about the injected coarse-space solution, $\mathbf{U}_h^H$.

Writing Eqn. 2.98 in terms of $\boldsymbol{\Psi}_h^T$ then gives

$$J_h(\mathbf{U}_h) \;\approx\; J_h(\mathbf{U}_h^H) \;-\; \boldsymbol{\Psi}_h^T\,\mathbf{R}_h(\mathbf{U}_h^H)\,. \tag{2.99}$$

Finally, since $J_h(\mathbf{U}_h^H) = J_H(\mathbf{U}_H)$ as long as the mesh geometry does not change between the coarse and fine spaces, this equation can be rewritten as

$$J_h(\mathbf{U}_h) - J_H(\mathbf{U}_H) \;\approx\; -\boldsymbol{\Psi}_h^T\,\mathbf{R}_h(\mathbf{U}_h^H)\,. \tag{2.100}$$

The left-hand side of this equation is exactly the error estimate we wished to compute, $\delta J_{\text{est}}$ (Eqn. 2.90). Thus, we have

$$\boxed{\delta J_{\text{est}} \;\approx\; -\boldsymbol{\Psi}_h^T\,\mathbf{R}_h(\mathbf{U}_h^H)}\,. \tag{2.101}$$

We see that, as in the continuous case (Eqn. 2.83), the output error estimate involves the product of a "fine-space" adjoint and the residuals evaluated with the coarse-space solution. This $\mathbf{R}_h(\mathbf{U}_h^H)$ will in general be nonzero, and indicates regions of the domain where local truncation errors are generated. Weighting this term by the adjoint then determines to what extent each of these local truncations errors contributes to the final output error.

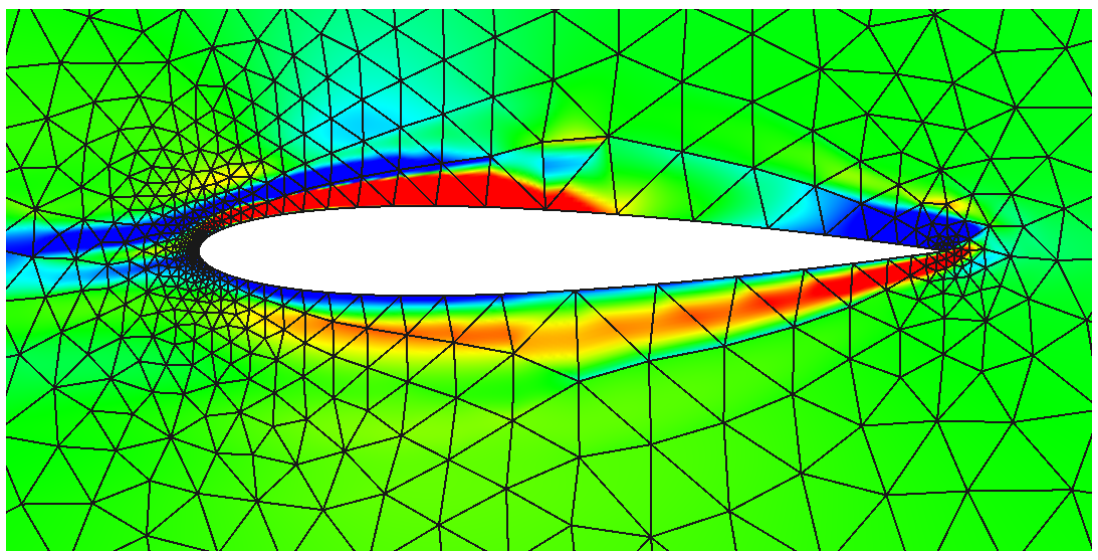

(a) Lift adjoint, $\boldsymbol{\Psi}_h$

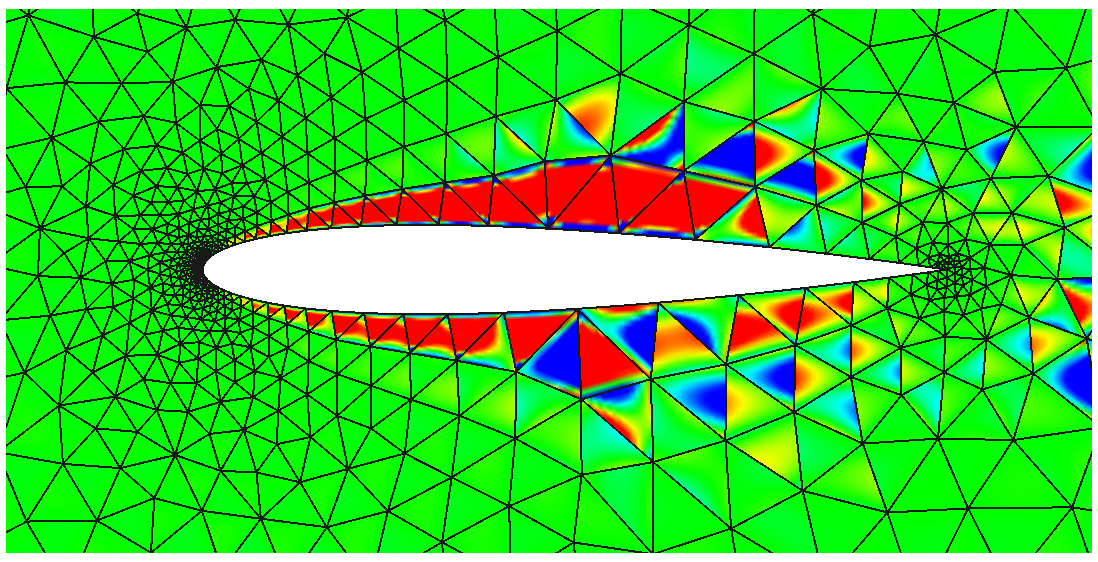

(b) Residual, $\mathbf{R}_h(\mathbf{U}_h^H)$

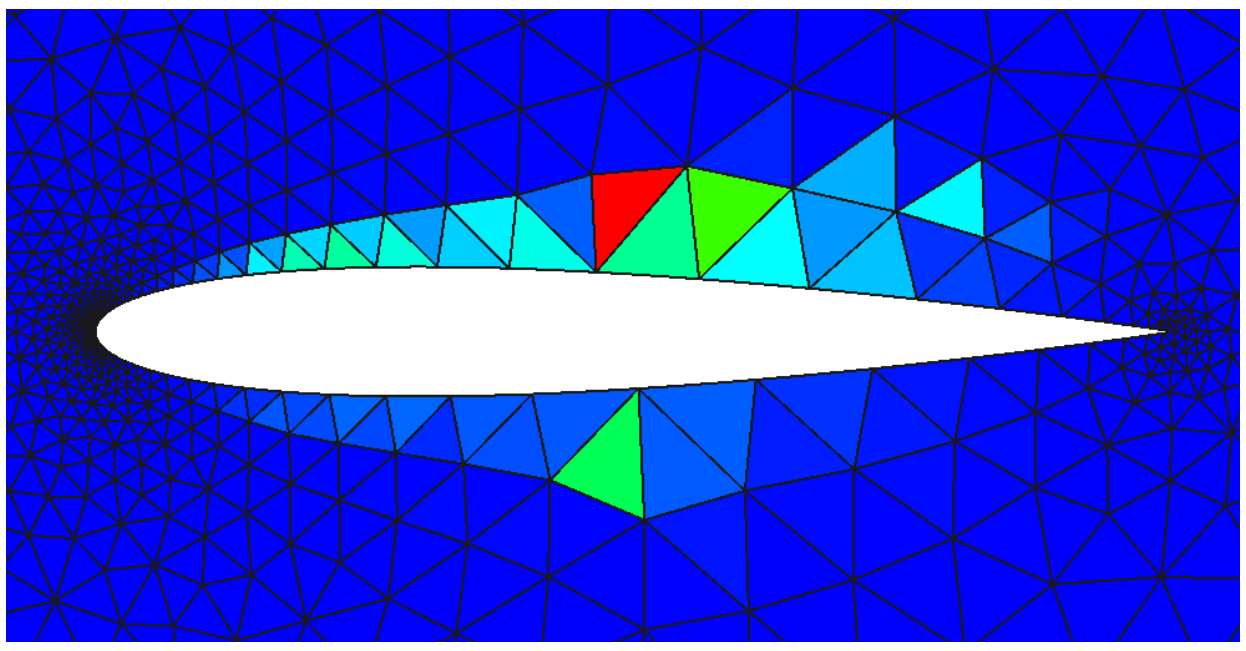

(c) Error indicator

Figure 2.4: (a) Fine-space adjoint for the lift on the airfoil. (b) Fine-space residual evaluated with the injected coarse state (a measure of local truncation error). (c) Error indicators, representing the amount of error each element contributes to the lift. This is given by the product of the adjoint and residual, i.e. (a)·(b). (Figures reproduced from [5].)

#### 2.4.2.1 Error Localization and Mesh Adaptation

If an element-based method (such as a finite element or finite volume method) is used to solve for $\mathbf{U}_H$, the above error estimate can be localized to individual elements in the mesh. If there are $N_e^h$ total elements on the fine space, it can be rewritten as simply

$$\delta J_{\text{est}} \approx \sum_{e=1}^{N_e^h} -\boldsymbol{\Psi}_{h,e}^T \mathbf{R}_{h,e}(\mathbf{U}_h^H) , \tag{2.102}$$

where $\boldsymbol{\Psi}_{h,e}^T$ and $\mathbf{R}_{h,e}(\mathbf{U}_h^H)$ are the components of the adjoint and residual associated with (i.e. restricted to) a given element $e$. A local **error indicator** can then be defined as the absolute value of this product on each fine-space element, i.e. as

$$\varepsilon_e = \left|\boldsymbol{\Psi}_{h,e}^T \mathbf{R}_{h,e}(\mathbf{U}_h^H)\right| . \tag{2.103}$$

If order enrichment is used to obtain the fine space, then $N_e^H = N_e^h$, and $\varepsilon_e$ directly indicates how much error a given coarse-space element contributes to the output. Alternatively, if uniform $h$-refinement is used to obtain the fine space, then $\varepsilon_e$ can be summed over all fine-space elements within a given coarse-space element to arrive at a final coarse-space indicator. Figure 2.4 shows an example of the error indicator computation around an airfoil, where order enrichment was used for the fine space.

Once an error indicator is computed for each element in the coarse-space mesh, the elements with the highest indicators (or a related "figure of merit") can be selected for refinement. Refining these elements then drives down their local residuals, leading to a reduction in the amount of output error generated. This error estimation and adaptation procedure can be performed in an iterative fashion to efficiently drive the output error toward zero.

#### 2.4.2.2 Finite Element Methods

If a finite element method is used to compute $\mathbf{U}_H$, a coarse-space adjoint can be subtracted from the above $\delta J_{\text{est}}$ (and corresponding error indicators) with no effect. This is due to the orthogonality property of finite element methods, as discussed previously in the continuous context (Eqn. 2.86).

If we call the coarse-space adjoint $\mathbf{\Psi}_H$ and denote its injection into the fine space as $\mathbf{\Psi}_h^H$, then for finite element methods, Eqns. 2.101, 2.102, and 2.103 become:

$$\delta J_{\text{est}} \approx -\left(\mathbf{\Psi}_h^T - \left(\mathbf{\Psi}_h^H\right)^T\right) \mathbf{R}_h(\mathbf{U}_h^H) \quad , \tag{2.104}$$

$$\delta J_{\text{est}} \approx \sum_{e=1}^{N_e^h} -\left(\mathbf{\Psi}_{h,e}^T - \left(\mathbf{\Psi}_{h,e}^H\right)^T\right) \mathbf{R}_{h,e}(\mathbf{U}_h^H) \quad , \tag{2.105}$$

and

$$\varepsilon_e = \left|\left(\mathbf{\Psi}_{h,e}^T - \left(\mathbf{\Psi}_{h,e}^H\right)^T\right) \mathbf{R}_{h,e}(\mathbf{U}_h^H)\right| \quad . \tag{2.106}$$

In practice, it is not necessary to subtract off a coarse-space adjoint, but these expressions illustrate that for a finite element method, output errors will be generated not where the adjoint *values* are large, but instead where the adjoint is not *well-approximated*. In other words, the important regions are those where the difference between fine- and coarse-space adjoints, $\left(\mathbf{\Psi}_{h,e}^T - \left(\mathbf{\Psi}_{h,e}^H\right)^T\right)$, is large. Thus, if the fine-space adjoint were (e.g.) large but constant in a given region, then since this constant could also be represented in the coarse space (assuming it includes the $p = 0$ mode), no output errors would be generated in this region.

A related point to emphasize is that, while a coarse-space adjoint $\mathbf{\Psi}_H$ may be useful for optimization applications, for error estimation some type of fine-space adjoint is required. If we were to simply inject $\mathbf{\Psi}_H$ to the fine space and use that as our effective "$\mathbf{\Psi}_h$," then the above formulas would reduce to zero, providing no useful information. In practice, however, we do not need to solve the fine-space adjoint equations exactly. Oftentimes, injecting $\mathbf{\Psi}_H$ to the fine space and performing a few smoothing iterations is enough to provide meaningful error estimates.

#### 2.4.2.3 Linearization Error

Finally, we note that for nonlinear problems – where the residuals $\mathbf{R}(\mathbf{U})$ and/or the output $J(\mathbf{U})$ are nonlinear – there is a so-called ***linearization* error** associated with the above error estimate, $\delta J_{\text{est}}$.

Looking back at the derivation of $\delta J_{\text{est}}$ (Eqn. 2.101), we see that when performing the Taylor expansion of $J_h(\mathbf{U}_h)$ in Eqn. 2.94, we ignored the $\mathcal{O}(\delta\mathbf{U}^2)$ remainder term. Likewise, we ignored a similar $\mathcal{O}(\delta\mathbf{U}^2)$ term in the Taylor expansion of $\mathbf{R}_h(\mathbf{U}_h)$ in Eqn. 2.96. If we had carried these terms throughout the derivation, they would have appeared in the final error estimate, giving:

$$\boxed{\delta J_{\text{est}} \;\approx\; -\mathbf{\Psi}_h^T \mathbf{R}_h(\mathbf{U}_h^H) \;+\; \mathcal{O}(\delta\mathbf{U}^2)}\,. \tag{2.107}$$

Thus, for nonlinear problems, even if the adjoint $\mathbf{\Psi}_h$ were computed on an infinitely refined mesh, there would still be an $\mathcal{O}(\delta\mathbf{U}^2)$ *error* in the error estimate, due to the fact that the adjoint problem represents a linearization about the coarse-space state, $\mathbf{U}_h^H$. Since $\delta\mathbf{U}$ typically converges at order $p_H+1$ (where $p_H$ is the coarse-space approximation order), this $\mathcal{O}(\delta\mathbf{U}^2)$ term is of order $2(p_H+1) = 2p_H+2$. Thus, for nonlinear problems, the accuracy of the output error estimate overall is limited to $\mathcal{O}(h^{2p+2})$. (Though we note that, if desired, the accuracy of this error estimate can be improved to $\mathcal{O}(\delta\mathbf{U}^3) = \mathcal{O}(h^{3p+3})$ by averaging Eqn. 2.107 with a dual form of the error estimate. See Appendix A (Sec. A.3.2) for a discussion of this topic.)

#### 2.4.2.4 Summary: Discrete Error Estimation and Mesh Adaptation

Here, we summarize the steps involved in the discrete error estimation and mesh adaptation process:

1. Solve $\mathbf{R}_H(\mathbf{U}_H) = 0$ on the coarse space, $\mathcal{V}_H$, to obtain $\mathbf{U}_H$.

2. Evaluate the output of interest, $J(\mathbf{U}_H)$.

3. Inject $\mathbf{U}_H$ to an order-incremented or uniformly refined space, $\mathcal{V}_h$. (Compute $\mathbf{U}_h^H$.)

4. Evaluate the fine-space residuals with the injected solution. (Compute $\mathbf{R}_h(\mathbf{U}_h^H)$.)

5. Solve (or approximate) the linear adjoint equation

$$\left.\frac{\partial \mathbf{R}_h}{\partial \mathbf{U}_h}^T\right|_{\mathbf{U}_h^H} \boldsymbol{\Psi}_h = \left.\frac{\partial J_h}{\partial \mathbf{U}_h}^T\right|_{\mathbf{U}_h^H} \tag{2.108}$$

   to find $\boldsymbol{\Psi}_h$ on the fine space.

6. Compute the error estimate $\delta J_{\text{est}} \approx -\boldsymbol{\Psi}_h^T \mathbf{R}_h(\mathbf{U}_h^H)$.

7. Correct the original $J(\mathbf{U}_H)$ with this error estimate. (Compute $J_{\text{corrected}} = J(\mathbf{U}_H) + \delta J_{\text{est}}$.) This corrected output is more accurate than $J(\mathbf{U}_H)$.

8. Localize $\delta J_{\text{est}}$ to individual elements in the mesh. (Compute the indicators $\varepsilon_e$ from Eqn. 2.103.)

9. Select a certain percentage of elements with the highest error indicators (or a related "figure of merit") and refine them.

10. Solve the primal problem on the new mesh and repeat steps 2-10 until the output error is driven below a desired tolerance.

## 2.5 Example: Adjoint-based Error Estimation and Mesh Adaptation for a 3D Wing

Here, we give an example of how the error estimation and adaptation procedures described above can provide improved output accuracy and reduced computational cost for a problem of engineering interest: transonic flow around a three-dimensional wing.

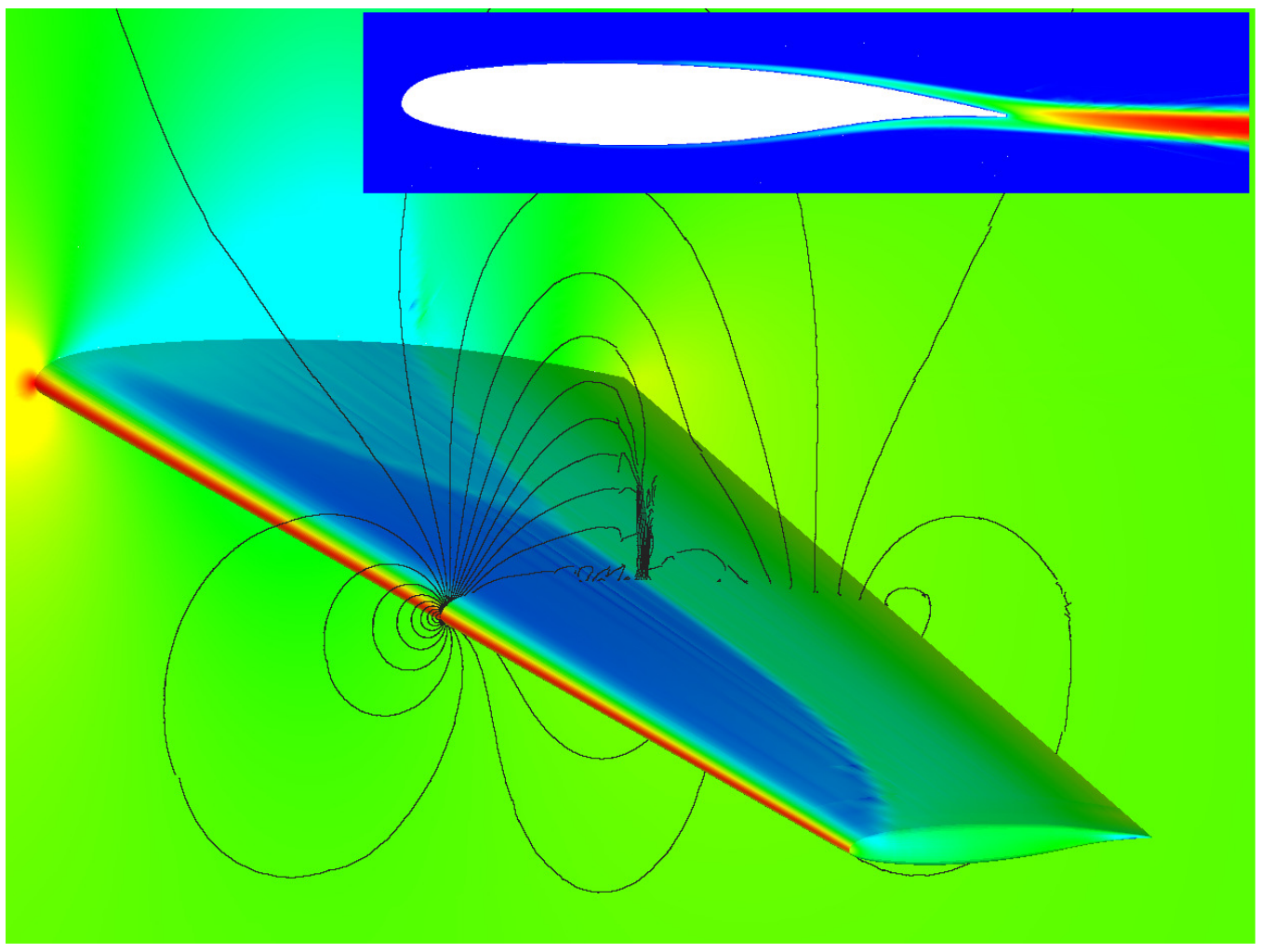

Figure 2.5: Pressure contours for $Ma = 0.76$, $Re = 5 \times 10^6$ flow around a wing.

Pressure contours around the wing, which has a free-stream Mach number of 0.76 and a Reynolds number of $5 \times 10^6$, are shown in Figure 2.5.

As described in [2], the goal for this problem is to predict the drag coefficient as accurately as possible, with as little computational expense as possible. To achieve this, the adjoint-based error estimation and adaptation procedures described above are followed. That is, the problem is first solved on an inexpensive (coarse) mesh using a Discontinuous Galerkin finite element method. The fine-space drag adjoint is then computed and used to generate both adaptive indicators and an error estimate. The elements in the mesh with the largest adaptive indicators are then refined, and the problem is solved again. This cycle is repeated until the error estimate is driven down to a desired level.

Figure 2.6 shows how the drag coefficent converges as the adaptive iterations progress. The drag values corresponding to the adjoint-based adaptations are given by the solid blue curve, while the "corrected" drag values (i.e. the drag values plus the adjoint-based error estimates) are given by the dashed blue lines. The red lines show the same adjoint-based adaptation, but with a slightly different figure-of-merit used to adapt the mesh. For our purposes, the red and blue curves can be treated as the same.

As can be seen, as the adaptive iterations progress (i.e. as we move to the right on the plot), the drag coefficient converges toward 0.021 and the error estimate steadily decreases, eventually reaching a small enough value that the adaptive process is terminated. Our final estimate for the drag coefficient is thus $\sim 0.0208$, with an uncertainty

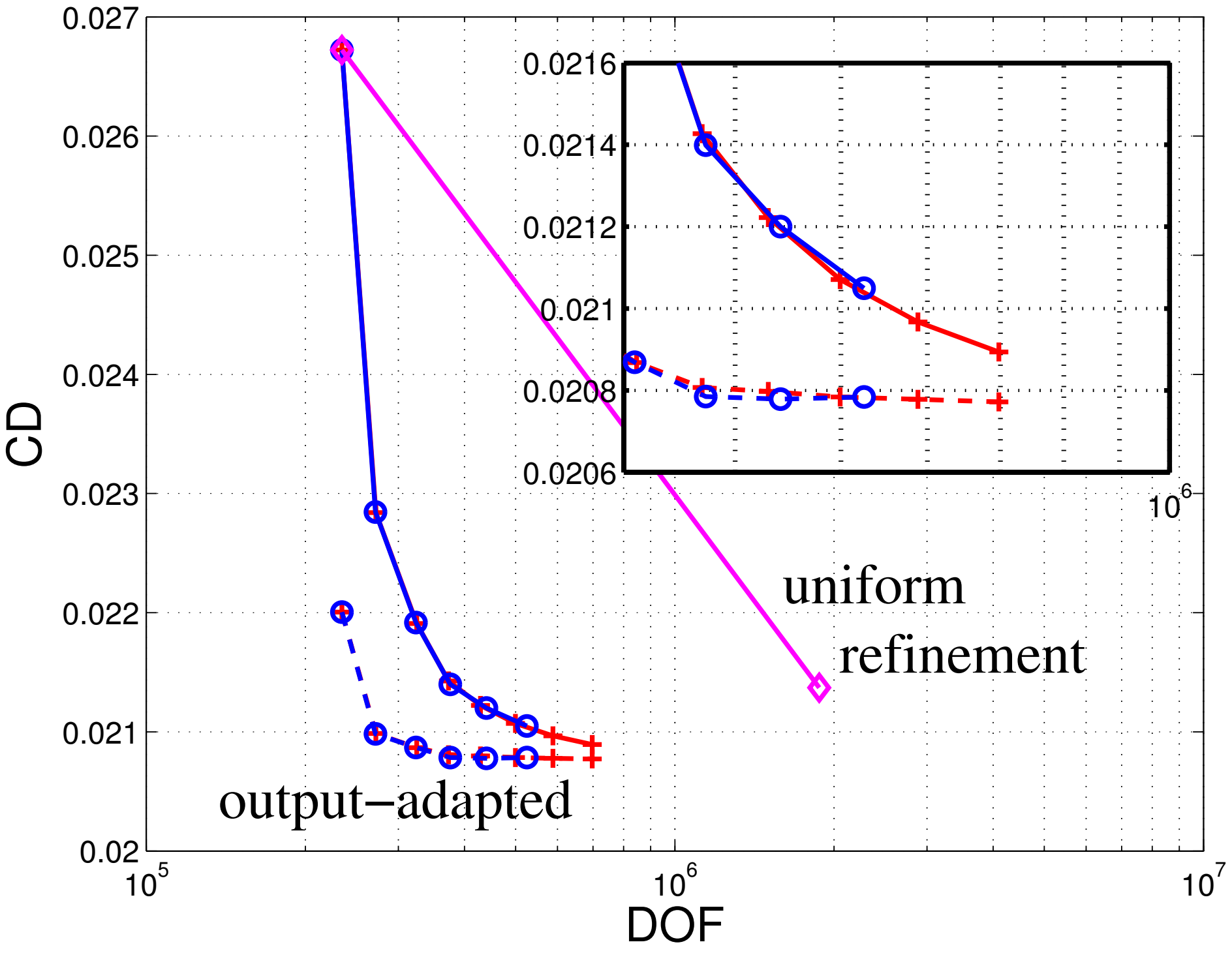

Figure 2.6: Comparison of drag coefficient convergence for different adaptation methods with respect to total mesh degrees-of-freedom (DOF). The **magenta** curve shows the convergence obtained by uniformly $h$-refining the mesh. The **solid** **blue**/**red** curves show the convergence obtained when adaptation is performed based on the drag adjoint (note that for our purposes the blue and red curves are essentially the same – both are based on the drag adjoint, but use slightly different "figures of merit" to refine the mesh [2]). Finally, the **dashed** **blue**/**red** curves show the convergence of the "corrected" output for the adjoint-based methods: i.e., the raw drag output plus the adjoint-based error estimate. The difference between the solid and dashed curves is thus the magnitude of the output error estimate, $\delta J_{\text{est}}$. Overall, we see that the adjoint-based methods achieve greater output accuracy than the uniform-refinement method for much less computational expense.

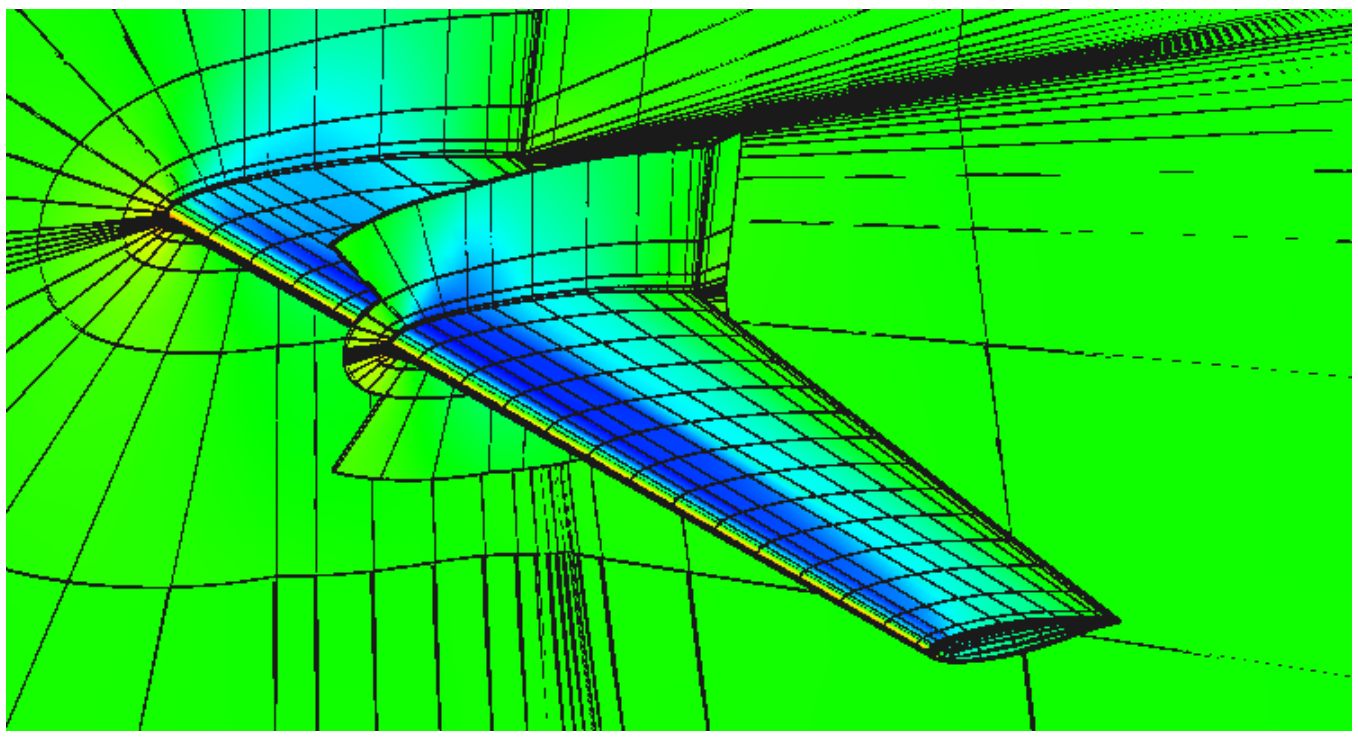

(a) Initial Mesh

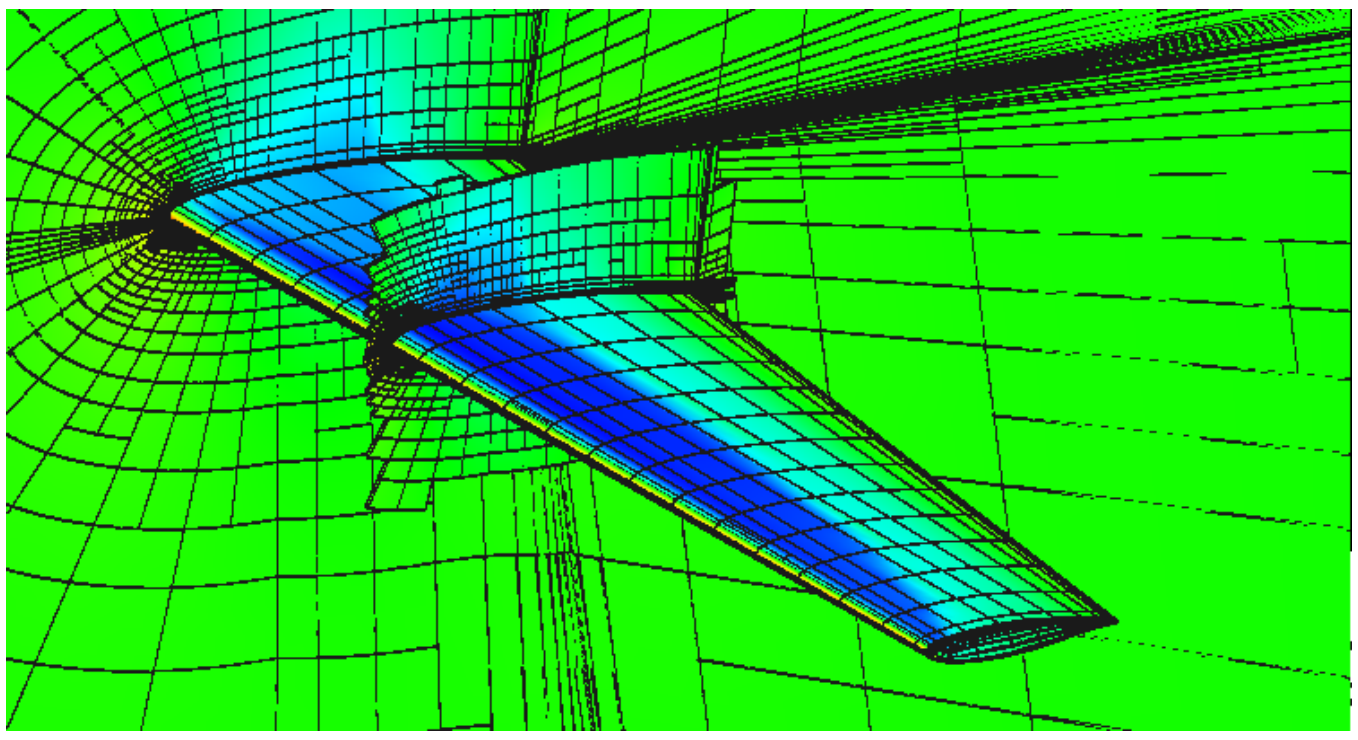

(b) Final drag-adjoint-adapted mesh

Figure 2.7: Initial and final meshes associated with the drag-adjoint-based adaptation. The adjoint-based method refines elements near the leading edge and in the boundary layer, leading to improved drag prediction.

of $\sim 0.002$.

The magenta curve, on the other hand, shows the drag values that are obtained when the initial mesh is uniformly (as opposed to adaptively) refined. While uniformly refining the mesh also improves the output accuracy, it is computationally intensive – so much so that only one refinement was feasible for this problem. However, to be confident in the final output value, we can see that at least one additional uniform refinement would be required beyond the final magenta point shown in Figure 2.6.

Thus, obtaining an accurate drag value with uniform refinement would ultimately require a mesh with over $10^7$ degrees-of-freedom. On the other hand, the adjoint-based adaptation provides an accurate drag value with fewer than $10^6$ degrees-of-freedom. Overall then, adjoint-based adaptation results in a nearly factor-of-10 reduction in the required mesh size for this case, with corresponding reductions in CPU time [2].

# CHAPTER III

# Unsteady Problems

While adjoint-based error estimation and mesh adaptation are often discussed in the context of steady problems, these techniques can be extended to unsteady problems as well. This extension is straightforward from a mathematical perspective: the time dimension can be treated as another "space-like" dimension, and the adjoint can be defined in a similar manner as before. The differences lie primarily in the details of the adjoint, error estimation, and mesh adaptation procedures, as well as in the fact that – unlike space – information propagates in a specific *direction* in time.

## 3.1 Unsteady Adjoints

As in the previous chapter, let us start by assuming we have a linear problem. While in the steady section we considered the system $\mathbf{AU} = \mathbf{F}$, here we will include a time derivative, so that the governing equations become

$$\mathbf{M}\frac{d\mathbf{U}}{dt} + \mathbf{AU} = \mathbf{F}\,. \tag{3.1}$$

Here, $\mathbf{M}$ is a "mass matrix," which is often just the identity matrix, but may in general differ (e.g. for finite element methods). As before, $\mathbf{A}$ represents a discrete spatial operator, while $\mathbf{U}$ represents the state and $\mathbf{F}$ is a prescribed source term. Many physical phenomena satisfy this form of equation, including heat diffusion, acoustic propagation, and the linearized Euler equations, to name a few.

To make the discussion more concrete, let us assume that the temporal derivative is discretized using a backward Euler method, such that

$$\mathbf{M}\,\frac{\mathbf{U}^m - \mathbf{U}^{m-1}}{\Delta t} + \mathbf{AU}^m = \mathbf{F}^m\,. \tag{3.2}$$

Here, a superscript on a vector denotes a time index, so that e.g. $\mathbf{U}^m$ represents the state at time level $m$. As before, the number of spatial degrees of freedom is assumed to be $N$, while the temporal index $m$ ranges from 1 to $N_t$, with $N_t$ being the total number of temporal degrees of freedom. (In addition, from now on, when we write a variable such as $\mathbf{U}$ or $\mathbf{F}$ *without* a time index, we are referring to the column vector containing all space-time components of that variable, so that e.g. $\mathbf{U} \equiv \{\mathbf{U}^m\}_{\forall m} \in \mathbb{R}^{N \times N_t}$.)

Returning to the problem, we see that since Eqn. 3.2 is linear, we can rewrite it as simply

$$\bar{\mathbf{A}}\mathbf{U} = \mathbf{F}\,, \tag{3.3}$$

where $\bar{\mathbf{A}}$ is the matrix representing the full space-time operator, i.e.

$$\left(\bar{\mathbf{A}}\mathbf{U}\right)^m \;\equiv\; \mathbf{M}\,\frac{\mathbf{U}^m - \mathbf{U}^{m-1}}{\Delta t} + \mathbf{A}\mathbf{U}^m\,. \tag{3.4}$$

Here, the bar is used to distinguish the space-time operator $\bar{\mathbf{A}}$ from the spatial operator, $\mathbf{A}$.

Written out in matrix form, Eqn. 3.3 looks like

$$\underbrace{\begin{bmatrix} \bullet & & & & & \\ \bullet & \bullet & & & & \\ & \bullet & \bullet & & & \\ & & \ddots & \ddots & & \\ & & & & \bullet & \\ & & & & \bullet & \bullet \end{bmatrix}}_{\bar{\mathbf{A}}} \underbrace{\begin{bmatrix} \mathbf{U}^1 \\ \mathbf{U}^2 \\ \mathbf{U}^3 \\ \vdots \\ \mathbf{U}^{N_t-1} \\ \mathbf{U}^{N_t} \end{bmatrix}}_{\mathbf{U}} = \underbrace{\begin{bmatrix} \mathbf{F}^1 \\ \mathbf{F}^2 \\ \mathbf{F}^3 \\ \vdots \\ \mathbf{F}^{N_t-1} \\ \mathbf{F}^{N_t} \end{bmatrix}}_{\mathbf{F}} \tag{3.5}$$

where each dot in the matrix has the dimension of the spatial Jacobian, $\mathbf{A}$:

$\bullet =$ 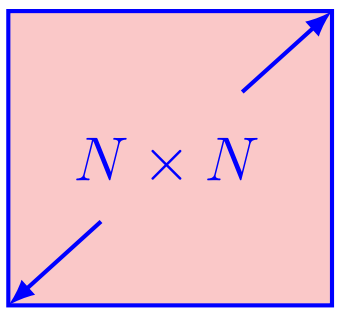

$=$ Size of spatial $\mathbf{A}$ matrix (3.6)

From Eqn. 3.5 we see that, due to the nature of the backward Euler discretization, the $\bar{\mathbf{A}}$ matrix has a block diagonal structure in which the equations at a given time level (i.e. row of $\bar{\mathbf{A}}$) depend on the states at both the current and previous time levels. This structure arises due to the forward-in-time propagation of physical information, and would be similar for other temporal discretizations as well.

Now, if we were interested in solving for the entire $\mathbf{U}$ vector, this could be accomplished by performing a forward time march, which amounts to inverting the entire $\bar{\mathbf{A}}$ matrix:

$$\mathbf{U} = \bar{\mathbf{A}}^{-1}\mathbf{F}\,. \tag{3.7}$$

However, as in the previous chapter (Sec. 2.1), we can again ask: what if, rather than the entire solution $\mathbf{U}$, we are interested in just a single component of $\mathbf{U}$? Let us again take as an example the last component of $\mathbf{U}$, i.e. $U_N^{N_t}$, which represents the last spatial unknown at the final time. Physically, this could represent a "point" output at a certain location in space, evaluated at the end of the simulation.

The solution vector (given by Eqn. 3.7) can be written out explicitly as

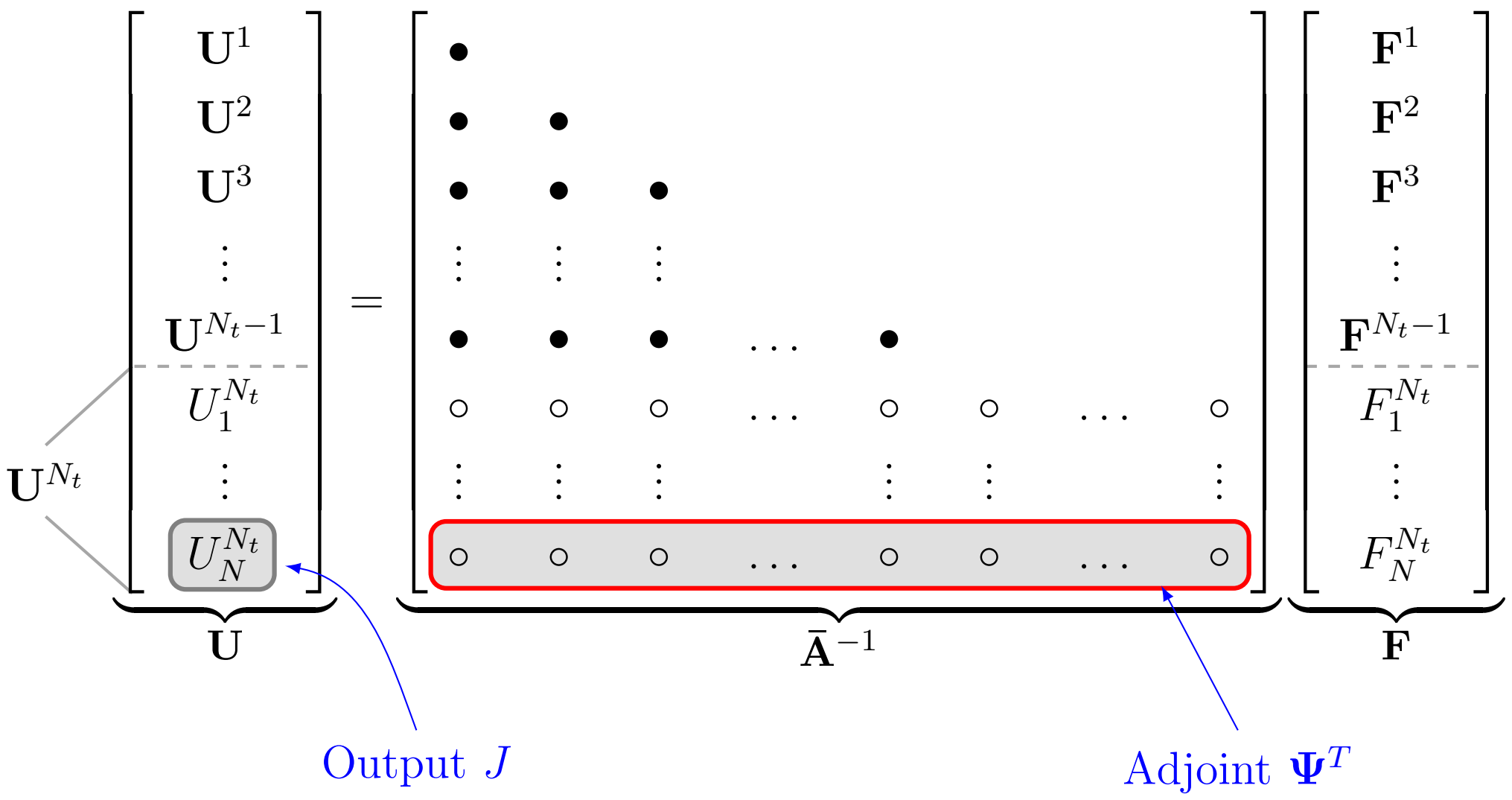

(3.8)

(3.9)

Here, we have expanded the state at the last timestep, $\mathbf{U}^{N_t}$, into its spatial components in order to reveal the desired output, $J = U_N^{N_t}$. Also, note that $\bar{\mathbf{A}}^{-1}$ has a lower block-triangular structure due to the fact that the solution at a given timestep depends only on the source terms associated with earlier times.

From this equation, we see that (as in the steady case) the only information required to compute $U_N^{N_t}$ is the highlighted row of $\bar{\mathbf{A}}^{-1}$. Likewise, from our earlier argument, this row also represents the *sensitivity* of the output $J$ to perturbations in the source term $\mathbf{F}$, since it multiplies the components of $\mathbf{F}$ during the computation of $U_N^{N_t}$. Thus, as before, this row is exactly the **adjoint** of $J$, which we can again denote by $\mathbf{\Psi}^T$. Finally, note that for unsteady problems, $\mathbf{\Psi}^T$ is now a vector spanning the entire space-time domain, meaning it can be thought of as a quantity that evolves in time (similar to the state).

As before, $J$ need not be a point output – it could be any linear combination of the components of $\mathbf{U}$, and could then be represented in **dual form** as

$$J = \mathbf{\Psi}^T \mathbf{F} \ , \tag{3.10}$$

where the adjoint $\mathbf{\Psi}^T$ is defined to be an output-specific weighted average of the rows of $\bar{\mathbf{A}}^{-1}$:

$$\mathbf{\Psi}^T = \frac{\partial J}{\partial \mathbf{U}} \bar{\mathbf{A}}^{-1} \ . \tag{3.11}$$

Finally, to write this equation in a more common form, we can define the **space-time residual** as

$$\overline{\mathbf{R}} = \bar{\mathbf{A}}\mathbf{U} - \mathbf{F} = \mathbf{0} \, , \tag{3.12}$$

so that

$$\frac{\partial \overline{\mathbf{R}}}{\partial \mathbf{U}} = \bar{\mathbf{A}} \, . \tag{3.13}$$

Making this replacement in Eqn. 3.11 and rearranging then gives the following form of the **adjoint equation**:

$$\frac{\partial \overline{\mathbf{R}}}{\partial \mathbf{U}}^T \mathbf{\Psi} = \frac{\partial J}{\partial \mathbf{U}}^T \ . \tag{3.14}$$

This is virtually identical to the adjoint equation derived in the steady case (Eqn. 2.15). Comparing the two, we see that the extension of the adjoint to unsteady problems consists primarily in drawing a bar over the residual. Indeed, the true differences lie not in the theory but in how this adjoint equation is *solved*. We will discuss this solution procedure later on. First, let us generalize to nonlinear problems.

### 3.1.1 Nonlinear Unsteady Problems

In practice, we are often interested in problems where both the governing equations and output are nonlinear. For example, we may wish to solve the unsteady Navier-Stokes equations around e.g. an airfoil and compute a time-averaged lift or drag output.

In that case, instead of the equation

$$\mathbf{M}\frac{d\mathbf{U}}{dt} + \underbrace{\mathbf{A}\mathbf{U} - \mathbf{F}}_{\mathbf{R}} = \mathbf{0}\,, \tag{3.15}$$

where the spatial residual $\mathbf{R}$ is linear, we would instead write the governing equations as

$$\mathbf{M}\frac{d\mathbf{U}}{dt} + \mathbf{R}(\mathbf{U}) = \mathbf{0}\,, \tag{3.16}$$

where $\mathbf{R}(\mathbf{U})$ is a general nonlinear function of $\mathbf{U}$.

Next, if we assume for simplicity that a backward Euler method is used, the unsteady residual associated with the $m$th temporal degree of freedom can be written as

$$\underbrace{\mathbf{M}\,\frac{\mathbf{U}^m - \mathbf{U}^{m-1}}{\Delta t} + \mathbf{R}(\mathbf{U}^m)}_{\overline{\mathbf{R}}^m(\mathbf{U})} = \mathbf{0}\,. \tag{3.17}$$

Here, the bar over the space-time residual $\overline{\mathbf{R}}^m(\mathbf{U})$ is used to distinguish it from the spatial residual, $\mathbf{R}(\mathbf{U})$.

Stepping back one level further, the entire set of space-time residuals can then be written:

$$\overline{\mathbf{R}}(\mathbf{U}) = \mathbf{0}\,. \tag{3.18}$$

Now, recall that our goal in the end is to define the adjoint equation, which requires the derivative of the residual with respect to $\mathbf{U}$. While in the above section this derivative was just the constant operator $\bar{\mathbf{A}}$ (from Eqn. 3.12), for a nonlinear $\overline{\mathbf{R}}(\mathbf{U})$, this derivative will be non-constant. Instead, it will be a function of the particular state about which it is computed. We can write this derivative as the

space-time Jacobian

$$\left.\frac{\partial \overline{\mathbf{R}}}{\partial \mathbf{U}}\right|_{\mathbf{U}}, \tag{3.19}$$

which, as shown, is evaluated at a given state $\mathbf{U}$. Note that while the individual entries in this matrix will depend on $\mathbf{U}$, its sparsity pattern is identical to that of $\bar{\mathbf{A}}$ in Eqn. 3.5.

Likewise, as with the residual, if the output $J(\mathbf{U})$ is nonlinear, its derivative $\partial J/\partial \mathbf{U}$ will also depend on the state $\mathbf{U}$, and can be written as

$$\left.\frac{\partial J}{\partial \mathbf{U}}\right|_{\mathbf{U}}. \tag{3.20}$$

By substituting these residual and output linearizations into Eqn. 3.14, we then obtain the definition of the adjoint for nonlinear problems:

$$\boxed{\left.\frac{\partial \overline{\mathbf{R}}}{\partial \mathbf{U}}^T\right|_{\mathbf{U}} \mathbf{\Psi} = \left.\frac{\partial J}{\partial \mathbf{U}}^T\right|_{\mathbf{U}}}. \tag{3.21}$$

For a given output of interest, this is a linear system of space-time equations that can be solved for $\mathbf{\Psi}$. If the problem were steady, we would use a standard iterative method to find the solution. However, for unsteady problems, we can solve this problem more efficiently by taking advantage of the direction of information flow. We discuss this below.

### 3.1.2 Solution of the Unsteady Adjoint Equation

As mentioned, regardless of whether the residual is linear or nonlinear, the primal Jacobian $\partial\overline{\mathbf{R}}/\partial\mathbf{U}$ will have the sparsity pattern of the $\bar{\mathbf{A}}$ matrix in Eqn. 3.5. (Assuming a two-step temporal discretization is used.) This structure arises due to the fact that primal information propagates only forward in time. Consequently, the primal equation is most efficiently solved by performing a forward time march.

Just like the primal equation, the adjoint equation (Eqn. 3.21) also spans the entire space-time domain, and can therefore be thought of as representing the evolution of adjoint information in time. This raises the question: can we also solve the *adjoint* equation by using a forward time march? To answer this question, we take a closer look at the sparsity pattern of the discrete primal and adjoint operators.

$$
\underbrace{\begin{bmatrix} \bullet & & & & \\ \bullet & \bullet & & & \\ & \bullet & \bullet & & \\ & & \ddots & \ddots & \\ & & & \bullet & \\ & & & \bullet & \bullet \end{bmatrix}}_{\frac{\partial \overline{\mathbf{R}}}{\partial \overline{\mathbf{U}}}}
\qquad
\underbrace{\begin{bmatrix} \bullet & \bullet & & & \\ & \bullet & \bullet & & \\ & & \bullet & \ddots & \\ & & & \ddots & \\ & & & \bullet & \bullet \\ & & & & \bullet \end{bmatrix}}_{\frac{\partial \overline{\mathbf{R}}}{\partial \overline{\mathbf{U}}}^T}
\tag{3.22}
$$

**Primal Unsteady Jacobian** (left), **Adjoint Unsteady Jacobian** (right)

The above diagram shows the primal Jacobian on the left and its transpose on the right. From Eqn. 3.21, this transpose operator is exactly the one that appears in the adjoint equation. As before, each "dot" in these matrices represents a spatial Jacobian matrix (plus any temporal discretization terms), i.e.

$$
\bullet = \boxed{N \times N} = \text{Size of spatial } \frac{\partial \mathbf{R}}{\partial \mathbf{U}} \text{ matrix} \tag{3.23}
$$

In the primal Jacobian, since there is a single unknown (vector) in the first row, it is clear that we should first solve for this unknown, then perform forward-substitution (i.e. a forward time march) to obtain the remaining unknowns. On the other hand, due to the transposed nature of the adjoint operator, we see that it has a single unknown in the *last* row, which corresponds to the final time in the simulation. Therefore, unlike the primal problem, the most efficient way to solve the **adjoint problem** is to perform a *back*-substitution – in other words, a **backward time march**.

To see this more clearly, we can write out the adjoint equation (Eqn. 3.21) explic-

itly as

$$\underbrace{\begin{bmatrix} \bullet & \bullet & & & & \\ & \bullet & \bullet & & & \\ & & \bullet & \ddots & & \\ & & & \ddots & & \\ & & & & \bullet & \bullet \\ & & & & & \bullet \end{bmatrix}}_{\frac{\partial \bar{\mathbf{R}}}{\partial \mathbf{U}}^T} \underbrace{\begin{bmatrix} \boldsymbol{\Psi}^1 \\ \boldsymbol{\Psi}^2 \\ \boldsymbol{\Psi}^3 \\ \vdots \\ \boldsymbol{\Psi}^{N_t-1} \\ \boldsymbol{\Psi}^{N_t} \end{bmatrix}}_{\boldsymbol{\Psi}} = \underbrace{\begin{bmatrix} (\partial J/\partial \mathbf{U}^1)^T \\ (\partial J/\partial \mathbf{U}^2)^T \\ (\partial J/\partial \mathbf{U}^3)^T \\ \vdots \\ \left(\partial J/\partial \mathbf{U}^{N_t-1}\right)^T \\ \left(\partial J/\partial \mathbf{U}^{N_t}\right)^T \end{bmatrix}}_{\frac{\partial J}{\partial \mathbf{U}}^T} \tag{3.24}$$

The right-hand side of this equation is the output linearization, which is a known quantity that depends on the particular output of interest. (For example, for a final-time output, all terms on the right would be zero except for $\partial J/\partial \mathbf{U}^{N_t}$.) On the left-hand side, we have the Jacobian transpose weighting the adjoint vector $\boldsymbol{\Psi}$. As suggested, it is clear from this diagram that to find $\boldsymbol{\Psi}$, we should first compute $\boldsymbol{\Psi}^{N_t}$, then perform a back-substitution to find the remaining adjoint values. Since the backward Euler method is adjoint-consistent[1], this is equivalent to performing a backward time march of the adjoint problem, starting from a final rather than initial condition.

#### 3.1.2.1 State-Dependence of the Adjoint Equation

Note that for nonlinear problems, both $\frac{\partial \bar{\mathbf{R}}}{\partial \mathbf{U}}^T$ and $\frac{\partial J}{\partial \mathbf{U}}^T$ in the above equation must be evaluated at a particular primal state, $\mathbf{U}$. Thus, for these problems, the adjoint equation cannot be solved until some approximation of the primal solution is obtained. For this reason, the procedure (as depicted in Fig. 3.1) is typically to:

1. March the primal problem forward in time to compute the state $\mathbf{U}$.
2. Save $\mathbf{U}$ to disk.
3. March the adjoint equation backward in time, evaluating $\frac{\partial \bar{\mathbf{R}}}{\partial \mathbf{U}}^T\Big|_{\mathbf{U}}$ and $\frac{\partial J}{\partial \mathbf{U}}^T\Big|_{\mathbf{U}}$ with the corresponding $\mathbf{U}$ values at each time level.

[1] Note that not all temporal discretizations are adjoint-consistent. For example, the second-order backward difference (BDF2) method is adjoint-inconsistent if non-uniform time steps are used [13]. However, Runge-Kutta methods [12] – as well as the DG-in-time method employed in this work – are adjoint-consistent regardless of time-step size.

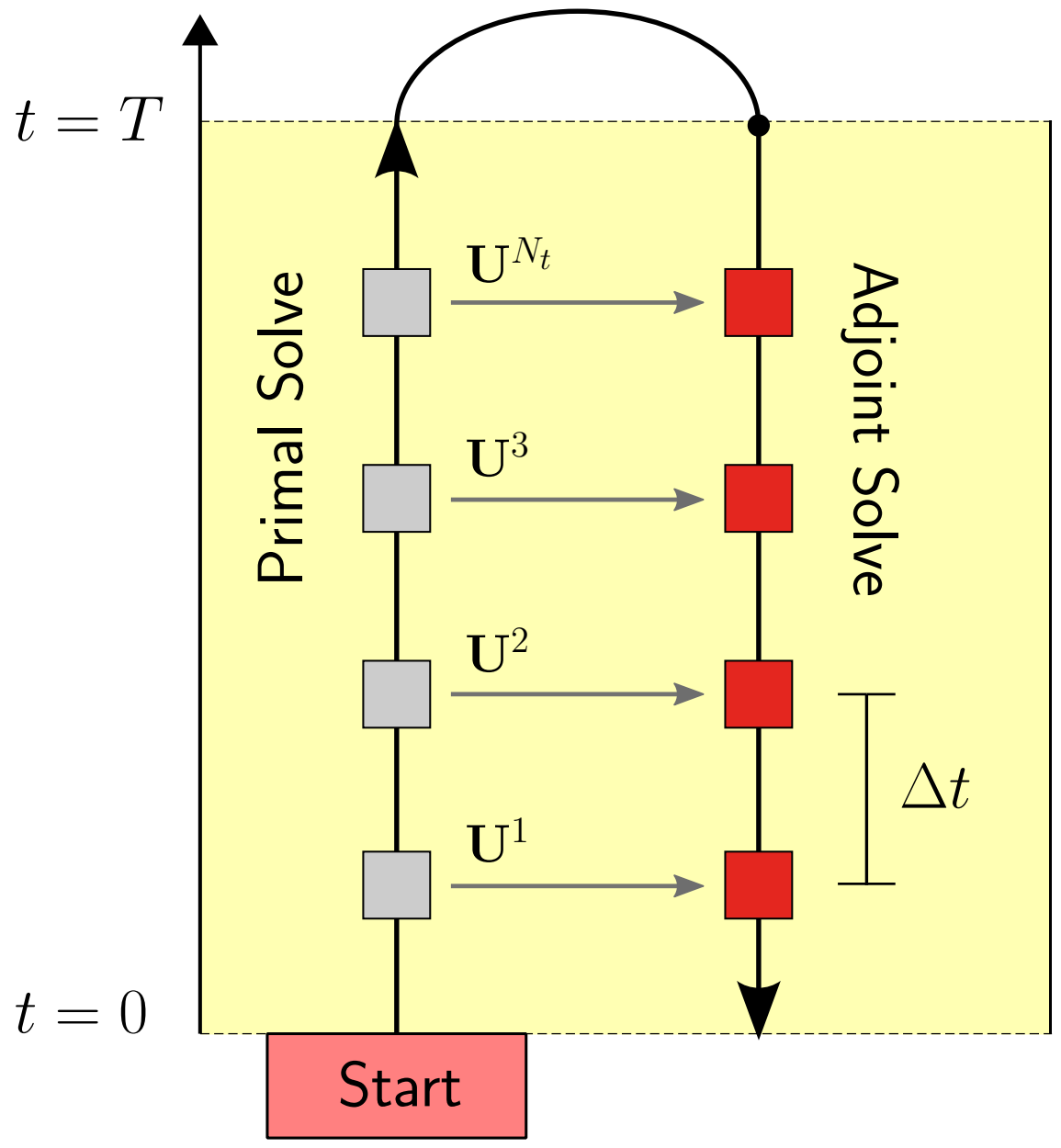

Figure 3.1: Unsteady primal and adjoint solution procedure. For nonlinear problems, the state $\mathbf{U}$ is first computed via a forward time march. It is then stored and used in the adjoint problem, which is solved by marching backward in time.

**Remark 5.** **(Solution Checkpointing)** In certain cases, storing the entire space-time state $\mathbf{U}$ may be prohibitive in terms of memory. In that case, a procedure known as "solution checkpointing" can be performed. This procedure consists of saving the solution at a relatively small number of "checkpoints" in time, and then re-solving for the solution between the checkpoints as the backward-in-time adjoint solve progresses. While this results in additional computational expense, it alleviates the memory requirements associated with storing the entire $\mathbf{U}$ vector.

#### 3.1.2.2 Continuous Unsteady Adjoint

Note that the backward-in-time propagation of adjoint information can also be revealed by studying the continuous adjoint equation for unsteady problems. In fact, we have already discussed a continuous unsteady adjoint problem without recognizing it. In our original steady advection example (i.e. Eqn. 2.39), information flows in only one direction (to the right), just as in an unsteady problem it would flow only forward in time. Thus, a steady advection problem is actually "time-like," so that we could replace all instances of $x$ with $t$ to no effect. The primal problem would then become an ordinary differential equation in time, and the adjoint equation given by

Eqn. 2.46 would become:

$$L^*\psi = -a\frac{d\psi}{dt} = g(t) \quad \text{and} \quad \psi\big|_T = 0 \quad , \tag{3.25}$$

where $T$ denotes the final time. We see then that we have a *final-time* condition on $\psi$ (instead of an initial condition) as well as a negative sign in front of the $a\,d\psi/dt$ term – both of which indicate a backward flow of information in time.

## 3.2 Unsteady Error Estimation and Mesh Adaptation

With the unsteady adjoint defined, we now give an overview of how it can be used to perform output-based error estimation and mesh adaptation for unsteady CFD simulations.

### 3.2.1 Unsteady Error Estimation

Assume that we have computed an unsteady output $J(\mathbf{U}_H)$ on a coarse space $\mathcal{V}_H$, and would like to estimate the amount of error in this output. As in the steady case, we can estimate the error with respect to a "fine space," $\mathcal{V}_h$. For unsteady problems, we can choose $\mathcal{V}_h$ to be a uniformly refined version of $\mathcal{V}_H$ in both space and time. For example, the space-time grid could be uniformly h-refined, or (if the numerical method allows) the solution order could be uniformly incremented in both space and time.

In either case, we would then derive the output error estimate

$$\delta J_{\text{est}} = J_h(\mathbf{U}_h) - J_H(\mathbf{U}_H) \tag{3.26}$$

in the same manner as for steady problems, by performing Taylor expansions of both the fine-space output and residuals about the coarse-space state. Indeed, looking back at Sec. 2.4.2, we see that we made no assumptions about whether the problem was steady or unsteady. Thus, the results from that section carry over directly, with the only difference being that we must now place a bar over the residual to denote that it is unsteady. For unsteady problems, the error estimate given by Eqn. 2.107 then becomes

$$\delta J_{\text{est}} \;\approx\; -\mathbf{\Psi}_h^T\,\overline{\mathbf{R}}_h(\mathbf{U}_h^H) \;+\; \mathcal{O}(\delta\mathbf{U}^2) \quad . \tag{3.27}$$

This is just the product of the adjoint (computed on the fine space) and the fine-space residuals, $\overline{\mathbf{R}}_h(\mathbf{U}_h^H)$.

#### 3.2.1.1 Fine-Space Unsteady Adjoint Solve

For unsteady problems, the cost of a full fine-space adjoint solve can be prohibitive, due to the fact that it involves a full backward time-march. In practice, rather than computing the fine-space adjoint $\mathbf{\Psi}_h$ exactly, we can either (1) compute a coarse space adjoint $\mathbf{\Psi}_H$ and perform several (inexact) smoothing iterations on the fine space, or (2) compute $\mathbf{\Psi}_H$ and perform a nearest-neighbors reconstruction in space and time to obtain an approximation for $\mathbf{\Psi}_h$. Since these procedures depend upon the numerical discretization, they will be discussed in more detail later on.

### 3.2.2 Unsteady Mesh Adaptation

In a similar manner as for steady problems, the error estimate in Eqn. 3.27 can be localized to individual space-time "elements" in the mesh. This enables us to determine which regions of the mesh are contributing most to the output error, and to then selectively adapt those regions.

For example, for a backward Euler discretization in time (combined with, e.g., a finite volume or finite element method in space), the error estimate can be written as a sum over all space-time elements in $\mathcal{V}_h$ as follows:

$$\delta J_{\text{est}} \approx \sum_{k=1}^{N_t^h} \sum_{e=1}^{N_e^h} -\left(\mathbf{\Psi}_{h,\,e}^k\right)^T \overline{\mathbf{R}}_{h,\,e}^k(\mathbf{U}_h^H) \quad . \tag{3.28}$$

Here, $N_t^h$ is the number of time steps on the fine space $\mathcal{V}_h$, while $N_e^h$ is the number of spatial elements (which could theoretically depend on time). A pair of indices $(e, k)$ then corresponds to a single space-time element, which in this case would mean a spatial element $e$ over a particular time step, $[k-1, k]$. If $\mathcal{V}_H \subset \mathcal{V}_h$, the output error generated on each element $(e, k)$ in $\mathcal{V}_h$ could then be summed over the "parent" element in $\mathcal{V}_H$ to obtain a coarse-space error indicator.

#### 3.2.2.1 Space-Time Anisotropy

Unlike steady problems, having an estimate for the total amount of output error generated on each space-time element does not give us enough information to adapt the mesh. This is because, for unsteady problems, not only do we need to determine *which* space-time elements to adapt, but also whether to adapt them in space, time, or both. Thus, we need to determine how much of the output error generated on a given space-time element is due to the spatial vs. temporal discretization. In other words, we need a measure of **space-time *anisotropy***.

This anisotropy indicator can be computed by projecting the space-time adjoint onto semi-refined meshes in space and time. We will not discuss the details of this procedure here, but once computed, the anisotropy indicator provides us with the fraction of output error on each element due to the spatial and temporal discretizations, which we will call

$$\beta_{e,k}^{\text{space}} \quad \text{and} \quad \beta_{e,k}^{\text{time}}, \tag{3.29}$$

respectively, where

$$\beta_{e,k}^{\text{time}} = 1 - \beta_{e,k}^{\text{space}}. \tag{3.30}$$

These error fractions can then be multiplied by the output error indicator on each element to determine the individual spatial and temporal errors.

#### 3.2.2.2 Adaptation Mechanics

Finally, once we have computed an approximation of $\mathbf{\Psi}_h$ and the spatial/temporal errors on each element, we are ready to adapt the mesh. The next question is: *how* should we adapt the mesh?

In unsteady problems, critical flow features such as vortices move throughout the domain as time progresses. In order to maintain resolution of these (and other) features, we would like our mesh adaptation algorithm to "track" them as they propagate, provided they are deemed important by the adjoint. Thus, we would ideally like the spatial mesh resolution to change **dynamically** in time (i.e. to be different at each time step). Furthermore, in order to address temporal errors, we would like to selectively **refine** or **coarsen** time step sizes. By combining these two procedures, we can arrive at an algorithm that is capable of eliminating errors as they propagate in both space and time.

Note that when deciding which elements and time steps to adapt, we do not use the error estimates directly as our adaptive indicator. Since in the end we are interested in reducing the most error for the least computational cost, this cost must be factored into the adaptive indicator as well. Thus, as our final adaptive indicator, we compute a "**figure of merit**" that represents the amount of **output error on a** **given element or time slab** *divided by* the **cost of refinement** (i.e. the number of new degrees of freedom introduced). Since refining a single spatial element introduces a different number of degrees of freedom than refining an entire time step, this figure of merit will give a different (and ideally, more computationally efficient) result than refining based on error values alone.

#### 3.2.2.3 Summary: Unsteady Error Estimation and Mesh Adaptation

Figure 3.2 provides an overview of the unsteady error estimation and mesh adaptation process, the steps of which are listed below.

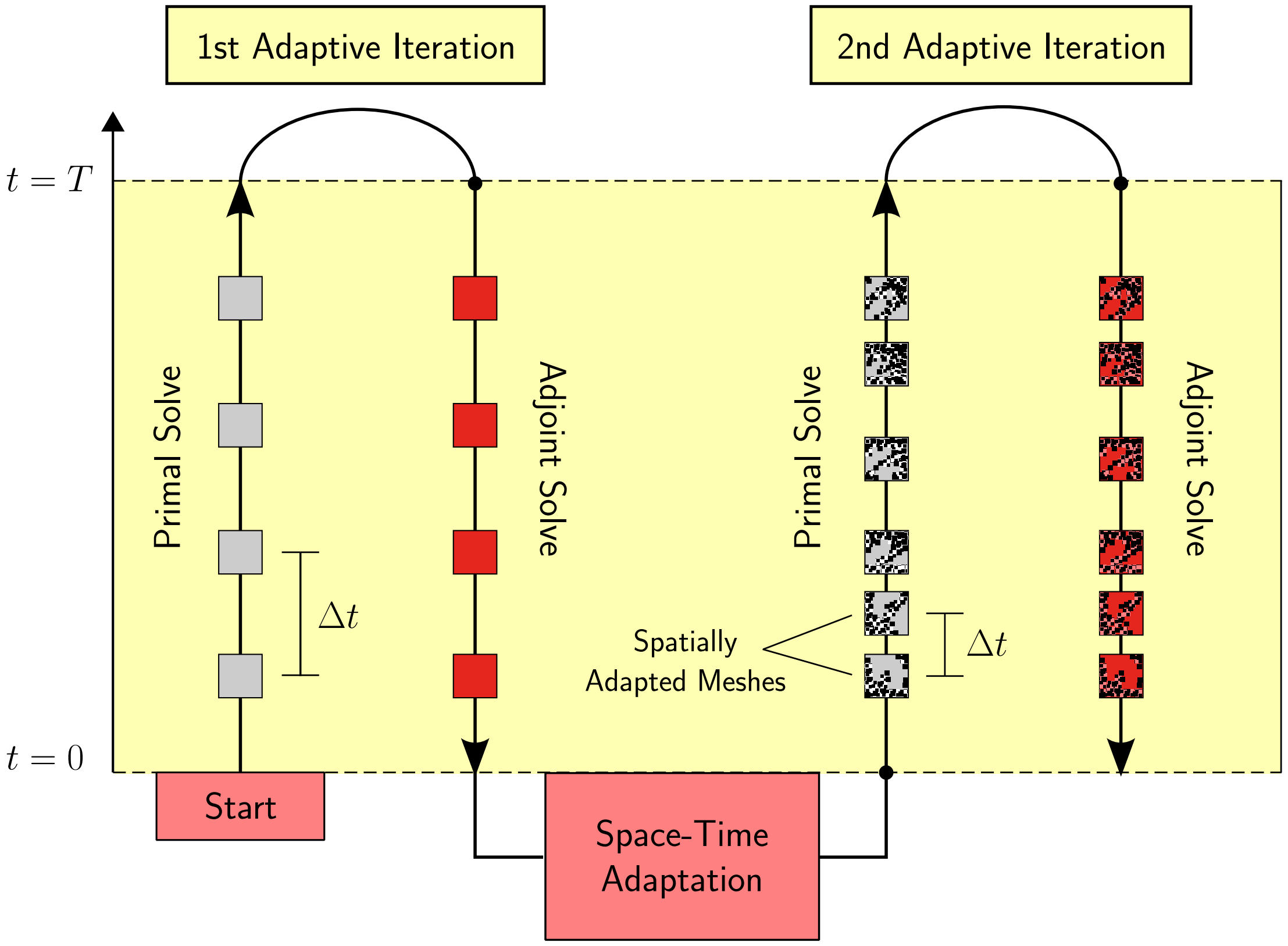

Figure 3.2: For each adaptive iteration, the primal problem is marched forward in time, and the adjoint problem is marched backward in time (on a uniformly refined space-time mesh). An output error estimate is then computed and mesh adaptation in space and time is performed. The spatial mesh is adapted differently at each time-step (as indicated by the variable shading in the 2nd iteration above), and time steps are selectively refined or coarsened.

**Procedure**:

1. Solve $\overline{\mathbf{R}}_H(\mathbf{U}_H) = 0$ on the coarse space, $\mathcal{V}_H$, to obtain $\mathbf{U}_H$.

2. Evaluate the output of interest, $J(\mathbf{U}_H)$.

3. Inject $\mathbf{U}_H$ to an order-incremented or uniformly refined space-time mesh, $\mathcal{V}_h$. (Compute $\mathbf{U}_h^H$.)

4. Evaluate the space-time residuals on $\mathcal{V}_h$ with the injected solution. (Compute $\overline{\mathbf{R}}_h(\mathbf{U}_h^H)$.)

5. Solve (or approximate) the fine-space adjoint equation

$$\left.\frac{\partial \overline{\mathbf{R}}_h}{\partial \mathbf{U}_h}^T\right|_{\mathbf{U}_h^H} \mathbf{\Psi}_h = \left.\frac{\partial J_h}{\partial \mathbf{U}_h}^T\right|_{\mathbf{U}_h^H} \tag{3.31}$$

   for $\mathbf{\Psi}_h$ by marching backward in time.

6. Compute the error estimate $\delta J_{\text{est}} \approx -\mathbf{\Psi}_h^T \overline{\mathbf{R}}_h(\mathbf{U}_h^H)$.

7. Correct the original $J(\mathbf{U}_H)$ with this error estimate. (Compute $J_{\text{corrected}} = J(\mathbf{U}_H) + \delta J_{\text{est}}$.) This corrected output is more accurate than $J(\mathbf{U}_H)$.

8. Localize $\delta J_{\text{est}}$ to individual space-time elements in the mesh.

9. Compute **space-time anisotropy** fractions $\beta_{e,k}^{\text{space}}$ and $\beta_{e,k}^{\text{time}}$, which indicate how much of $\delta J_{\text{est}}$ on each space-time element is due to the spatial vs. temporal discretization.

10. Compute a **figure of merit** representing the amount of output error eliminated by refining a given elment/step divided by the additional degrees of freedom introduced by the refinement.

11. Select a certain percentage of elements or time steps with the highest figure of merit and refine them. Coarsen a certain percentage of elements/steps with the lowest figure of merit.

12. Solve the primal problem on the new mesh and repeat steps 2-12 until the output error is driven below a desired tolerance.

## 3.3 Example: Dynamic Mesh Adaptation for Pitching and Plunging Airfoils

To demonstrate the unsteady error estimation and adaptation procedures described above, we show an example case consisting of two airfoils pitching and plunging in series [10]. The motion of the airfoils is similar to that of dragonfly wings in flight.

The airfoils start from an impulsive free-stream condition and undergo three periods of motion. The plunge amplitude is 0.25 chords, the pitch amplitude is 30°, and the period of both motions is $T = 2.5$. The Strouhal, Mach, and Reynolds numbers are 2/3, 0.3, and 1200, respectively. The airfoils are offset 4.5 chords horizontally and 1 chord vertically, and are situated in a 60 x 60 chord-length mesh. An Arbitrary Lagrangian-Eulerian (ALE) DG formulation of the compressible Navier-Stokes equations is used to model the problem in space, combined with an $r = 1$ (linear basis function) DG method in time.

Entropy contours at various phases of the motion are shown in Fig. 3.3. A reverse Kármán vortex street develops behind each airfoil, and the second airfoil interacts with the wake from the first airfoil near the end of the simulation. We take our output of interest to be the lift on the second airfoil integrated from time $t = 7.25$ to $t = 7.5$ (the final time).

To compute this output, four different refinement strategies are compared: adjoint-based adaptation, residual-based adaptation, uniform $h$-refinement, and uniform $p$-refinement. All methods start from an initial $p = 1$, 90 time step solution. During each adaptation, the uniform $h$- and $p$-refinement methods refine the entire space-time mesh by bisecting and order incrementing all spatial elements (respectively) and doubling the number of time steps. On the other hand, the adjoint- and residual-based methods refine $\sim 35\%$ of space-time elements while coarsening $\sim 5\%$. Adaptation for these methods consists of incrementing or decrementing the solution order $p$ on each element, while bisecting or coarsening time steps as appropriate.

Note that the residual-based method refines the mesh wherever the residuals (i.e. truncation errors) are large, and is a relatively common adaptation strategy found in the literature. Finally, note that while in theory the spatial order $p$ of the DG solution could be taken as high as desired, for this problem we constrain it to lie between 0 and 5 for all methods.

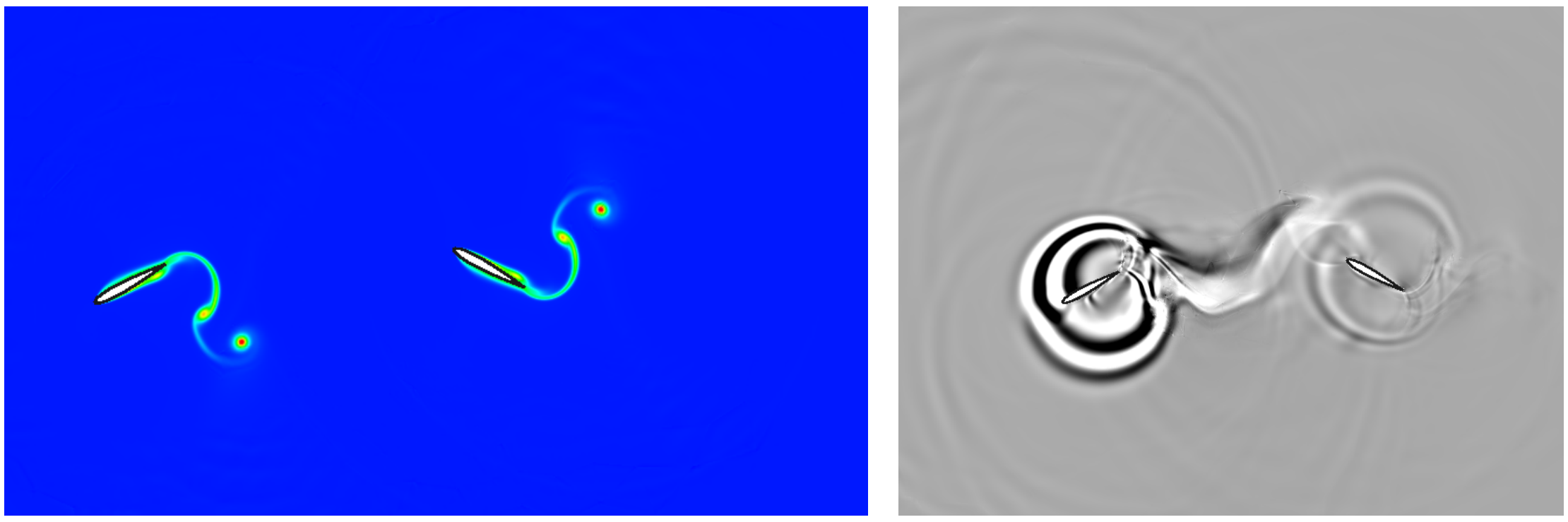

(a) $t = 0.70T$

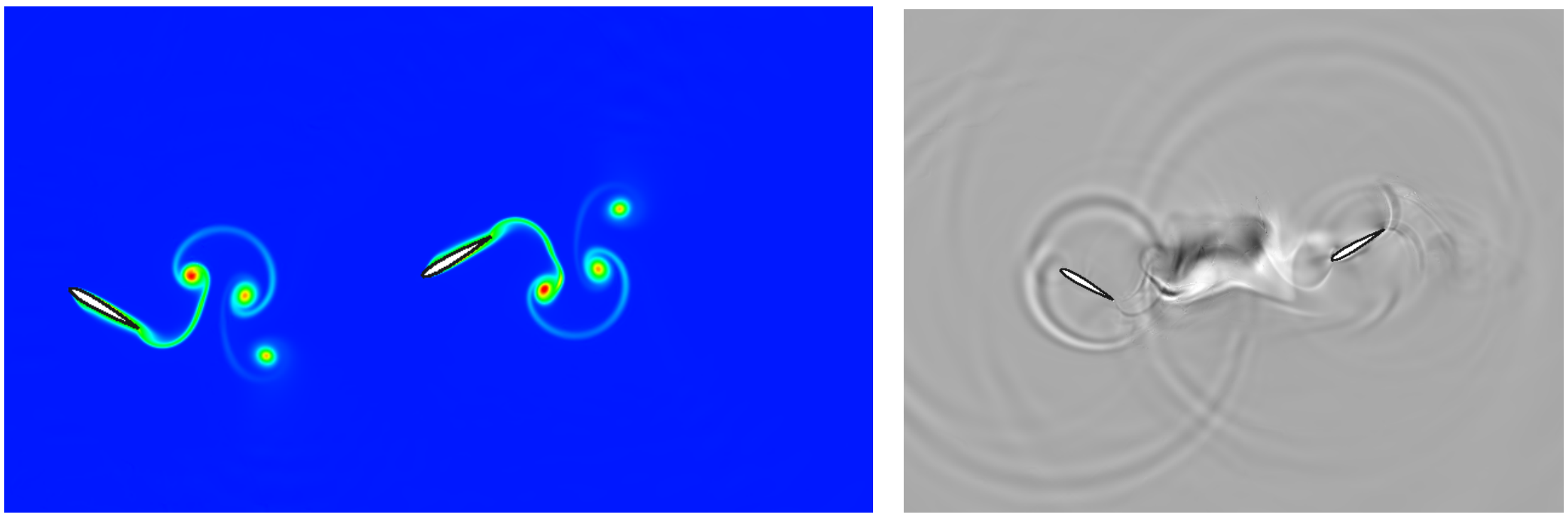

(b) $t = 1.25T$

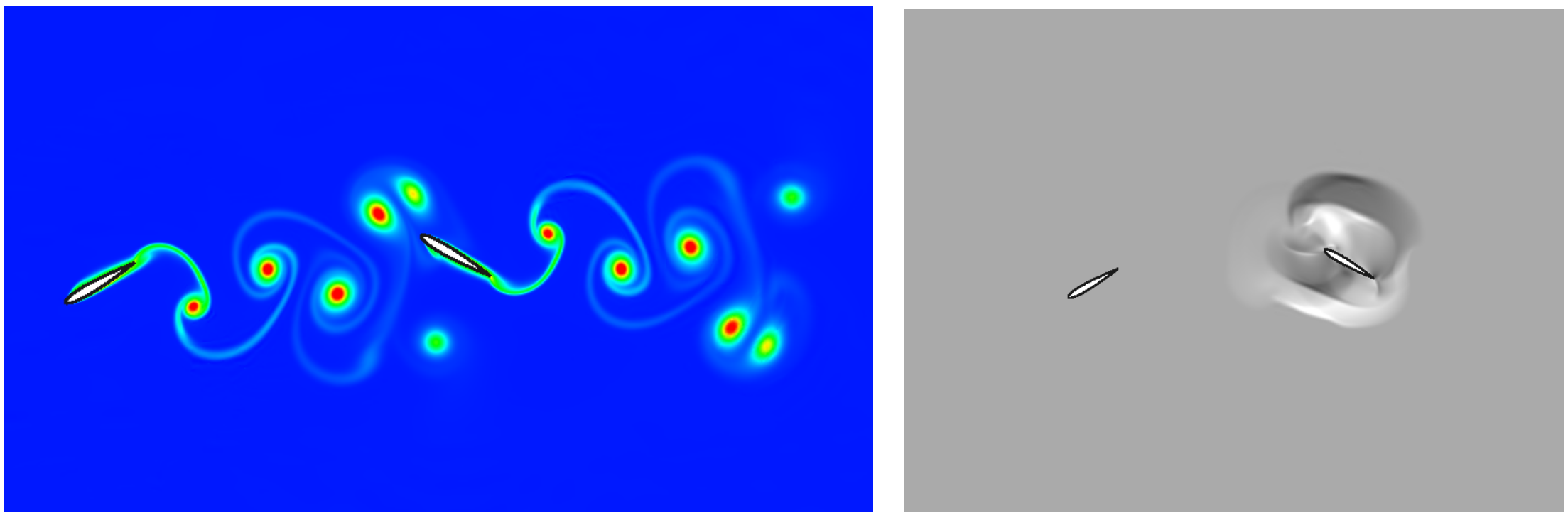

(c) $t = 2.75T$

Figure 3.3: Entropy (left) and adjoint (right) contours at various stages of the motion on a fine mesh. The adjoint contours have been scaled such that black corresponds to +2, white corresponds to -1, and medium gray is 0. Both acoustic and convective modes of error propagation can be seen in the adjoint contours. At the final time, the adjoint field collapses onto the second airfoil, since errors made far away from this airfoil no longer have time to reach it, making its sensitivity to them zero.

### 3.3.1 Results

The output convergence for each adaptive method as a function of total space-time degrees of freedom is shown in Fig. 3.4. We see that the adjoint-based (i.e. output-based) adaptation converges significantly faster than uniform refinement, requiring nearly two orders of magnitude fewer degrees of freedom to obtain an accurate output value. These gains relative to uniform refinement are impressive, but equally interesting is the performance of the output-based method relative to the residual-based method.

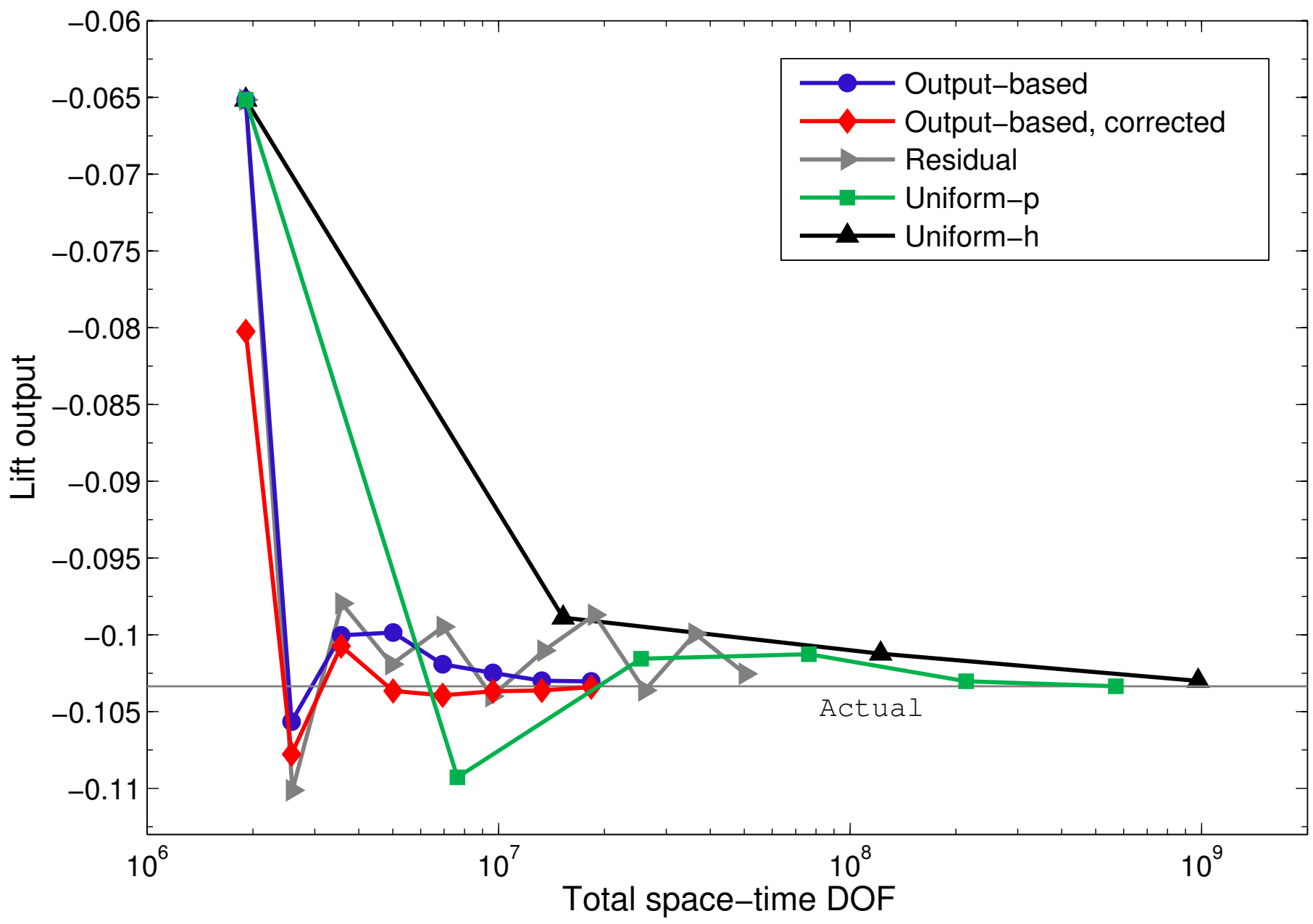

Figure 3.4: Output convergence for various adaptive methods. The output-based method converges significantly faster than the other methods. Note that the red curve is simply the blue curve plus the adjoint-based error correction, so that the gap between blue and red curves is the output error estimate.

The residual indicator targets regions of the domain where the governing equations are not well-satisfied, and hence usually performs well for static problems. However, in this case, its performance is erratic and no better than uniform refinement. The erratic behavior is primarily a consequence of the acoustic waves that emanate from the airfoils as they pitch back and forth. The residual indicator becomes distracted by these waves and exhausts degrees of freedom trying to resolve them as they propagate throughout the domain. The output-based method, on the other hand, deems the majority of these waves irrelevant to the output and does not expend resources on them.

The spatial and temporal meshes from the final output-based adaptation are shown in Figs. 3.5 and 3.6, respectively. We see that the near-airfoil and vortex shedding regions are targeted for adaptation, as well as a group of larger elements further away from the airfoils. While somewhat difficult to observe in the still-frames, the initial vortex shed from the first airfoil is heavily targeted throughout the simulation, since this vortex later collides with the second airfoil near the final time.

Contours of the adjoint (technically, contours of one of its state components) are shown alongside the entropy contours in Fig. 3.3. The time $t = 0.70T$ is the instant before the initial vortex is shed, and the large sensitivity of the output to this event can be seen in the adjoint contours. As the simulation proceeds, the output sensitivity gradually shifts from the first airfoil to the second, before collapsing upon the second airfoil at the final time. This collapse occurs because as the final time approaches, any errors made far away from the second airfoil no longer have time to propagate to it, making the output's sensitivity to them zero.

Some other aspects of the adjoint are worth pointing out. In the first two contours, the near-circular rings represent inward-moving (adjoint) acoustic waves, which converge upon a particular region as the simulation proceeds. The existence of a ring implies that an important event in space-time is about to occur, and any errors made within the circumference of the ring have the ability to influence this event. In this simulation, the important events tend to be instances of vortex shedding, and the rings converge on the trailing edge regions from which the vortices are shed. Lastly, between the two airfoils, a path can be seen tethering them together. This path appears because any errors generated within it ultimately reach the second airfoil via convection and thus directly affect the output.

Overall, this case demonstrates the reductions in computational cost that can be achieved for unsteady problems if an adjoint-based adaptive method is used. For the same level of output accuracy, the adjoint-based method requires a mesh that is nearly 100 times smaller than that required by the uniform-refinement strategies. Unsurprisingly, this reduction in required mesh size translates into a corresponding reduction in CPU time as well [10].

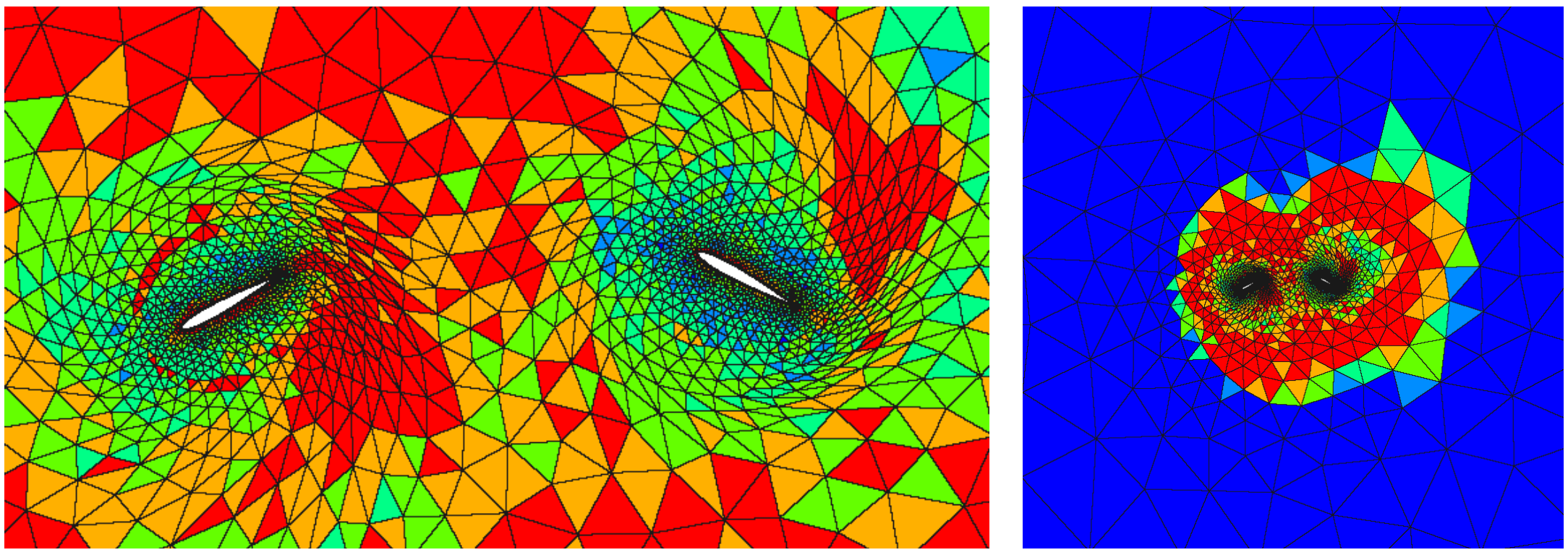

(a) $t = 0.70T$

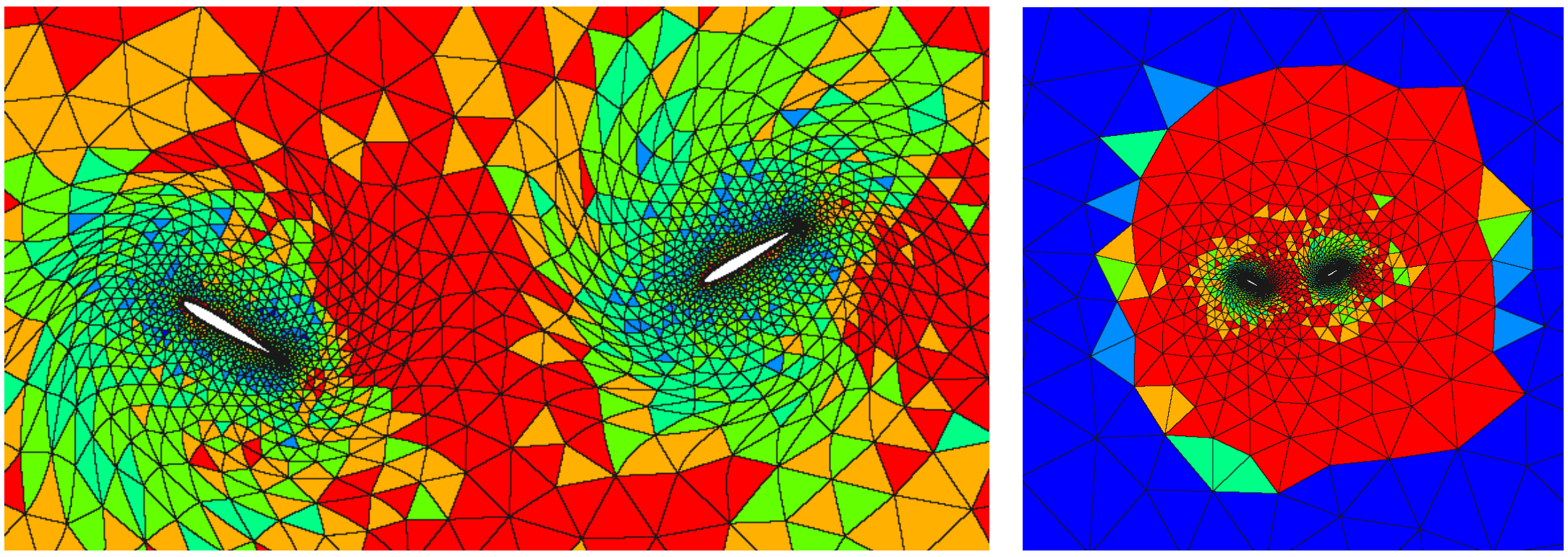

(b) $t = 1.25T$

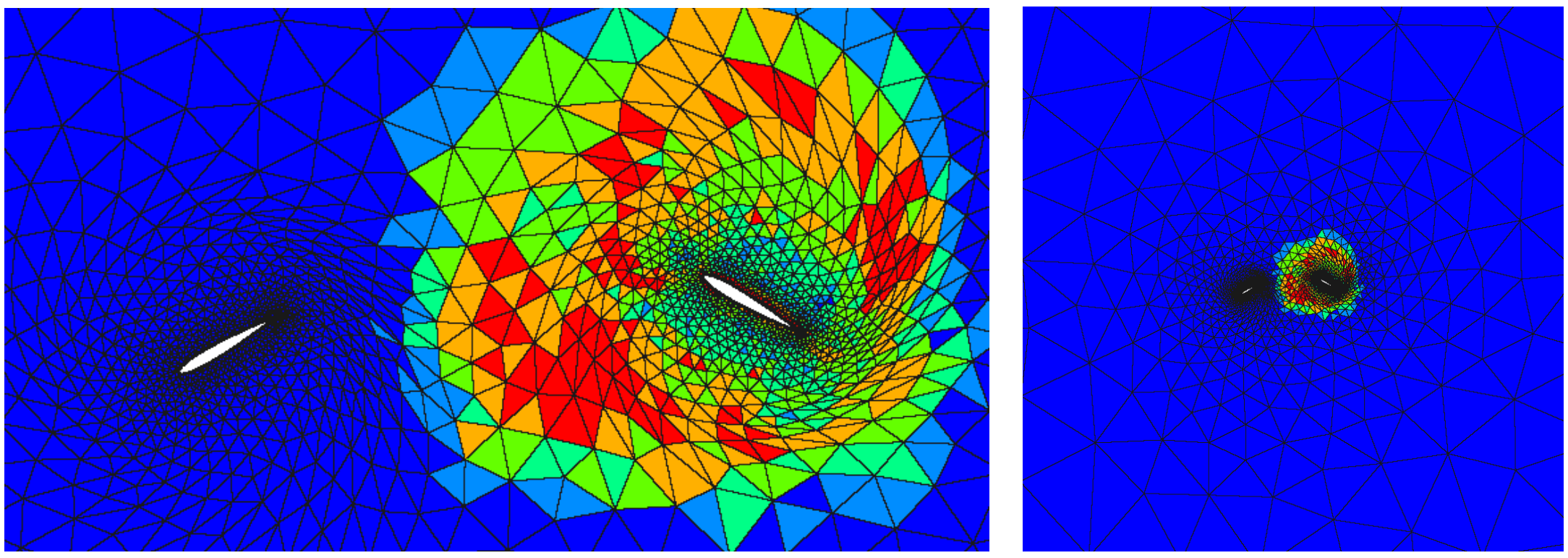

(c) $t = 2.75T$

Figure 3.5: Mesh during the final adpative iteration of the adjoint-based method, shown at various stages of the motion. Blue is $p = 0$, red is $p = 5$. The red regions are those deemed most important for obtaining an accurate lift output on the second airfoil.

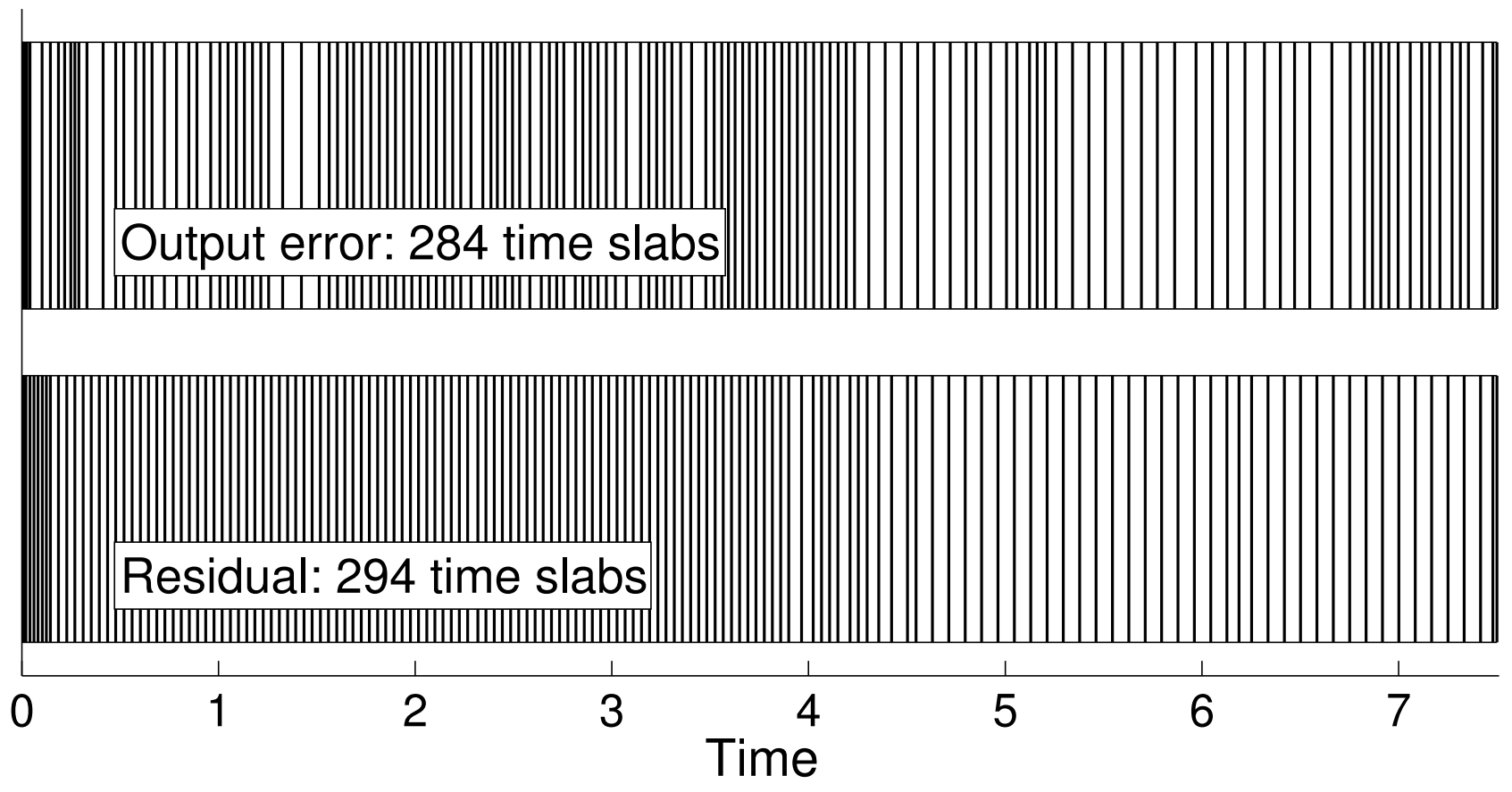

Figure 3.6: Temporal grids from one of the last adaptations of both output-based and residual-based methods. For clarity, only every other time slab is plotted. We see that the residual-based method refines primarily the early times, whereas the output-based method refines both early and final times. The final times are important since the output of interest is defined to be near the end of the simulation.

# APPENDICES

# APPENDIX A

# Additional Output Error Details, a Third-Order Estimate, Dual and Primal-Dual Forms

In this appendix, we derive some additional forms of adjoint-based output error estimates, including a version that has third-order (rather than second-order) accuracy.

## A.1 Dual Form of Output Error Estimate

Assume we are solving a differential equation $Lu = f$ and are interested in an output written as

$$J = (g, u)\,. \tag{A.1}$$

If we have some approximate solution $u_H$, the output is approximately

$$J(u_H) = (g, u_H)\,. \tag{A.2}$$

Recall from Sec. 2.4.1 that the amount of error in this output is then given by

$$\delta J(u_H) = -(r(u_H), \psi) \;=\; -\int_\Omega \psi\, r(u_H) dx\,, \tag{A.3}$$

where $r(u_H) = Lu_H - f$ is the residual evaluated with the approximate solution and $\psi$ is the (exact) adjoint for $J$, which satisfies the adjoint equation

$$L^*\psi = g\,. \tag{A.4}$$

While Eqn. A.3 is often the most useful, we can derive a slightly different form of the output error. In Chapter II, we showed how an output $J = (g, u)$ can be equivalently written in "dual form" as $J = (f, \psi)$. It stands to reason that if we can write the output itself in two ways, we should also be able to write the output *error* in two ways.

When deriving the so-called "primal" form of the output error (Eqn. A.3), we started with the output definition $J = (g, u)$ and proceeded from there. If we instead start from the relation $J = (f, \psi)$, we can derive a corresponding "dual" form of the output error.

Note that if we knew $\psi$ exactly, the error in the output would be zero, since we could compute the exact output value from the above dual form. However, assume that we do not know the exact adjoint, but instead know only an approximation, $\psi_h$. This $\psi_h$ could come from numerically approximating the adjoint equations on a given mesh. In general, $\psi_h$ may be computed in a different space (or on a different mesh) than $u_H$, so we use the subscript $h$ rather than $H$ to emphasize this potential difference. If $\psi_h$ happened to be computed in the same space as $u_H$, we would write $\psi_h = \psi_H$.

With $\psi_h$ in hand, the output can be computed as $J(\psi_h) = (f, \psi_h)$. The amount of error in this output can then be expressed as:

$$\begin{aligned}
\delta J(\psi_h) &= J(\psi) - J(\psi_h) \\
&= (f, \psi) - (f, \psi_h) && \text{(dual form of output)} \\
&= (f, \psi - \psi_h) && \text{(linearity of inner product)} \\
&= (Lu, \psi - \psi_h) && \text{(primal equation)} \\
&= (u, L^*(\psi - \psi_h)) && \text{(adjoint identity)} \\
&= (u, L^*\psi) - (u, L^*\psi_h) && \text{(linearity of inner product)} \\
&= (u, g) - (u, L^*\psi_h) && \text{(adjoint equation)} \\
&= -(u, L^*\psi_h - g) && \text{(linearity of inner product).} \qquad \text{(A.5)}
\end{aligned}$$

Now, we know that $\psi$ satisfies $L^*\psi - g = 0$, so the expression $L^*\psi_h - g$ is an ***adjoint residual***, in the same way that the quantitiy $Lu_H - f$ is a primal residual. If we define this adjoint residual to be:

$$r^*(\psi_h) \equiv L^*\psi_h - g\,,$$

then we can write the above form of the output error as

$$\delta J(\psi_h) = -(u, r^*(\psi_h)) \, .$$

Or, by symmetry of the inner product:

$$\boxed{\delta J(\psi_h) = -(r^*(\psi_h), u)} \, . \tag{A.6}$$

This is known as the **dual form of the output error**.

For comparison, recall that the primal form of the output error is

$$\boxed{\delta J(u_H) = -(r(u_H), \psi)} \, . \tag{A.7}$$

We see that the above expressions have the roles of $\psi$ and $u$ switched, but otherwise have the same form. Thus, there is a certain "duality" between them. In summary then: if we have computed a given output using $\psi_h$, the corresponding output error is given by Eqn. A.6. On the other hand, if we have computed the output using $u_H$, the corresponding output error is given by Eqn. A.7. Depending on how $\psi_h$ and $u_H$ were obtained, these errors could potentially differ.

In the next section, we will show that if $\psi_h$ is computed in the same space as $u_H$ (so that $\psi_h = \psi_H$) and a Galerkin-type finite element method is used, then the primal and dual forms of the output error are identical. In other words, $\delta J(\psi_H) = \delta J(u_H)$.

## A.2 Galerkin Methods: Equivalence of Primal and Dual Forms of Output Error

Above, we derived both a primal and dual form of the output error. For a primal problem $Lu = f$ with corresponding adjoint problem $L^*\psi = g$, we have the output errors:

$$\delta J(u_H) = J(u) - J(u_H) = -(r(u_H), \psi) \, ,$$

and

$$\delta J(\psi_h) = J(\psi) - J(\psi_h) = -(r^*(\psi_h), u) \, ,$$

where the residuals are given by $r(u_H) = Lu_H - f$ and $r^*(\psi_h) = L^*\psi_h - g$.

In general, since the $h$ and $H$ spaces may differ, there is no reason to think that these error values will be the same. Even if we use the same mesh to compute both primal and adjoint solutions, Eqns. A.6 and A.7 may give different values for $\delta J$ depending on the method used. However, it turns out that for Galerkin methods in particular, the two error values *are* in fact identical. In other words, if we compute $\psi_H$ and $u_H$ in the same space using a Galerkin method, then we will have $\delta J(u_H) = \delta J(\psi_H)$.

This is straightforward to show. The Galerkin formulation of both primal and adjoint problems is (respectively):

$$(Lu_H, v_H) = (f, v_H) \qquad \forall v_H \in \mathcal{V}_H \tag{A.8}$$
$$(L^*\psi_H, v_H) = (g, v_H) \qquad \forall v_H \in \mathcal{V}_H, \tag{A.9}$$

where $\mathcal{V}_H$ is a discrete space of choice.

Furthermore, the *exact* solutions $u$ and $\psi$ will satisfy

$$(Lu, v) = (f, v) \qquad \forall v \in \mathcal{V}, \tag{A.10}$$
$$(L^*\psi, v) = (g, v) \qquad \forall v \in \mathcal{V}, \tag{A.11}$$

where $\mathcal{V}$ is an appropriate continuous space. Note that in general we will have $\mathcal{V}_H \subset \mathcal{V}$.

Define the primal error to be $e = u - u_H$ and the adjoint error to be $e^* = \psi - \psi_H$. Next, since $\mathcal{V}_H$ is contained in $\mathcal{V}$, we can choose the test functions in A.10 and A.11 to *be* the discrete ones – i.e. we can take $v = v_H$. We can then subtract equations A.8 and A.9 from A.10 and A.11, respectively. This gives, for the primal problem:

$$\begin{aligned} (Lu, v_H) - (Lu_H, v_H) &= (f, v_H) - (f, v_H) && \forall v_H \in \mathcal{V}_H \\ (L(u - u_H), v_H) &= 0 && \forall v_H \in \mathcal{V}_H \\ (Le, v_H) &= 0 && \forall v_H \in \mathcal{V}_H, \end{aligned} \tag{A.12}$$

and for the adjoint problem:

$$\begin{aligned} (L^*\psi, v_H) - (L^*\psi_H, v_H) &= (g, v_H) - (g, v_H) && \forall v_H \in \mathcal{V}_H \\ (L^*(\psi - \psi_H), v_H) &= 0 && \forall v_H \in \mathcal{V}_H \\ (L^*e^*, v_H) &= 0 && \forall v_H \in \mathcal{V}_H. \end{aligned} \tag{A.13}$$

The relations A.12 and A.13 are statements of primal and adjoint Galerkin orthogonality. They say that the errors $e$ and $e^*$, after application of the primal and adjoint operators (respectively), are orthogonal to the discrete space $\mathcal{V}_H$.

Now, our goal is to show that for Galerkin methods, we have $\delta J(u_H) = \delta J(\psi_H)$. Since $\delta J(u_H) = (g, u - u_H) = (g, e)$ and $\delta J(\psi_H) = (f, \psi - \psi_H) = (f, e^*)$, we need to show that $(g, e) = (f, e^*)$.

Noting that both $e$ and $e^*$ are $\in \mathcal{V}$ (and hence can be used as test functions in Eqns. A.10 and A.11), we have:

$$
\begin{aligned}
(g, e) &= (L^*\psi, e) && \text{(adjoint equation)} \\
&= (Le, \psi) && \text{(adjoint identity and symmetry)} \\
&= (Le, \psi) - \underbrace{(Le, \psi_H)}_{0} && \text{(Galerkin orthogonality)} \\
&= (Le, \psi - \psi_H) && \text{(linearity of inner product)} \\
&= (Le, e^*) && \text{(definition of } e^*\text{)} \\
&= (L(u - u_H), e^*) && \text{(definition of } e\text{)} \\
&= (Lu, e^*) - (Lu_H, e^*) && \text{(linearity of inner product)} \\
&= (Lu, e^*) - \underbrace{(u_H, L^*e^*)}_{0} && \text{(adjoint identity)} \\
&= (Lu, e^*) && \text{(adjoint Galerkin orthogonality)} \\
&= (f, e^*) && \text{(primal equation)} \\
\implies \quad (g, e) &= (f, e^*)\,. && \text{(A.14)}
\end{aligned}
$$

Thus, we have shown that for Galerkin methods, both primal and dual forms of the error estimate are identical if $\psi_H$ and $u_H$ are computed in the same space. So we can write the single error esimate $\delta J$ as

$$\boxed{\delta J = -(r(u_H), \psi) \;=\; -(r^*(\psi_H), u)}\;. \tag{A.15}$$

Furthermore, as mentioned in Chapter II, by Galerkin orthogonality we can subtract coarse-space approximations $\psi_H$ and $u_H$ from the above estimates, resulting in the following equivalent forms:

$$\boxed{\delta J = -(r(u_H), \psi - \psi_H) \;=\; -(r^*(\psi_H), u - u_H)}\;. \tag{A.16}$$

## A.3 Continuous Error Estimation: Nonlinear Problems

Up until now, we have focused on linear problems. However, in practice, the most relevant problems are nonlinear. This raises the question: how do we perform error estimation for nonlinear problems? Can we use a similar adjoint-based strategy to approximate the error?

The answer is yes. We will show how this is done in the following sections.

### A.3.1 A Second-Order Nonlinear Error Estimate

Recall that the generalized adjoint equation (Eqn. 2.35) defines the adjoint as the function $\psi$ satisfying

$$J'_u(\delta u) = \int_\Omega \psi \, r'_u(\delta u) \, d\Omega \qquad \forall \, (\text{permissible}) \, \delta u \, , \tag{A.17}$$

where $J'_u$ and $r'_u$ denote the variations of the output and residual with respect to $u$, respectively. For a nonlinear problem, these variations must be taken about a particular state. If we assume that this state is an approximation denoted by $u_H$, then we can rewrite the adjoint equation as

$$J'_u[u_H](\delta u) = \int_\Omega \psi \, r'_u[u_H](\delta u) \, d\Omega \qquad \forall \, (\text{permissible}) \, \delta u \, . \tag{A.18}$$

We can use this adjoint definition to derive a second-order error estimate for a given output of interest, $J(u_H)$.

To obtain an output error estimate, we Taylor expand our true output $J(u)$ about the current state $u_H$ as follows:

$$J(u) \; \approx \; J(u_H) \; + \; J'_u[u_H](\delta u) \; + \; O(\delta u^2) \, . \tag{A.19}$$

From Eqn. A.18, the first-order term in this expansion can be written as

$$\begin{aligned} J'_u[u_H](\delta u) &= (\psi, \, r'_u[u_H](\delta u)) && \text{(inner product notation)} \\ &= (r'_u[u_H](\delta u), \, \psi) && \text{(symmetry of inner product)} \, . \end{aligned} \tag{A.20}$$

Thus, from Eqns. A.19 and A.20, we can write the output error as

$$J(u) - J(u_H) \; \approx \; (r'_u[u_H](\delta u), \, \psi) \; + \; O(\delta u^2) \, . \tag{A.21}$$

Our final step is to write the quantity $r'_u[u_H](\delta u)$ in a simpler form. To do this, we expand the true residual $r(u)$ about the current solution $u_H$, just as we did with $J$:

$$
\begin{aligned}
r(u) = 0 \;\approx\; r(u_H) \,+\, r'_u[u_H](\delta u) \,+\, O(\delta u^2) \\
\Longrightarrow\; r'_u[u_H](\delta u) \;\approx\; -r(u_H) \,+ O(\delta u^2)\,.
\end{aligned}
\tag{A.22}
$$

Substituting Eqn. A.22 into A.21 gives the second-order error estimate:

$$
\boxed{J(u) - J(u_H) \;\approx\; -(r(u_H), \psi) \,+\, R^{(2)}}\;, \tag{A.23}
$$

where $R^{(2)} = O(\delta u^2)$ is the remainder.

Now, once again, for a Galerkin discretization the weak form of the primal equation is

$$
\int_\Omega r(u_H)\, v_H \, dx \;=\; (r(u_H), v_H) \;=\; 0 \qquad \forall v_H \in \mathcal{V}_H\,. \tag{A.24}
$$

Therefore, if we have a coarse-space approximation to the adjoint, denoted by $\psi_H$, the quantity $(r(u_H), \psi_H)$ will be 0 and can be subtracted from the above error estimate to no effect, allowing us to write it as:

$$
\boxed{J(u) - J(u_H) \;\approx\; -(r(u_H), \psi - \psi_H) \,+\, R^{(2)}}\;. \tag{A.25}
$$

Overall, this error estimate looks exactly like the linear one derived in Chapter II, except that we now have an $O(\delta u^2)$ ***error*** in the error estimate.

A few points worth emphasizing are:

- It is important to note that the $\psi$ in the above expressions is the exact adjoint for the *inexact* linearization $r'_u[u_H](\delta u)$. In other words, for the current state $u_H$, it is the adjoint solution that would be obtained by solving the adjoint equations on an infinitely fine mesh. It is *not*, however, the "true" adjoint solution, unless the approximate solution $u_H$ happens to be the exact solution $u$.

- The $O(\delta u^2)$ error that comprises the remainder is a *linearization* error, which arises due to the fact that we kept only the first two terms in the Taylor expansions of $J(u)$ and $r(u)$. If $J(u)$ and $r(u)$ are both linear, this error will disappear, and we will recover our earlier (linear) form of the error estimate.

However, if either $J(u)$ or $r(u)$ is nonlinear, then $R^{(2)}$ will be nonzero, and the error estimate will be inexact.

- The question then arises: can we eliminate the $R^{(2)}$ remainder and obtain a more accurate error estimate? One obvious way to do this is to simply keep more terms in the expansions of $J(u)$ and $r(u)$. However, this is an expensive proposition and would require computation of the primal Hessian $r''_u[u_H](\delta u)$. It turns out that there is a better way to eliminate $R^{(2)}$. This is described in the next section.

### A.3.2 A Third-Order Nonlinear Error Estimate for Galerkin Methods

In this section, which follows the work of Becker and Rannacher [1], we describe how to eliminate the second-order remainder $R^{(2)}$ and hence obtain a third-order error estimate. We assume that a Galerkin discretization method is used, so that the equations being solved can be written as

$$\int_{\Omega} v_H \, r(u_H) \, d\Omega = 0 \qquad \forall v_H \in \mathcal{V}_H = \mathcal{U}_H \,. \tag{A.26}$$

In order to set the stage for the derivation, we will first need to review some calculus.

From basic calculus, the change of a function $f(x)$ over an interval $[a, b]$ can be written as:

$$f(b) - f(a) = \int_a^b f'(s) ds \,, \tag{A.27}$$

where the prime denotes a standard derivative. If we take $a = x$ and $b = x + \Delta x$, this becomes:

$$f(x + \Delta x) - f(x) = \int_x^{x+\Delta x} f'(s) ds. \tag{A.28}$$

Now, we can parameterize the variable $s$ by a new variable $t$, which goes from 0 to 1 over the interval $[x, x + \Delta x]$. Taking $s = x + t\Delta x$ and rewriting the above equation

in terms of $t$ gives:

$$f(x+\Delta x) - f(x) = \left[ \int_0^1 f'(\underbrace{x + t\Delta x}_{s})\, dt \right] \underbrace{\Delta x}_{\frac{ds}{dt}} . \tag{A.29}$$

Thus, we get the following representation for the change in the function $f$:

$$f(x+\Delta x) - f(x) = \left[ \int_0^1 f'(x + t\Delta x)\, dt \right] \Delta x \quad . \tag{A.30}$$

The term in brackets is the "mean value" of the slope of $f$ between $x$ and $x+\Delta x$. In other words, in a plot of $f$ vs. $x$, it is just the slope of the line connecting $f(x)$ and $f(x+\Delta x)$.

Above, $f$ was a function of the single variable $x$. If instead $f$ depends on two variables (say $x$ and $y$), we can write a similar formula for the change of $f$ between the points $(x, y)$ and $(x + \Delta x, y + \Delta y)$:

$$\begin{aligned} f(x+\Delta x, y+\Delta y) - f(x,y) &= \left[ \int_0^1 \frac{\partial f}{\partial x}(x + t\Delta x, y + t\Delta y)\, dt \right] \Delta x \\ &+ \left[ \int_0^1 \frac{\partial f}{\partial y}(x + t\Delta x, y + t\Delta y)\, dt \right] \Delta y \quad . \end{aligned} \tag{A.31}$$

Here, the variable $t$ again goes from 0 to 1 as we traverse the diagonal between $(x, y)$ and $(x + \Delta x, y + \Delta y)$.

We now need to make one further generalization. What if, rather than a function, $f$ is actually a *functional*, which depends now on two *functions* $x$ and $y$ (as opposed to coordinates)? Let us call this functional $F$ rather than $f$. Then our new goal is to find how $F$ changes when we perturb the functions $x$ and $y$ from the original "point" $(x, y)$ to some new "point" $(x + \delta x, y + \delta y)$. Note that we use a lowercase $\delta$ to denote that these perturbations are now *variations* rather than scalar values.

While functionals cannot be differentiated in the classic sense, they *do* have Fréchet derivatives. And we can use these Fréchet derivatives of $F(x, y)$ to determine how it changes when we go from $(x, y)$ to $(x + \delta x, y + \delta y)$. We denote the

Fréchet derivative of $F$ with respect to $x$ as

$$F'_x[\cdot,\cdot](\delta x), \tag{A.32}$$

and the Fréchet derivative of $F$ with respect to $y$ as

$$F'_y[\cdot,\cdot](\delta y), \tag{A.33}$$

where the terms in brackets denote the "point" in state space about which $F$ is being linearized. For example, the Fréchet derivative of $F$ with respect to $x$ at the point $(x_0, y_0)$ would be denoted by $F'_x[x_0, y_0](\delta x)$.

Since the Fréchet derivatives play a similar role as the multi-dimensional derivatives in Eqn. A.31, we can write our formula for the change in $F$ as follows:

$$\begin{aligned} F(x+\delta x, y+\delta y) - F(x,y) &= \int_0^1 F'_x\left[x + t\,\delta x, y + t\,\delta y\right](\delta x)\,dt \\ &+ \int_0^1 F'_y\left[x + t\,\delta x, y + t\,\delta y\right](\delta y)\,dt \end{aligned}\quad, \tag{A.34}$$

where again $t$ parameterizes the space between $(x, y)$ and $(x + \delta x, y + \delta y)$. We will use this formula to obtain a **third-order output error estimate**.

Assume that we have obtained a state approximation $u_H \in \mathcal{U}_H$ and an adjoint approximation $\psi_H \in \mathcal{U}_H$, which have been computed using Galerkin discretizations of the primal and adjoint problems, respectively. Note that the Galerkin discretization of the adjoint problem (linearized about $u_H$) is defined by

$$J'_u[u_H](\delta v) = \int_\Omega \psi_H\, r'_u[u_H](\delta v)\,d\Omega \qquad \forall\, \delta v \in \mathcal{U}_H\,, \tag{A.35}$$

for some discrete space $\mathcal{U}_H$.

In the following derivation, we will also make use of the *true* adjoint, $\psi$, which is associated with the exact solution $u$ and defined to satisfy

$$J'_u[u](\delta v) = \int_\Omega \psi\, r'_u[u](\delta v)\,d\Omega \qquad \forall\, \delta v \in \mathcal{U}\,. \tag{A.36}$$

Note that the definition of $\psi$ here is different than in the previous section, where $\psi$

represented the infinite-dimensional adjoint for the *inexact* state $u_H$.

To obtain an error estimate for some output of interest $J(u_H)$, we then construct two functionals, $F(u, \psi)$ and $F(u_H, \psi_H)$:

$$F(u,\psi) \equiv J(u) - \underbrace{\int_\Omega \psi\, r(u)\, d\Omega}_{0 \text{ since } r(u)=0} \tag{A.37}$$

$$F(u_H,\psi_H) \equiv J(u_H) - \underbrace{\int_\Omega \psi_H\, r(u_H)\, d\Omega}_{0 \text{ by Galerkin orthogonality}} \tag{A.38}$$

Since the adjoint-weighted residual terms in the above equations are exactly zero, we have that:

$$F(u,\psi) - F(u_H,\psi_H) = J(u) - J(u_H)\,.$$

Next, we define the difference between the true and approximate adjoints to be $\delta\psi = \psi - \psi_H$ and the difference between the true and approximate states to be $\delta u = u - u_H$. We then have that $\psi = \psi_H + \delta\psi$ and $u = u_H + \delta u$, which means the above formula can be written:

$$F(u_H + \delta u, \psi_H + \delta\psi) - F(u_H,\psi_H) = J(u) - J(u_H)\,.$$

We have now expressed the output error in terms of a change in some functional $F$, and from Eqn. A.34, we know a formula for computing this change. Just swapping $[x, y] \to [u_H, \psi_H]$ and $[\delta x, \delta y] \to [\delta u, \delta\psi]$ in Eqn. A.34 allows us to write:

$$\begin{aligned} J(u) - J(u_H) = &\int_0^1 F'_u\,[u_H + t\,\delta u, \psi_H + t\,\delta\psi]\,(\delta u)\,dt \\ &+ \int_0^1 F'_\psi\,[u_H + t\,\delta u, \psi_H + t\,\delta\psi]\,(\delta\psi)\,dt \end{aligned}\quad . \tag{A.39}$$

Now the question is: how should we compute the integrals in this expression? Since in general we cannot evaluate them analytically, we will instead approximate them using the trapezoidal rule. For a classic function $f$, the trapezoidal rule over the interval $[a, b]$ is shown in blue below:

$$\int_a^b f(x)\,dx = \frac{1}{2}\left[f(a) + f(b)\right] \underbrace{-\frac{1}{2}\int_a^b f''(x)(b-x)(x-a)\,dx}_{\text{remainder}} \tag{A.40}$$

$$= \frac{1}{2}\left[f(a) + f(b)\right] \underbrace{-\frac{\Delta x^3}{12} f''(\bar{x})}_{\text{remainder}} \tag{A.41}$$

(Here, either form of the remainder may be used, and $\Delta x = b - a$ while $\bar{x}$ is a particular point inside the interval.) From Eqn. A.41, we see that the trapezoidal rule gives a **third-order** approximation to the integral by using only the "endpoint" values of $f$.

We can use the trapezoidal rule to approximate each of the Fréchet derivatives in the above output error equation. The "endpoints" of the intervals correspond to parameter values of $t = 0$ and $t = 1$, so we have:

$$\int_0^1 F'_u\left[u_H + t\,\delta u, \psi_H + t\,\delta\psi\right](\delta u)\,dt \approx \frac{1}{2}\Big[F'_u[u_H, \psi_H](\delta u) + F'_u[u, \psi](\delta u)\Big]$$

$$\int_0^1 F'_\psi\left[u_H + t\,\delta u, \psi_H + t\,\delta\psi\right](\delta\psi)\,dt \approx \frac{1}{2}\Big[F'_\psi[u_H, \psi_H](\delta\psi) + F'_\psi[u, \psi](\delta\psi)\Big]$$

With these equations, we have converted our integrals into simple point evaluations of the Fréchet derivatives at $(u_H, \psi_H)$ and $(u, \psi)$. The only thing that remains is to actually write out the derivatives at these points. We can obtain these by just taking variations of Eqns. A.37 and A.38 with respect to $u$ and $\psi$, then inserting our particular values of $\delta u$ and $\delta\psi$. The four terms in the above trapezoidal approxima-

tions then become:

$$F'_u[u_H, \psi_H](\delta u) = J'_u[u_H](\delta u) - \int_\Omega \psi_H \, r'_u[u_H](\delta u) \, d\Omega \tag{A.42}$$

$$F'_u[u, \psi](\delta u) = J'_u[u](\delta u) - \int_\Omega \psi \, r'_u[u](\delta u) \, d\Omega \tag{A.43}$$

$$F'_\psi[u_H, \psi_H](\delta \psi) = - \int_\Omega r(u_H) \, \delta\psi \, d\Omega \tag{A.44}$$

$$F'_\psi[u, \psi](\delta \psi) = - \int_\Omega r(u) \, \delta\psi \, d\Omega \tag{A.45}$$

Now, looking at these expressions, we see that the last term (Eqn. A.45) is exactly zero, since the residual $r(\cdot)$ evaluated with the exact solution $u$ vanishes. Likewise, Eqn. A.43 is also zero. This is because the exact adjoint equation (Eqn. A.36) holds for *all* $\delta v \in \mathcal{U}$, so it must hold for our particular $\delta u \in \mathcal{U}$ as well.

So out of our four original terms, we are left with only two that are nonzero. Thus, our error estimate in Eqn. A.39, after approximation by trapezoidal rule, can be written as:

$$\begin{aligned} J(u) - J(u_H) &\approx \frac{1}{2}\Big[ F'_u[u_H, \psi_H](\delta u) + \underbrace{F'_u[u, \psi](\delta u)}_{=0} \Big] \\ &\quad + \frac{1}{2}\Big[ F'_\psi[u_H, \psi_H](\delta \psi) + \underbrace{F'_\psi[u, \psi](\delta \psi)}_{=0} \Big] \\ &\approx \frac{1}{2}\Big[ F'_u[u_H, \psi_H](\delta u) + F'_\psi[u_H, \psi_H](\delta \psi) \Big] \\ &\approx \frac{1}{2} F'_u[u_H, \psi_H](\delta u) \; - \; \frac{1}{2} \left( r(u_H), \delta\psi \right) \, . \end{aligned} \tag{A.46}$$

Now, let us look more closely at the $F'_u[u_H, \psi_H](\delta u)$ term in the above expression, which is given by Eqn. A.42 as

$$F'_u[u_H, \psi_H](\delta u) = J'_u[u_H](\delta u) - (\psi_H, r'_u[u_H](\delta u)) \, . \tag{A.47}$$

We will show that this term can actually be written as an adjoint residual weighted by $\delta u$.

Since the $r'_u[u_H](\delta u)$ term in the above expression represents the linearized residual, it can be rewritten as simply some linear operator $L$ applied to the perturbation

$\delta u$. Thus, we have that

$$r'_u[u_H](\delta u) = L\delta u \,. \tag{A.48}$$

Now, we can define the adjoint operator ($L^*$) to this linear operator $L$ in exactly the same way as usual, i.e. via the identity

$$(Lu, v) \;=\; (u, L^* v) \qquad\qquad \forall u, v \in \mathcal{U} \,. \tag{A.49}$$

Furthermore, since $\delta u \in \mathcal{U}$, this implies that

$$(L\delta u, v) \;=\; (\delta u, L^* v) \qquad\qquad \forall v \in \mathcal{U} \,. \tag{A.50}$$

Inserting $r'_u[u_H](\delta u)$ back in for $L\delta u$ then gives

$$(r'_u[u_H](\delta u), v) \;=\; (\delta u, L^* v) \qquad\qquad \forall v \in \mathcal{U} \,. \tag{A.51}$$

Finally, since $\psi_H \in \mathcal{U}$, using the above relation, the last term in Eqn. A.47 can be rewritten as

$$(r'_u[u_H](\delta u), \psi_H) \;=\; (\delta u, L^* \psi_H) \;, \tag{A.52}$$

or

$$\boxed{(\psi_H, r'_u[u_H](\delta u)) \;=\; (L^* \psi_H, \delta u)} \;. \tag{A.53}$$

Next, let us rewrite the $J'_u[u_H](\delta u)$ term in Eqn. A.47 in a slightly different form as well. If we assume the output $J$ is defined as

$$J(u) = \int_\Omega j(u)\, d\Omega \,, \tag{A.54}$$

where $j(u)$ is some nonlinear function of our choosing (e.g. $j(u) = u^2$), then the Fréchet linearization of this output (about $u_H$) can be written as

$$\boxed{J'_u[u_H](\delta u) = (j'[u_H], \delta u)} \;. \tag{A.55}$$

Recall that for linear problems, our output was defined to be $J = (g, u)$ and its variation was just $J'_u(\delta u) = (g, \delta u)$. Then, comparing to the above equation, we see that $j'[u_H]$ is effectively our "$g$" for a nonlinear problem.

Inserting Eqns. A.55 and A.53 back into Eqn. A.47 gives

$$\begin{aligned} F'_u[u_H, \psi_H](\delta u) &= (j'[u_h], \delta u) - (L^* \psi_H, \delta u) \\ &= -(L^* \psi_H - j'[u_H], \delta u) \,. \end{aligned} \tag{A.56}$$

Now, just as the continuous adjoint equation is defined as

$$L^* \psi = g \tag{A.57}$$

for a linear problem, it is straightforward to verify that the continuous adjoint equation is defined as

$$L^* \psi = j'_u[u_H] \tag{A.58}$$

for a nonlinear problem. (This follows from the fact that, as discussed, $j'_u[u_H]$ plays effectively the same role as $g$.) Thus, for a nonlinear problem, the adjoint *residual* can be defined as

$$r^*[u_H](\psi) \equiv L^* \psi - j'_u[u_H] \,. \tag{A.59}$$

Looking back at Eqn. A.56, we see then that it can be rewritten as

$$\boxed{F'_u[u_H, \psi_H](\delta u) = -(r^*[u_H](\psi_H), \delta u)} \,, \tag{A.60}$$

i.e. as an *adjoint residual* weighted by $\delta u$. This is exactly what we sought to show.

Inserting the above expression (Eqn. A.60) back into the original output error estimate (Eqn. A.46) and including the remainder term from the trapezoidal rule then yields

$$J(u) - J(u_H) \approx -\frac{1}{2} (r(u_H), \delta\psi) \; - \; \frac{1}{2} (r^*[u_H](\psi_H), \delta u) \; + \; R^{(3)} \,. \tag{A.61}$$

Finally, expanding $\delta u$ and $\delta\psi$, the output error estimate becomes:

$$\boxed{J(u) - J(u_H) \approx -\frac{1}{2} (r(u_H), \psi - \psi_H) - \frac{1}{2} (r^*[u_H](\psi_H), u - u_H) + R^{(3)}} \,. \tag{A.62}$$

We have now done what we set out to do – we have derived a third-order accurate error estimate for our output of interest. By including both primal and adjoint terms in the error estimate, we have taken advantage of the duality between these two vectors, allowing us to obtain a more accurate estimate than if we had used the primal form alone.

Some further observations about the error estimate are:

- The error estimate depends on the perturbations $\psi - \psi_H$ and $u - u_H$. In practice, we do not know $\psi$ or $u$, so we must approximate them in some way. This can be done, for example, by smoothing coarse approximations to the state and adjoint on a finer mesh. Note that for linear Galerkin problems, our error estimate also involved a factor of $\psi - \psi_H$. In that case, we could approximate $\psi$ by just solving the adjoint problem on a finer mesh. However, for nonlinear problems, the adjoint equations themselves depend on the current value of the primal state. Thus, to approximate the true $\psi$ in the $\psi - \psi_H$ term, we could consider approximating the adjoint equations on *both* a finer mesh and using a better approximation to the state.

- In the above analysis, we have claimed that the trapezoidal rule is third-order accurate. However, it is typically thought of as a second-order method. So which is correct? The answer is that it is second-order *globally*, but third-order *locally*. In other words, if we partition our domain of integration into subintervals, then the trapezoidal rule will be third-order accurate over each subinterval, but when we sum all of the errors, the *total* error will be second-order. This is the same thing that happens, for example, when numerically integrating ODEs in time. The local accuracy on each time step is one order higher than the global accuracy of the time scheme. In the case of our error estimate, the important point is that we have only one interval (from $(u_H, \psi_H)$ to $(u_H + \delta u, \psi_H + \delta\psi)$), so we obtain third-order accuracy over that interval.

- We have said that the remainder $R^{(3)}$ is third-order in the adjoint/state perturbations, but we can give a more detailed description of this term. From the trapezoidal rule in Eqn. A.40, the remainder is given by:

$$-\frac{1}{2}\int_a^b f''(x)(b-x)(x-a)\,dx.$$

Now, in our output error estimation, the term we approximated with the trapezoidal rule (in Eqn. A.39) was $F'_u[\cdot,\cdot](\delta u) + F'_\psi[\cdot,\cdot](\delta\psi)$. So this is our effective

"$f(x)$" in the above formula. Thus, to compute the remainder, we need to take two more derivatives of this quantity. Noting that our interval ranges from $a = 0$ to $b = 1$, we can write our output error remainder as:

$$R^{(3)} = -\frac{1}{2}\int_0^1 \underbrace{\mathcal{D}^{(2)}\left[F'_u[\cdot,\cdot](\delta u) + F'_\psi[\cdot,\cdot](\delta\psi)\right]}_{\text{effective } f''}(1-t)\,t\,dt. \tag{A.63}$$

Here, for ease of notation, we simply use dots $[\cdot,\cdot]$ to denote the states about which the linearization is performed. These dots actually represent the states $[u_H + t\delta u, \psi_H + t\delta\psi]$. Furthermore, note that $\mathcal{D}^{(2)}$ denotes a full second Fréchet derivative, which requires linearizations with respect to both $u$ and $\psi$. In other words, we can write the **first** Fréchet derivative of our "$f(x)$" term as:

$$\begin{aligned} f'[\cdot,\cdot] \equiv \mathcal{D}^{(1)}\left[F'_u[\cdot,\cdot](\delta u) + F'_\psi[\cdot,\cdot](\delta\psi)\right] =& F''_{uu}[\cdot,\cdot](\delta u^2) \\ &+F''_{u\psi}[\cdot,\cdot](\delta u\,\delta\psi) \\ &+F''_{\psi u}[\cdot,\cdot](\delta\psi\,\delta u) \\ &+F''_{\psi\psi}[\cdot,\cdot](\delta\psi^2), \end{aligned} \tag{A.64}$$

and the **second** Fréchet derivative as:

$$\begin{aligned} f''[\cdot,\cdot] \equiv \mathcal{D}^{(2)}\left[F'_u[\cdot,\cdot](\delta u) + F'_\psi[\cdot,\cdot](\delta\psi)\right] =& F'''_{uuu}[\cdot,\cdot](\delta u^3) \\ &+F'''_{uu\psi}[\cdot,\cdot](\delta u^2\,\delta\psi) \\ &+F'''_{u\psi u}[\cdot,\cdot](\delta u^2\,\delta\psi) \\ &+F'''_{u\psi\psi}[\cdot,\cdot](\delta u\,\delta\psi^2) \\ &+F'''_{\psi uu}[\cdot,\cdot](\delta\psi\,\delta u^2) \\ &+F'''_{\psi u\psi}[\cdot,\cdot](\delta\psi^2\,\delta u) \\ &+F'''_{\psi\psi u}[\cdot,\cdot](\delta\psi^2\,\delta u) \\ &+F'''_{\psi\psi\psi}[\cdot,\cdot](\delta\psi^3). \end{aligned} \tag{A.65}$$

This second Fréchet derivative is the term we are interested in. Now, recall that from Eqns. A.37 and A.38, our functional $F$ depends only *linearly* on $\psi$. This means that any terms above containing more than one derivative with respect to $\psi$ will be zero. Eliminating these terms and combining equivalent mixed partials causes our second derivative to reduce to:

$$f''[\cdot,\cdot] \;=\; F'''_{uuu}[\cdot,\cdot](\delta u^3) \,+\, 3F'''_{uu\psi}[\cdot,\cdot](\delta u^2\,\delta\psi). \tag{A.66}$$

Finally, writing out these variations of $F$ explicitly (based on its definition in Eqn. A.37) gives:

$$f''[u,\psi] \;=\; \underbrace{J'''_{uuu}[u](\delta u^3) - \int\limits_{\Omega} \psi\, r'''_{uuu}[u](\delta u^3)\, dx}_{F'''_{uuu}} \; \underbrace{-3\int\limits_{\Omega} \delta\psi\, r''_{uu}[u](\delta u^2)\, dx}_{3F'''_{uu\psi}} \tag{A.67}$$

Here, we have inserted $[u,\psi]$ as the state about which we are linearizing, just to avoid confusion.

Finally, inserting this $f''$ into our expression for the remainder gives:

$$R^{(3)} = -\frac{1}{2}\int\limits_0^1 \Bigg[J'''_{uuu}[u](\delta u^3) - \int\limits_{\Omega} \psi\, r'''_{uuu}[u](\delta u^3)\, dx \\ -3\int\limits_{\Omega} \delta\psi\, r''_{uu}[u](\delta u^2)\, dx\Bigg](1-t)\,t\,dt \tag{A.68}$$

Again, for simplicity here we are using the states $[u,\psi]$ to represent the actual parameterized states $[u_H + t\delta u, \psi_H + t\delta\psi]$.

With the form of the remainder derived, it is useful to consider what it means in practice. First, it says that if the output and primal equations are *linear* (so that the above derivatives vanish), then the remainder is zero and the output error estimate is exact. This is expected, since we previously showed that the error estimates for linear Galerkin problems are exact. Also, it follows from this (and from the equivalence of primal and dual forms for Galerkin methods) that for linear problems, the combined primal-dual error estimate will reduce precisely to the single adjoint-weighted residual estimate derived previously.

We also see that, if $J$ is quadratic in $u$ (so that its third derivative vanishes) then the corresponding term in the remainder will vanish. Thus, if our output is **quadratic** and the primal equations **linear**, the third-order error estimate will again be exact. However, for any nonlinear primal residual, there will always be at least one nonzero remainder term, and the error estimate will not be exact.

Note also that since all terms in the above equation involve $\delta u$, but only one involves $\delta\psi$, there is an asymmetry. This asymmetry means that if we have the exact *primal* solution, then the remainder (as well as the error estimate as a whole) will reduce to zero; however, if we have the exact *adjoint* solution, then both the output error and the remainder may still be nonzero. This reflects the fact that, while for linear problems the adjoint and primal problems were independent (and therefore of equal "importance"), for nonlinear problems the adjoint is strictly dependent on the primal problem, and hence of less fundamental importance for computing the output. (Another way of saying this is that for nonlinear problems, outputs cannot be written in an equivalent "dual" form, so more information is required to compute the output than just the adjoint.)

- Finally, after deriving the third-order error estimate, we might wonder: **is it actually worth using?** Or is the second-order form typically good enough?

  The remainder terms of both the second- and third-order error estimates depend on the perturbations $\delta u = u - u_H$ and $\delta\psi = \psi - \psi_H$. Since $u_H$ and $\psi_H$ are assumed to be approximated on a coarse mesh with order $p$, we expect them to converge at order $p+1$. Therefore, the exected rates for the remainder terms are:

$$R^{(2)} = O((\delta u)^2) = O((\delta\psi)^2) \sim ((\Delta x)^{p+1})^2 \sim (\Delta x)^{2p+2} \quad \text{(A.69)}$$
$$R^{(3)} = O((\delta u)^3) = O((\delta\psi)^3) \sim ((\Delta x)^{p+1})^3 \sim (\Delta x)^{3p+3}. \quad \text{(A.70)}$$

  Now, given some coarse-space output and corresponding error estimate, we can write "corrected" forms of the output as:

$$J^{(2)}_{\text{corrected}} = J(u_H) + \delta J^{(2)}_{\text{est}} = J(u_h) + R^{(2)} \quad \text{(A.71)}$$
$$J^{(3)}_{\text{corrected}} = J(u_H) + \delta J^{(3)}_{\text{est}} = J(u_h) + R^{(3)} \quad \text{(A.72)}$$

  Here, $\delta J^{(2)}_{\text{est}}$ refers to the second-order form of the error estimate, while $\delta J^{(3)}_{\text{est}}$ refers to the third-order form. Also, we assume that in computing these error estimates, a "fine" space has been used to approximate the true $u$ and $\psi$ (the fine-space state is denoted by $u_h$ above). For this reason, our corrected output is actually an approximation to $J(u_h)$ rather than the true $J(u)$.

  Looking at the above equations, we see that in general, the convergence rate of the corrected output will be limited by the rate of *either* $J(u_h)$ or $R$. So the question is, which of these quantities is more restrictive?

Consider the case where the output converges at a **standard** $p+1$ rate – i.e. it does not superconverge. Also, assume that our fine space is a $p+1$ space, as is often used in practice. Then we have the following expected rates:

$$\begin{aligned} J(u_h) &\sim (\Delta x)^{p+2} \\ R^{(2)} &\sim (\Delta x)^{2p+2} \\ R^{(3)} &\sim (\Delta x)^{3p+3} \end{aligned}$$

Here, we see that both $R^{(2)}$ and $R^{(3)}$ converge at a higher order than $J(u_h)$. Thus, $J_{\text{corrected}}$ will achieve the same convergence rate as the fine-space output, regardless of whether the second- or third-order form of the error estimate is used. So we see that in this case, it is not necessarily worth using the third-order estimate.

However, now consider the case where our output **superconverges** at a rate of $2p+1$, as is often true of integral outputs. Then our new expected rates are:

$$\begin{aligned} J(u_h) &\sim (\Delta x)^{2(p+1)+1} \sim (\Delta x)^{2p+3} \\ R^{(2)} &\sim (\Delta x)^{2p+2} \\ R^{(3)} &\sim (\Delta x)^{3p+3} \end{aligned}$$

In this case, we see that the rate of $R^{(2)}$ is one order lower than $J(u_h)$. This means that if we were to use the second-order form of the error estimate, our corrected output would converge sub-optimally, at a lower order than $J(u_h)$. On the other hand, using the third-order error estimate would allow us to achieve the optimal rate of $J(u_h)$.